%
\expandafter\ifx\csname fmtname\endcsname\relax\input plain\fi
\def\texinfoversion{2005-01-30.17}

\message{Loading texinfo [version \texinfoversion]:}

\everyjob{\message{[Texinfo version \texinfoversion]}%
  \catcode`+=\active \catcode`\_=\active}

\message{Basics,}
\chardef\other=12

\let\+ = \relax

\let\ptexb=\b
\let\ptexbullet=\bullet
\let\ptexc=\c
\let\ptexcomma=\,
\let\ptexdot=\.
\let\ptexdots=\dots

\let\ptexequiv=\equiv
\let\ptexexclam=\!

\let\ptexgtr=>
\let\ptexhat=^
\let\ptexi=\i
\let\ptexindent=\indent
\let\ptexinsert=\insert
\let\ptexlbrace=\{
\let\ptexless=<

\let\ptexnoindent=\noindent
\let\ptexplus=+
\let\ptexrbrace=\}
\let\ptexslash=\/
\let\ptexstar=\*
\let\ptext=\t

\newlinechar = `^^J

%
\ifx\inputlineno\thisisundefined
  \let\linenumber = \empty 
\else
  \def\linenumber{l.\the\inputlineno:\space}
\fi

\ifx\putwordAppendix\undefined  \gdef\putwordAppendix{Appendix}\fi
\ifx\putwordChapter\undefined   \gdef\putwordChapter{Chapter}\fi
\ifx\putwordfile\undefined      \gdef\putwordfile{file}\fi
\ifx\putwordin\undefined        \gdef\putwordin{in}\fi
\ifx\putwordIndexIsEmpty\undefined     \gdef\putwordIndexIsEmpty{(Index is empty)}\fi
\ifx\putwordIndexNonexistent\undefined \gdef\putwordIndexNonexistent{(Index is nonexistent)}\fi
\ifx\putwordInfo\undefined      \gdef\putwordInfo{Info}\fi
\ifx\putwordInstanceVariableof\undefined \gdef\putwordInstanceVariableof{Instance Variable of}\fi
\ifx\putwordMethodon\undefined  \gdef\putwordMethodon{Method on}\fi
\ifx\putwordNoTitle\undefined   \gdef\putwordNoTitle{No Title}\fi
\ifx\putwordof\undefined        \gdef\putwordof{of}\fi
\ifx\putwordon\undefined        \gdef\putwordon{on}\fi
\ifx\putwordpage\undefined      \gdef\putwordpage{page}\fi
\ifx\putwordsection\undefined   \gdef\putwordsection{section}\fi
\ifx\putwordSection\undefined   \gdef\putwordSection{Section}\fi
\ifx\putwordsee\undefined       \gdef\putwordsee{see}\fi
\ifx\putwordSee\undefined       \gdef\putwordSee{See}\fi
\ifx\putwordShortTOC\undefined  \gdef\putwordShortTOC{Short Contents}\fi
\ifx\putwordTOC\undefined       \gdef\putwordTOC{Table of Contents}\fi
\ifx\putwordMJan\undefined \gdef\putwordMJan{January}\fi
\ifx\putwordMFeb\undefined \gdef\putwordMFeb{February}\fi
\ifx\putwordMMar\undefined \gdef\putwordMMar{March}\fi
\ifx\putwordMApr\undefined \gdef\putwordMApr{April}\fi
\ifx\putwordMMay\undefined \gdef\putwordMMay{May}\fi
\ifx\putwordMJun\undefined \gdef\putwordMJun{June}\fi
\ifx\putwordMJul\undefined \gdef\putwordMJul{July}\fi
\ifx\putwordMAug\undefined \gdef\putwordMAug{August}\fi
\ifx\putwordMSep\undefined \gdef\putwordMSep{September}\fi
\ifx\putwordMOct\undefined \gdef\putwordMOct{October}\fi
\ifx\putwordMNov\undefined \gdef\putwordMNov{November}\fi
\ifx\putwordMDec\undefined \gdef\putwordMDec{December}\fi
\ifx\putwordDefmac\undefined    \gdef\putwordDefmac{Macro}\fi
\ifx\putwordDefspec\undefined   \gdef\putwordDefspec{Special Form}\fi
\ifx\putwordDefvar\undefined    \gdef\putwordDefvar{Variable}\fi
\ifx\putwordDefopt\undefined    \gdef\putwordDefopt{User Option}\fi
\ifx\putwordDeffunc\undefined   \gdef\putwordDeffunc{Function}\fi

\chardef\colonChar = `\:
\chardef\commaChar = `\,
\chardef\dotChar   = `\.
\chardef\exclamChar= `\!
\chardef\questChar = `\?
\chardef\semiChar  = `\;
\chardef\underChar = `\_

\chardef\spaceChar = `\ %
\chardef\spacecat = 10
\def\spaceisspace{\catcode\spaceChar=\spacecat}

%
\def\gobble#1{}

\def\makecsname#1{\expandafter\noexpand\csname#1\endcsname}

\hyphenation{
  Flor-i-da Ghost-script Ghost-view Mac-OS Post-Script
  ap-pen-dix bit-map bit-maps
  data-base data-bases eshell fall-ing half-way long-est man-u-script
  man-u-scripts mini-buf-fer mini-buf-fers over-view par-a-digm
  par-a-digms rath-er rec-tan-gu-lar ro-bot-ics se-vere-ly set-up spa-ces
  spell-ing spell-ings
  stand-alone strong-est time-stamp time-stamps which-ever white-space
  wide-spread wrap-around
}

\newdimen\bindingoffset
\newdimen\normaloffset
\newdimen\pagewidth \newdimen\pageheight

%

%
\def\|{%
  \leavevmode
  %
  \vadjust{%
    \vskip-\baselineskip
    %
    \llap{%
      %
      \vrule height\baselineskip width1pt
      %
      \hskip 12pt
    }%
  }%
}

%
%
\def\loggingall{%
  \tracingstats2
  \tracingpages1
  \tracinglostchars2  
  \tracingparagraphs1
  \tracingoutput1
  \tracingmacros2
  \tracingrestores1
  \showboxbreadth\maxdimen \showboxdepth\maxdimen
  \ifx\eTeXversion\undefined\else 
    \tracingscantokens1
    \tracingifs1
    \tracinggroups1
    \tracingnesting2
    \tracingassigns1
  \fi
  \tracingcommands3  
  \errorcontextlines16
}%

%
\def\smallbreak{\ifnum\lastpenalty<10000\par\ifdim\lastskip<\smallskipamount
  \removelastskip\penalty-50\smallskip\fi\fi}
\def\medbreak{\ifnum\lastpenalty<10000\par\ifdim\lastskip<\medskipamount
  \removelastskip\penalty-100\medskip\fi\fi}
\def\bigbreak{\ifnum\lastpenalty<10000\par\ifdim\lastskip<\bigskipamount
  \removelastskip\penalty-200\bigskip\fi\fi}

%
\newif\ifcropmarks

%
%
\newdimen\outerhsize \newdimen\outervsize 
\newdimen\cornerlong  \cornerlong=1pc
\newdimen\cornerthick \cornerthick=.3pt
\newdimen\topandbottommargin \topandbottommargin=.75in

\chardef\PAGE = 255
\output = {\onepageout{\pagecontents\PAGE}}

\newbox\headlinebox
\newbox\footlinebox

\def\onepageout#1{%
  \ifcropmarks \hoffset=0pt \else \hoffset=\normaloffset \fi
  \ifodd\pageno  \advance\hoffset by \bindingoffset
  \else \advance\hoffset by -\bindingoffset\fi
  %
  \setbox\headlinebox = \vbox{\let\hsize=\pagewidth \makeheadline}%
  \setbox\footlinebox = \vbox{\let\hsize=\pagewidth \makefootline}%
  {%
    %
    \escapechar = `\\     
    \indexdummies         
    \normalturnoffactive  
    \shipout\vbox{%
      \ifpdfmakepagedest \pdfdest name{\the\pageno} xyz\fi
      \ifcropmarks \vbox to \outervsize\bgroup
        \hsize = \outerhsize
        \vskip-\topandbottommargin
        \vtop to0pt{%
          \line{\ewtop\hfil\ewtop}%
          \nointerlineskip
          \line{%
            \vbox{\moveleft\cornerthick\nstop}%
            \hfill
            \vbox{\moveright\cornerthick\nstop}%
          }%
          \vss}%
        \vskip\topandbottommargin
        \line\bgroup
          \hfil 
          \ifodd\pageno\hskip\bindingoffset\fi
          \vbox\bgroup
      \fi
      \unvbox\headlinebox
      \pagebody{#1}%
      \ifdim\ht\footlinebox > 0pt
        \vskip 2\baselineskip
        \unvbox\footlinebox
      \fi
      \ifcropmarks
          \egroup 
        \hfil\egroup 
        \vskip\topandbottommargin plus1fill minus1fill
        \boxmaxdepth = \cornerthick
        \vbox to0pt{\vss
          \line{%
            \vbox{\moveleft\cornerthick\nsbot}%
            \hfill
            \vbox{\moveright\cornerthick\nsbot}%
          }%
          \nointerlineskip
          \line{\ewbot\hfil\ewbot}%
        }%
      \egroup 
      \fi
    }
  }
  \advancepageno
  \ifnum\outputpenalty>-20000 \else\dosupereject\fi
}

\newinsert\margin \dimen\margin=\maxdimen

\def\pagebody#1{\vbox to\pageheight{\boxmaxdepth=\maxdepth #1}}
{\catcode`\@ =11
\gdef\pagecontents#1{\ifvoid\topins\else\unvbox\topins\fi
\ifvoid\margin\else 
  \rlap{\kern\hsize\vbox to\z@{\kern1pt\box\margin \vss}}\fi
\dimen@=\dp#1 \unvbox#1
\ifvoid\footins\else\vskip\skip\footins\footnoterule \unvbox\footins\fi
\ifr@ggedbottom \kern-\dimen@ \vfil \fi}
}

%
\def\ewtop{\vrule height\cornerthick depth0pt width\cornerlong}
\def\nstop{\vbox
  {\hrule height\cornerthick depth\cornerlong width\cornerthick}}
\def\ewbot{\vrule height0pt depth\cornerthick width\cornerlong}
\def\nsbot{\vbox
  {\hrule height\cornerlong depth\cornerthick width\cornerthick}}

%
\def\parsearg{\parseargusing{}}
\def\parseargusing#1#2{%
  \def\next{#2}%
  \begingroup
    \obeylines
    \spaceisspace
    #1%
    \parseargline\empty
}

{\obeylines %
  \gdef\parseargline#1^^M{%
    \endgroup 
    \argremovecomment #1\comment\ArgTerm%
  }%
}

\def\argremovecomment#1\comment#2\ArgTerm{\argremovec #1\c\ArgTerm}
\def\argremovec#1\c#2\ArgTerm{\argcheckspaces#1\^^M\ArgTerm}

%
%
\def\argcheckspaces#1\^^M{\argcheckspacesX#1\^^M \^^M}
\def\argcheckspacesX#1 \^^M{\argcheckspacesY#1\^^M}
\def\argcheckspacesY#1\^^M#2\^^M#3\ArgTerm{%
  \def\temp{#3}%
  \ifx\temp\empty
    \let\temp\finishparsearg
  \else
    \let\temp\argcheckspaces
  \fi
  \temp#1 #3\ArgTerm
}

%
%
\def\finishparsearg#1 \ArgTerm{\expandafter\next\expandafter{#1}}

%

\def\parseargdef#1{%
  \expandafter \doparseargdef \csname\string#1\endcsname #1%
}
\def\doparseargdef#1#2{%
  \def#2{\parsearg#1}%
  \def#1##1%
}

{
  \obeyspaces
  \gdef\obeyedspace{ }

  %
  \gdef\sepspaces{\obeyspaces\let =\tie}

  \gdef\unsepspaces{\let =\space}
}

\def\flushcr{\ifx\par\lisppar \def\next##1{}\else \let\next=\relax \fi \next}

%
%
%

\def\startenvironment#1{\begingroup\def\thisenv{#1}}
\let\thisenv\empty

\long\def\envdef#1#2{\def#1{\startenvironment#1#2}}
\def\envparseargdef#1#2{\parseargdef#1{\startenvironment#1#2}}

\def\checkenv#1{%
  \def\temp{#1}%
  \ifx\thisenv\temp
  \else
    \badenverr
  \fi
}

\def\badenverr{%
  \errhelp = \EMsimple
  \errmessage{This command can appear only \inenvironment\temp,
    not \inenvironment\thisenv}%
}
\def\inenvironment#1{%
  \ifx#1\empty
    out of any environment%
  \else
    in environment \expandafter\string#1%
  \fi
}

%
\parseargdef\end{%
  \if 1\csname iscond.#1\endcsname
  \else
    \expandafter\checkenv\csname#1\endcsname
    \csname E#1\endcsname
    \endgroup
  \fi
}

\newhelp\EMsimple{Press RETURN to continue.}


\def\@{{\tt\char64}}


\def\mylbrace {{\tt\char123}}
\def\myrbrace {{\tt\char125}}
\let\{=\mylbrace
\let\}=\myrbrace
\begingroup
  \catcode`\{ = \other \catcode`\} = \other
  \catcode`\[ = 1 \catcode`\] = 2
  \catcode`\! = 0 \catcode`\\ = \other
  !gdef!lbracecmd[\{]%
  !gdef!rbracecmd[\}]%
  !gdef!lbraceatcmd[@{]%
  !gdef!rbraceatcmd[@}]%
!endgroup

\let\comma = ,

\let\, = \c
\let\dotaccent = \.

\def\questiondown{?`}
\def\exclamdown{!`}
\def\ordf{\leavevmode\raise1ex\hbox{\selectfonts\lllsize \underbar{a}}}
\def\ordm{\leavevmode\raise1ex\hbox{\selectfonts\lllsize \underbar{o}}}

\def\imacro{i}
\def\jmacro{j}
\def\dotless#1{%
  \def\temp{#1}%
  \ifx\temp\imacro \ptexi
  \else\ifx\temp\jmacro \j
  \else \errmessage{@dotless can be used only with i or j}%
  \fi\fi
}

%
\edef\TeX{\TeX \spacefactor=1000 }

%
\def\LaTeX{%
  L\kern-.36em
  {\setbox0=\hbox{T}%
   \vbox to \ht0{\hbox{\selectfonts\lllsize A}\vss}}%
  \kern-.15em
  \TeX
}

{\catcode`@ = 11
 \global\let\tiepenalty = \@M
 \gdef\tie{\leavevmode\penalty\tiepenalty\ }
}

\def\:{\spacefactor=1000 }

\def\*{\hfil\break\hbox{}\ignorespaces}

\let\/=\allowbreak

\def\.{.\spacefactor=3000 }

\def\!{!\spacefactor=3000 }

\def\?{?\spacefactor=3000 }


%
%
\newbox\groupbox
\def\vfilllimit{0.7}
\envdef\group{%
  \ifnum\catcode`\^^M=\active \else
    \errhelp = \groupinvalidhelp
    \errmessage{@group invalid in context where filling is enabled}%
  \fi
  \startsavinginserts
  \setbox\groupbox = \vtop\bgroup
    \comment
}
%
\def\Egroup{%
    \endgraf 
    \global\dimen1 = \prevdepth
  \egroup           
  \dimen0 = \ht\groupbox  \advance\dimen0 by \dp\groupbox
  \dimen2 = \pageheight   \advance\dimen2 by -\pagetotal
  \ifdim \dimen0 > \dimen2
    \ifdim \pagetotal < \vfilllimit\pageheight
      \page
    \fi
  \fi
  \box\groupbox
  \prevdepth = \dimen1
  \checkinserts
}
%
%
\newhelp\groupinvalidhelp{%
group can only be used in environments such as @example,^^J%
where each line of input produces a line of output.}


\newdimen\mil  \mil=0.001in


\parseargdef\need{%
  \par
  %
  \dimen0 = #1\mil
  \dimen2 = \ht\strutbox
  \advance\dimen2 by \dp\strutbox
  \ifdim\dimen0 > \dimen2
    %
    \vtop to #1\mil{\strut\vfil}%
    %
    %
    \penalty9999
    %
    \kern -#1\mil
    %
    \nobreak
  \fi
}


%
\def\page{\par\vfill\supereject}


\newskip\exdentamount

\parseargdef\exdent{\hfil\break\hbox{\kern -\exdentamount{\rm#1}}\hfil\break}

\parseargdef\nofillexdent{{\advance \leftskip by -\exdentamount
  \leftline{\hskip\leftskip{\rm#1}}}}

%
\newskip\inmarginspacing \inmarginspacing=1cm
\def\strutdepth{\dp\strutbox}
\def\doinmargin#1#2{\strut\vadjust{%
  \nobreak
  \kern-\strutdepth
  \vtop to \strutdepth{%
    \baselineskip=\strutdepth
    \vss
    \ifx#1l%
      \llap{\ignorespaces #2\hskip\inmarginspacing}%
    \else
      \rlap{\hskip\hsize \hskip\inmarginspacing \ignorespaces #2}%
    \fi
    \null
  }%
}}
\def\inleftmargin{\doinmargin l}
\def\inrightmargin{\doinmargin r}
%
%

\def\parseinmargin#1,#2,#3\finish{
  \setbox0 = \hbox{\ignorespaces #2}%
  \ifdim\wd0 > 0pt
    \def\lefttext{#1}
    \def\righttext{#2}%
  \else
    \def\lefttext{#1}
    \def\righttext{#1}%
  \fi
  \ifodd\pageno
    \def\temp{\inrightmargin\righttext}
  \else
    \def\temp{\inleftmargin\lefttext}%
  \fi
  \temp
}

%
\def\include{\parseargusing\filenamecatcodes\input }
\def\includezzz#1{%
  \pushthisfilestack
  \def\thisfile{#1}%
  {%
    \makevalueexpandable
    \def\temp{\input #1 }%
    \expandafter
  }\temp
  \popthisfilestack
}
\def\filenamecatcodes{%
  \catcode`\\=\other
  \catcode`~=\other
  \catcode`^=\other
  \catcode`_=\other
  \catcode`|=\other
  \catcode`<=\other
  \catcode`>=\other
  \catcode`+=\other
  \catcode`-=\other
}

\def\pushthisfilestack{%
  \expandafter\pushthisfilestackX\popthisfilestack\StackTerm
}
\def\pushthisfilestackX{%
  \expandafter\pushthisfilestackY\thisfile\StackTerm
}
\def\pushthisfilestackY #1\StackTerm #2\StackTerm {%
  \gdef\popthisfilestack{\gdef\thisfile{#1}\gdef\popthisfilestack{#2}}%
}

\def\popthisfilestack{\errthisfilestackempty}
\def\errthisfilestackempty{\errmessage{Internal error:
  the stack of filenames is empty.}}

\def\thisfile{}

%
\parseargdef\center{%
  \ifhmode
    \let\next\centerH
  \else
    \let\next\centerV
  \fi
  \next{\hfil \ignorespaces#1\unskip \hfil}%
}
\def\centerH#1{%
  {%
    \hfil\break
    \advance\hsize by -\leftskip
    \advance\hsize by -\rightskip
    \line{#1}%
    \break
  }%
}
\def\centerV#1{\line{\kern\leftskip #1\kern\rightskip}}


\parseargdef\sp{\vskip #1\baselineskip}


\def\comment{\begingroup \catcode`\^^M=\other%
\catcode`\@=\other \catcode`\{=\other \catcode`\}=\other%
\commentxxx}
{\catcode`\^^M=\other \gdef\commentxxx#1^^M{\endgroup}}

\let\c=\comment

%
\def\asisword{asis} 
\def\noneword{none}
\parseargdef\paragraphindent{%
  \def\temp{#1}%
  \ifx\temp\asisword
  \else
    \ifx\temp\noneword
      \defaultparindent = 0pt
    \else
      \defaultparindent = #1em
    \fi
  \fi
  \parindent = \defaultparindent
}

\parseargdef\exampleindent{%
  \def\temp{#1}%
  \ifx\temp\asisword
  \else
    \ifx\temp\noneword
      \lispnarrowing = 0pt
    \else
      \lispnarrowing = #1em
    \fi
  \fi
}

%
%
\def\suppressfirstparagraphindent{\dosuppressfirstparagraphindent}
\def\insertword{insert}
\parseargdef\firstparagraphindent{%
  \def\temp{#1}%
  \ifx\temp\noneword
    \let\suppressfirstparagraphindent = \dosuppressfirstparagraphindent
  \else\ifx\temp\insertword
    \let\suppressfirstparagraphindent = \relax
  \else
    \errhelp = \EMsimple
    \errmessage{Unknown @firstparagraphindent option `\temp'}%
  \fi\fi
}

%
%
\gdef\dosuppressfirstparagraphindent{%
  \gdef\indent{%
    \restorefirstparagraphindent
    \indent
  }%
  \gdef\noindent{%
    \restorefirstparagraphindent
    \noindent
  }%
  \global\everypar = {%
    \kern -\parindent
    \restorefirstparagraphindent
  }%
}

\gdef\restorefirstparagraphindent{%
  \global \let \indent = \ptexindent
  \global \let \noindent = \ptexnoindent
  \global \everypar = {}%
}

%
\def\asis#1{#1}

%
{
  \catcode\underChar = \active
  \gdef\mathunderscore{%
    \catcode\underChar=\active
    \def_{\ifnum\fam=\slfam \_\else\sb\fi}%
  }
}
%
\def\mathbackslash{\ifnum\fam=\ttfam \mathchar"075C \else\backslash \fi}
\def\math{%
  \tex
  \mathunderscore
  \let\\ = \mathbackslash
  \mathactive
  $\finishmath
}
\def\finishmath#1{#1$\endgroup}  

%
{
  \catcode`^ = \active
  \catcode`< = \active
  \catcode`> = \active
  \catcode`+ = \active
  \gdef\mathactive{%
    \let^ = \ptexhat
    \let< = \ptexless
    \let> = \ptexgtr
    \let+ = \ptexplus
  }
}

\def\bullet{$\ptexbullet$}
\def\minus{$-$}

%
\def\dots{%
  \leavevmode
  \hbox to 1.5em{%
    \hskip 0pt plus 0.25fil
    .\hfil.\hfil.%
    \hskip 0pt plus 0.5fil
  }%
}

%
\def\enddots{%
  \dots
  \spacefactor=3000
}

%
\let\comma = ,


%
\newif\iflinks \linkstrue 

\def\setfilename{%
   \fixbackslash  
   \iflinks
     \tryauxfile
     \immediate\openout\auxfile=\jobname.aux
   \fi 
   \openindices
   \let\setfilename=\comment 
   %
   \openin 1 texinfo.cnf
   \ifeof 1 \else \input texinfo.cnf \fi
   \closein 1
   \comment 
}

%
\def\openindices{%
  \newindex{cp}%
  \newcodeindex{fn}%
  \newcodeindex{vr}%
  \newcodeindex{tp}%
  \newcodeindex{ky}%
  \newcodeindex{pg}%
}


\message{pdf,}
\newcount\tempnum
\newcount\lnkcount
\newtoks\filename
\newcount\filenamelength
\newcount\pgn
\newtoks\toksA
\newtoks\toksB
\newtoks\toksC
\newtoks\toksD
\newbox\boxA
\newcount\countA
\newif\ifpdf
\newif\ifpdfmakepagedest

\ifx\pdfoutput\undefined
\else
  \ifx\pdfoutput\relax
  \else
    \ifcase\pdfoutput
    \else
      \pdftrue
    \fi
  \fi
\fi
\ifpdf
  \input pdfcolor
  \pdfcatalog{/PageMode /UseOutlines}%
  \def\dopdfimage#1#2#3{%
    \def\imagewidth{#2}%
    \def\imageheight{#3}%
    \ifnum\pdftexversion < 14
      \immediate\pdfimage
    \else
      \immediate\pdfximage
    \fi
      \ifx\empty\imagewidth\else width \imagewidth \fi
      \ifx\empty\imageheight\else height \imageheight \fi
      \ifnum\pdftexversion<13
         #1.pdf%
       \else
         {#1.pdf}%
       \fi
    \ifnum\pdftexversion < 14 \else
      \pdfrefximage \pdflastximage
    \fi}
  \def\pdfmkdest#1{{%
    \atdummies
    \normalturnoffactive
    \pdfdest name{#1} xyz%
  }}
  \def\pdfmkpgn#1{#1}
  \let\linkcolor = \Blue  
  \def\endlink{\Black\pdfendlink}
  \def\expnumber#1{\expandafter\ifx\csname#1\endcsname\relax 0%
    \else \csname#1\endcsname \fi}
  \def\advancenumber#1{\tempnum=\expnumber{#1}\relax
    \advance\tempnum by 1
    \expandafter\xdef\csname#1\endcsname{\the\tempnum}}
  %
  %
  \def\dopdfoutline#1#2#3#4{%
    \def\pdfoutlinedest{#3}%
    \ifx\pdfoutlinedest\empty \def\pdfoutlinedest{#4}\fi
    \pdfoutline goto name{\pdfmkpgn{\pdfoutlinedest}}#2{#1}%
  }
  \def\pdfmakeoutlines{%
    \begingroup
      \edef\mylbrace{\iftrue \string{\else}\fi}\let\{=\mylbrace
      \edef\myrbrace{\iffalse{\else\string}\fi}\let\}=\myrbrace
      %
      \def\numchapentry##1##2##3##4{%
	\def\thischapnum{##2}%
	\def\thissecnum{0}%
	\def\thissubsecnum{0}%
      }%
      \def\numsecentry##1##2##3##4{%
	\advancenumber{chap\thischapnum}%
	\def\thissecnum{##2}%
	\def\thissubsecnum{0}%
      }%
      \def\numsubsecentry##1##2##3##4{%
	\advancenumber{sec\thissecnum}%
	\def\thissubsecnum{##2}%
      }%
      \def\numsubsubsecentry##1##2##3##4{%
	\advancenumber{subsec\thissubsecnum}%
      }%
      \def\thischapnum{0}%
      \def\thissecnum{0}%
      \def\thissubsecnum{0}%
      %
      \def\appentry{\numchapentry}%
      \def\appsecentry{\numsecentry}%
      \def\appsubsecentry{\numsubsecentry}%
      \def\appsubsubsecentry{\numsubsubsecentry}%
      \def\unnchapentry{\numchapentry}%
      \def\unnsecentry{\numsecentry}%
      \def\unnsubsecentry{\numsubsecentry}%
      \def\unnsubsubsecentry{\numsubsubsecentry}%
      \input \jobname.toc
      %
      %
      \def\numchapentry##1##2##3##4{%
        \dopdfoutline{##1}{count-\expnumber{chap##2}}{##3}{##4}}%
      \def\numsecentry##1##2##3##4{%
        \dopdfoutline{##1}{count-\expnumber{sec##2}}{##3}{##4}}%
      \def\numsubsecentry##1##2##3##4{%
        \dopdfoutline{##1}{count-\expnumber{subsec##2}}{##3}{##4}}%
      \def\numsubsubsecentry##1##2##3##4{
        \dopdfoutline{##1}{}{##3}{##4}}%
      %
      %
      \indexnofonts
      \turnoffactive
      \input \jobname.toc
    \endgroup
  }
  \def\makelinks #1,{%
    \def\params{#1}\def\E{END}%
    \ifx\params\E
      \let\nextmakelinks=\relax
    \else
      \let\nextmakelinks=\makelinks
      \ifnum\lnkcount>0,\fi
      \picknum{#1}%
      \startlink attr{/Border [0 0 0]}
        goto name{\pdfmkpgn{\the\pgn}}%
      \linkcolor #1%
      \advance\lnkcount by 1%
      \endlink
    \fi
    \nextmakelinks
  }
  \def\picknum#1{\expandafter\pn#1}
  \def\pn#1{%
    \def\p{#1}%
    \ifx\p\lbrace
      \let\nextpn=\ppn
    \else
      \let\nextpn=\ppnn
      \def\first{#1}
    \fi
    \nextpn
  }
  \def\ppn#1{\pgn=#1\gobble}
  \def\ppnn{\pgn=\first}
  
  \def\skipspaces#1{\def\PP{#1}\def\D{|}%
    \ifx\PP\D\let\nextsp\relax
    \else\let\nextsp\skipspaces
      \ifx\p\space\else\addtokens{\filename}{\PP}%
        \advance\filenamelength by 1
      \fi
    \fi
    \nextsp}
  \def\getfilename#1{\filenamelength=0\expandafter\skipspaces#1|\relax}
  \ifnum\pdftexversion < 14
    \let \startlink \pdfannotlink
  \else
    \let \startlink \pdfstartlink
  \fi
  \def\pdfurl#1{%
    \begingroup
      \normalturnoffactive\def\@{@}%
      \makevalueexpandable
      \leavevmode\Red
      \startlink attr{/Border [0 0 0]}%
        user{/Subtype /Link /A << /S /URI /URI (#1) >>}%
    \endgroup}
  \def\pdfgettoks#1.{\setbox\boxA=\hbox{\toksA={#1.}\toksB={}\maketoks}}
  \def\addtokens#1#2{\edef\addtoks{\noexpand#1={\the#1#2}}\addtoks}
  \def\adn#1{\addtokens{\toksC}{#1}\global\countA=1\let\next=\maketoks}
  \def\poptoks#1#2|ENDTOKS|{\let\first=#1\toksD={#1}\toksA={#2}}
  \def\maketoks{%
    \expandafter\poptoks\the\toksA|ENDTOKS|\relax
    \ifx\first0\adn0
    \else\ifx\first1\adn1 \else\ifx\first2\adn2 \else\ifx\first3\adn3
    \else\ifx\first4\adn4 \else\ifx\first5\adn5 \else\ifx\first6\adn6
    \else\ifx\first7\adn7 \else\ifx\first8\adn8 \else\ifx\first9\adn9
    \else
      \ifnum0=\countA\else\makelink\fi
      \ifx\first.\let\next=\done\else
        \let\next=\maketoks
        \addtokens{\toksB}{\the\toksD}
        \ifx\first,\addtokens{\toksB}{\space}\fi
      \fi
    \fi\fi\fi\fi\fi\fi\fi\fi\fi\fi
    \next}
  \def\makelink{\addtokens{\toksB}%
    {\noexpand\pdflink{\the\toksC}}\toksC={}\global\countA=0}
  \def\pdflink#1{%
    \startlink attr{/Border [0 0 0]} goto name{\pdfmkpgn{#1}}
    \linkcolor #1\endlink}
  \def\done{\edef\st{\global\noexpand\toksA={\the\toksB}}\st}
\else
  \let\pdfmkdest = \gobble
  \let\pdfurl = \gobble
  \let\endlink = \relax
  \let\linkcolor = \relax
  \let\pdfmakeoutlines = \relax
\fi  

\message{fonts,}

%
\def\setfontstyle#1{%
  \def\curfontstyle{#1}
  \csname ten#1\endcsname  
}

%
\def\selectfonts#1{\csname #1fonts\endcsname \csname\curfontstyle\endcsname}

\def\rm{\fam=0 \setfontstyle{rm}}
\def\it{\fam=\itfam \setfontstyle{it}}
\def\sl{\fam=\slfam \setfontstyle{sl}}
\def\bf{\fam=\bffam \setfontstyle{bf}}\def\bfstylename{bf}
\def\tt{\fam=\ttfam \setfontstyle{tt}}

\newfam\sffam
\def\sf{\fam=\sffam \setfontstyle{sf}}

\def\ttsl{\setfontstyle{ttsl}}

\newdimen\textleading  \textleading = 13.2pt

%
\def\lineskipfactor{.08333}
\def\strutheightpercent{.70833}
\def\strutdepthpercent {.29167}
\def\setleading#1{%
  \normalbaselineskip = #1\relax
  \normallineskip = \lineskipfactor\normalbaselineskip
  \normalbaselines
  \setbox\strutbox =\hbox{%
    \vrule width0pt height\strutheightpercent\baselineskip
                    depth \strutdepthpercent \baselineskip
  }%
}

\def\setfont#1#2#3#4{\font#1=\fontprefix#2#3 scaled #4}

\ifx\fontprefix\undefined
\def\fontprefix{cm}
\fi
\def\rmshape{r}
\def\rmbshape{bx}               
\def\bfshape{b}

\def\ttshape{tt}
\def\ttbshape{tt}
\def\ttslshape{sltt}
\def\itshape{ti}
\def\itbshape{bxti}
\def\slshape{sl}
\def\slbshape{bxsl}
\def\sfshape{ss}
\def\sfbshape{ss}
\def\scshape{csc}
\def\scbshape{csc}


\edef\mainmagstep{\magstephalf}
\setfont\textrm\rmshape{10}{\mainmagstep}
\setfont\texttt\ttshape{10}{\mainmagstep}
\setfont\textbf\bfshape{10}{\mainmagstep}
\setfont\textit\itshape{10}{\mainmagstep}
\setfont\textsl\slshape{10}{\mainmagstep}
\setfont\textsf\sfshape{10}{\mainmagstep}
\setfont\textsc\scshape{10}{\mainmagstep}
\setfont\textttsl\ttslshape{10}{\mainmagstep}
\font\texti=cmmi10 scaled \mainmagstep
\font\textsy=cmsy10 scaled \mainmagstep

\setfont\defbf\bfshape{10}{\magstep1}
\setfont\deftt\ttshape{10}{\magstep1}
\setfont\defttsl\ttslshape{10}{\magstep1}
\def\df{\let\tentt=\deftt \let\tenbf = \defbf \let\tenttsl=\defttsl \bf}


\setfont\smallrm\rmshape{9}{1000}
\setfont\smalltt\ttshape{9}{1000}
\setfont\smallbf\bfshape{10}{900}
\setfont\smallit\itshape{9}{1000}
\setfont\smallsl\slshape{9}{1000}
\setfont\smallsf\sfshape{9}{1000}
\setfont\smallsc\scshape{10}{900}
\setfont\smallttsl\ttslshape{10}{900}
\font\smalli=cmmi9
\font\smallsy=cmsy9


\setfont\smallerrm\rmshape{8}{1000}
\setfont\smallertt\ttshape{8}{1000}
\setfont\smallerbf\bfshape{10}{800}
\setfont\smallerit\itshape{8}{1000}
\setfont\smallersl\slshape{8}{1000}
\setfont\smallersf\sfshape{8}{1000}
\setfont\smallersc\scshape{10}{800}
\setfont\smallerttsl\ttslshape{10}{800}
\font\smalleri=cmmi8
\font\smallersy=cmsy8


\setfont\titlerm\rmbshape{12}{\magstep3}
\setfont\titleit\itbshape{10}{\magstep4}
\setfont\titlesl\slbshape{10}{\magstep4}
\setfont\titlett\ttbshape{12}{\magstep3}
\setfont\titlettsl\ttslshape{10}{\magstep4}
\setfont\titlesf\sfbshape{17}{\magstep1}
\let\titlebf=\titlerm
\setfont\titlesc\scbshape{10}{\magstep4}
\font\titlei=cmmi12 scaled \magstep3
\font\titlesy=cmsy10 scaled \magstep4
\def\authorrm{\secrm}
\def\authortt{\sectt}


\setfont\chaprm\rmbshape{12}{\magstep2}
\setfont\chapit\itbshape{10}{\magstep3}
\setfont\chapsl\slbshape{10}{\magstep3}
\setfont\chaptt\ttbshape{12}{\magstep2}
\setfont\chapttsl\ttslshape{10}{\magstep3}
\setfont\chapsf\sfbshape{17}{1000}
\let\chapbf=\chaprm
\setfont\chapsc\scbshape{10}{\magstep3}
\font\chapi=cmmi12 scaled \magstep2
\font\chapsy=cmsy10 scaled \magstep3


\setfont\secrm\rmbshape{12}{\magstep1}
\setfont\secit\itbshape{10}{\magstep2}
\setfont\secsl\slbshape{10}{\magstep2}
\setfont\sectt\ttbshape{12}{\magstep1}
\setfont\secttsl\ttslshape{10}{\magstep2}
\setfont\secsf\sfbshape{12}{\magstep1}
\let\secbf\secrm
\setfont\secsc\scbshape{10}{\magstep2}
\font\seci=cmmi12 scaled \magstep1
\font\secsy=cmsy10 scaled \magstep2


\setfont\ssecrm\rmbshape{12}{\magstephalf}
\setfont\ssecit\itbshape{10}{1315}
\setfont\ssecsl\slbshape{10}{1315}
\setfont\ssectt\ttbshape{12}{\magstephalf}
\setfont\ssecttsl\ttslshape{10}{1315}
\setfont\ssecsf\sfbshape{12}{\magstephalf}
\let\ssecbf\ssecrm
\setfont\ssecsc\scbshape{10}{1315}
\font\sseci=cmmi12 scaled \magstephalf
\font\ssecsy=cmsy10 scaled 1315


\setfont\reducedrm\rmshape{10}{1000}
\setfont\reducedtt\ttshape{10}{1000}
\setfont\reducedbf\bfshape{10}{1000}
\setfont\reducedit\itshape{10}{1000}
\setfont\reducedsl\slshape{10}{1000}
\setfont\reducedsf\sfshape{10}{1000}
\setfont\reducedsc\scshape{10}{1000}
\setfont\reducedttsl\ttslshape{10}{1000}
\font\reducedi=cmmi10
\font\reducedsy=cmsy10

%
\def\resetmathfonts{%
  \textfont0=\tenrm \textfont1=\teni \textfont2=\tensy
  \textfont\itfam=\tenit \textfont\slfam=\tensl \textfont\bffam=\tenbf
  \textfont\ttfam=\tentt \textfont\sffam=\tensf
}

%
%
%
\def\textfonts{%
  \let\tenrm=\textrm \let\tenit=\textit \let\tensl=\textsl
  \let\tenbf=\textbf \let\tentt=\texttt \let\smallcaps=\textsc
  \let\tensf=\textsf \let\teni=\texti \let\tensy=\textsy
  \let\tenttsl=\textttsl
  \def\curfontsize{text}%
  \def\lsize{reduced}\def\lllsize{smaller}%
  \resetmathfonts \setleading{\textleading}}
\def\titlefonts{%
  \let\tenrm=\titlerm \let\tenit=\titleit \let\tensl=\titlesl
  \let\tenbf=\titlebf \let\tentt=\titlett \let\smallcaps=\titlesc
  \let\tensf=\titlesf \let\teni=\titlei \let\tensy=\titlesy
  \let\tenttsl=\titlettsl
  \def\curfontsize{title}%
  \def\lsize{chap}\def\lllsize{subsec}%
  \resetmathfonts \setleading{25pt}}

\def\chapfonts{%
  \let\tenrm=\chaprm \let\tenit=\chapit \let\tensl=\chapsl
  \let\tenbf=\chapbf \let\tentt=\chaptt \let\smallcaps=\chapsc
  \let\tensf=\chapsf \let\teni=\chapi \let\tensy=\chapsy
  \let\tenttsl=\chapttsl
  \def\curfontsize{chap}%
  \def\lsize{sec}\def\lllsize{text}%
  \resetmathfonts \setleading{19pt}}
\def\secfonts{%
  \let\tenrm=\secrm \let\tenit=\secit \let\tensl=\secsl
  \let\tenbf=\secbf \let\tentt=\sectt \let\smallcaps=\secsc
  \let\tensf=\secsf \let\teni=\seci \let\tensy=\secsy
  \let\tenttsl=\secttsl
  \def\curfontsize{sec}%
  \def\lsize{subsec}\def\lllsize{reduced}%
  \resetmathfonts \setleading{16pt}}
\def\subsecfonts{%
  \let\tenrm=\ssecrm \let\tenit=\ssecit \let\tensl=\ssecsl
  \let\tenbf=\ssecbf \let\tentt=\ssectt \let\smallcaps=\ssecsc
  \let\tensf=\ssecsf \let\teni=\sseci \let\tensy=\ssecsy
  \let\tenttsl=\ssecttsl
  \def\curfontsize{ssec}%
  \def\lsize{text}\def\lllsize{small}%
  \resetmathfonts \setleading{15pt}}

\def\reducedfonts{%
  \let\tenrm=\reducedrm \let\tenit=\reducedit \let\tensl=\reducedsl
  \let\tenbf=\reducedbf \let\tentt=\reducedtt \let\reducedcaps=\reducedsc
  \let\tensf=\reducedsf \let\teni=\reducedi \let\tensy=\reducedsy
  \let\tenttsl=\reducedttsl
  \def\curfontsize{reduced}%
  \def\lsize{small}\def\lllsize{smaller}%
  \resetmathfonts \setleading{10.5pt}}
\def\smallfonts{%
  \let\tenrm=\smallrm \let\tenit=\smallit \let\tensl=\smallsl
  \let\tenbf=\smallbf \let\tentt=\smalltt \let\smallcaps=\smallsc
  \let\tensf=\smallsf \let\teni=\smalli \let\tensy=\smallsy
  \let\tenttsl=\smallttsl
  \def\curfontsize{small}%
  \def\lsize{smaller}\def\lllsize{smaller}%
  \resetmathfonts \setleading{10.5pt}}
\def\smallerfonts{%
  \let\tenrm=\smallerrm \let\tenit=\smallerit \let\tensl=\smallersl
  \let\tenbf=\smallerbf \let\tentt=\smallertt \let\smallcaps=\smallersc
  \let\tensf=\smallersf \let\teni=\smalleri \let\tensy=\smallersy
  \let\tenttsl=\smallerttsl
  \def\curfontsize{smaller}%
  \def\lsize{smaller}\def\lllsize{smaller}%
  \resetmathfonts \setleading{9.5pt}}

\let\smallexamplefonts = \smallfonts

%
%

%
\textfonts \rm

\def\angleleft{$\langle$}
\def\angleright{$\rangle$}

\newcount\fontdepth \fontdepth=0

\setfont\shortcontrm\rmshape{12}{1000}
\setfont\shortcontbf\bfshape{10}{\magstep1}  
\setfont\shortcontsl\slshape{12}{1000}
\setfont\shortconttt\ttshape{12}{1000}


\def\smartitalicx{\ifx\next,\else\ifx\next-\else\ifx\next.\else
                    \ptexslash\fi\fi\fi}
\def\smartslanted#1{{\ifusingtt\ttsl\sl #1}\futurelet\next\smartitalicx}
\def\smartitalic#1{{\ifusingtt\ttsl\it #1}\futurelet\next\smartitalicx}

\def\ttslanted#1{{\ttsl #1}\futurelet\next\smartitalicx}

\def\cite#1{{\sl #1}\futurelet\next\smartitalicx}

\let\i=\smartitalic

\let\var=\smartslanted

\let\emph=\smartitalic

\def\b#1{{\bf #1}}


%
\def\nohyphenation{\hyphenchar\font = -1  \aftergroup\restorehyphenation}
\def\restorehyphenation{\hyphenchar\font = `- }

%
\catcode`@=11
  \def\frenchspacing{%
    \sfcode\dotChar  =\@m \sfcode\questChar=\@m \sfcode\exclamChar=\@m
    \sfcode\colonChar=\@m \sfcode\semiChar =\@m \sfcode\commaChar =\@m
  }
\catcode`@=\other

\def\t#1{%
  {\tt \rawbackslash \frenchspacing #1}%
  \null
}
\def\samp#1{`\tclose{#1}'\null}
\setfont\keyrm\rmshape{8}{1000}
\font\keysy=cmsy9
\def\key#1{{\keyrm\textfont2=\keysy \leavevmode\hbox{%
  \raise0.4pt\hbox{\angleleft}\kern-.08em\vtop{%
    \vbox{\hrule\kern-0.4pt
     \hbox{\raise0.4pt\hbox{\vphantom{\angleleft}}#1}}%
    \kern-0.4pt\hrule}%
  \kern-.06em\raise0.4pt\hbox{\angleright}}}}

\let\file=\samp

\def\tclose#1{%
  {%
    \spaceskip = \fontdimen2\font
    %
    \tt
    %
    \def\ {{\spaceskip = 0pt{} }}%
    %
    \nohyphenation
    \rawbackslash
    \frenchspacing
    #1%
  }%
  \null
}


{
  \catcode`\-=\active
  \catcode`\_=\active
  \global\def\code{\begingroup
    \catcode`\-=\active \let-\codedash
    \catcode`\_=\active \let_\codeunder
    \codex
  }
}

\def\realdash{-}
\def\codedash{-\discretionary{}{}{}}
\def\codeunder{%
  \ifusingtt{\ifmmode
               \mathchar"075F 
             \else\normalunderscore \fi
             \discretionary{}{}{}}%
            {\_}%
}
\def\codex #1{\tclose{#1}\endgroup}


\parseargdef\kbdinputstyle{%
  \def\arg{#1}%
  \ifx\arg\worddistinct
    \gdef\kbdfont{\ttsl}%
  \else\ifx\arg\wordexample
    \gdef\kbdfont{\tt}%
  \else\ifx\arg\wordcode
    \gdef\kbdfont{\tt}%
  \else
    \errhelp = \EMsimple
    \errmessage{Unknown @kbdinputstyle option `\arg'}%
  \fi\fi\fi
}
\def\worddistinct{distinct}
\def\wordexample{example}
\def\wordcode{code}

\kbdinputstyle distinct

\def\xkey{\key}
\def\kbdfoo#1#2#3\par{\def\one{#1}\def\three{#3}\def\threex{??}%
\ifx\one\xkey\ifx\threex\three \key{#2}%
\else{\tclose{\kbdfont\look}}\fi
\else{\tclose{\kbdfont\look}}\fi}


%
\def\uref#1{\douref #1,,,\finish}
\def\douref#1,#2,#3,#4\finish{\begingroup
  \unsepspaces
  \pdfurl{#1}%
  \setbox0 = \hbox{\ignorespaces #3}%
  \ifdim\wd0 > 0pt
    \unhbox0 
  \else
    \setbox0 = \hbox{\ignorespaces #2}%
    \ifdim\wd0 > 0pt
      \ifpdf
        \unhbox0             
      \else
        \unhbox0\ (\code{#1})
      \fi
    \else
      \code{#1}
    \fi
  \fi
  \endlink
\endgroup}

%
\let\url=\uref

%
\ifpdf
  
  \def\doemail#1,#2,#3\finish{\begingroup
    \unsepspaces
    \pdfurl{mailto:#1}%
    \setbox0 = \hbox{\ignorespaces #2}%
    \ifdim\wd0>0pt\unhbox0\else\code{#1}\fi
    \endlink
  \endgroup}
\else
  
\fi

%
\def\ifmonospace{\ifdim\fontdimen3\font=0pt }

%

\def\kbd#1{\def\look{#1}\expandafter\kbdfoo\look??\par}



%

\def\doacronym#1,#2,#3\finish{%
  {\selectfonts\lsize #1}%
  \def\temp{#2}%
  \ifx\temp\empty \else
    \space ({\unsepspaces \ignorespaces \temp \unskip})%
  \fi
}

%

\def\doabbr#1,#2,#3\finish{%
  {\frenchspacing #1}%
  \def\temp{#2}%
  \ifx\temp\empty \else
    \space ({\unsepspaces \ignorespaces \temp \unskip})%
  \fi
}

%
\def\pounds{{\it\$}}

%
%
%
%
%
%
\def\euro{{\eurofont e}}
\def\eurofont{%
  %
  %
  %
  \def\eurosize{\csname\curfontsize nominalsize\endcsname}%
  \ifx\curfontstyle\bfstylename 
    \font\thiseurofont = \ifusingit{feybo10}{feybr10} at \eurosize
  \else 
    \font\thiseurofont = \ifusingit{feymo10}{feymr10} at \eurosize
  \fi
  \thiseurofont
}

%
\def\registeredsymbol{%
  $^{{\ooalign{\hfil\raise.07ex\hbox{\selectfonts\lllsize R}%
               \hfil\crcr\Orb}}%
    }$%
}

%
\ifx\Orb\undefined
\def\Orb{\mathhexbox20D}
\fi

\message{page headings,}

\newskip\titlepagetopglue \titlepagetopglue = 1.5in
\newskip\titlepagebottomglue \titlepagebottomglue = 2pc

\newif\ifseenauthor
\newif\iffinishedtitlepage

%
\newif\ifsetcontentsaftertitlepage
 
\newif\ifsetshortcontentsaftertitlepage

\parseargdef\shorttitlepage{\begingroup\hbox{}\vskip 1.5in \chaprm \centerline{#1}%
        \endgroup\page\hbox{}\page}

\envdef\titlepage{%
  \begingroup
    \parindent=0pt \textfonts
    \vglue\titlepagetopglue
    \finishedtitlepagetrue
    %
    \let\oldpage = \page
    \def\page{%
      \iffinishedtitlepage\else
	 \finishtitlepage
      \fi
      \let\page = \oldpage
      \page
      \null
    }%
}

\def\Etitlepage{%
    \iffinishedtitlepage\else
	\finishtitlepage
    \fi
    \oldpage
  \endgroup
  %
  \HEADINGSon
  %
  \ifsetshortcontentsaftertitlepage
    \shortcontents
    \contents
    \global\let\shortcontents = \relax
    \global\let\contents = \relax
  \fi
  \ifsetcontentsaftertitlepage
    \contents
    \global\let\contents = \relax
    \global\let\shortcontents = \relax
  \fi
}

\def\finishtitlepage{%
  \vskip4pt \hrule height 2pt width \hsize
  \vskip\titlepagebottomglue
  \finishedtitlepagetrue
}


\let\subtitlerm=\tenrm
\def\subtitlefont{\subtitlerm \normalbaselineskip = 13pt \normalbaselines}

\def\authorfont{\authorrm \normalbaselineskip = 16pt \normalbaselines
		\let\tt=\authortt}

\parseargdef\title{%
  \checkenv\titlepage
  \leftline{\titlefonts\rm #1}
  \finishedtitlepagefalse
  \vskip4pt \hrule height 4pt width \hsize \vskip4pt
}

\parseargdef\subtitle{%
  \checkenv\titlepage
  {\subtitlefont \rightline{#1}}%
}

%
\parseargdef\author{%
  \def\temp{\quotation}%
  \ifx\thisenv\temp
    \def\quotationauthor{#1}
  \else
    \checkenv\titlepage
    \ifseenauthor\else \vskip 0pt plus 1filll \seenauthortrue \fi
    {\authorfont \leftline{#1}}%
  \fi
}


\newtoks\evenheadline    
\newtoks\oddheadline     
\newtoks\evenfootline    
\newtoks\oddfootline     

\headline={{\textfonts\rm \ifodd\pageno \the\oddheadline
                            \else \the\evenheadline \fi}}
\footline={{\textfonts\rm \ifodd\pageno \the\oddfootline
                            \else \the\evenfootline \fi}\HEADINGShook}
\let\HEADINGShook=\relax


\def\evenheadingxxx #1{\evenheadingyyy #1\|\|\|\|\finish}
\def\evenheadingyyy #1\|#2\|#3\|#4\finish{%
\global\evenheadline={\rlap{\centerline{#2}}\line{#1\hfil#3}}}

\def\oddheadingxxx #1{\oddheadingyyy #1\|\|\|\|\finish}
\def\oddheadingyyy #1\|#2\|#3\|#4\finish{%
\global\oddheadline={\rlap{\centerline{#2}}\line{#1\hfil#3}}}

\parseargdef\everyheading{\oddheadingxxx{#1}\evenheadingxxx{#1}}%

\def\evenfootingxxx #1{\evenfootingyyy #1\|\|\|\|\finish}
\def\evenfootingyyy #1\|#2\|#3\|#4\finish{%
\global\evenfootline={\rlap{\centerline{#2}}\line{#1\hfil#3}}}

\def\oddfootingxxx #1{\oddfootingyyy #1\|\|\|\|\finish}
\def\oddfootingyyy #1\|#2\|#3\|#4\finish{%
  \global\oddfootline = {\rlap{\centerline{#2}}\line{#1\hfil#3}}%
  %
  \global\advance\pageheight by -\baselineskip
  \global\advance\vsize by -\baselineskip
}

\parseargdef\everyfooting{\oddfootingxxx{#1}\evenfootingxxx{#1}}


\def\headings #1 {\csname HEADINGS#1\endcsname}

\def\HEADINGSoff{%
\global\evenheadline={\hfil} \global\evenfootline={\hfil}
\global\oddheadline={\hfil} \global\oddfootline={\hfil}}
\HEADINGSoff
\def\HEADINGSdouble{%
\global\pageno=1
\global\evenfootline={\hfil}
\global\oddfootline={\hfil}
\global\evenheadline={\line{\folio\hfil\thistitle}}
\global\oddheadline={\line{\thischapter\hfil\folio}}
\global\let\contentsalignmacro = \chapoddpage
}
\let\contentsalignmacro = \chappager

\def\HEADINGSsingle{%
\global\pageno=1
\global\evenfootline={\hfil}
\global\oddfootline={\hfil}
\global\evenheadline={\line{\thischapter\hfil\folio}}
\global\oddheadline={\line{\thischapter\hfil\folio}}
\global\let\contentsalignmacro = \chappager
}
\def\HEADINGSon{\HEADINGSdouble}

\def\HEADINGSafter{\let\HEADINGShook=\HEADINGSdoublex}

\def\HEADINGSdoublex{%
\global\evenfootline={\hfil}
\global\oddfootline={\hfil}
\global\evenheadline={\line{\folio\hfil\thistitle}}
\global\oddheadline={\line{\thischapter\hfil\folio}}
\global\let\contentsalignmacro = \chapoddpage
}

\def\HEADINGSsingleafter{\let\HEADINGShook=\HEADINGSsinglex}
\def\HEADINGSsinglex{%
\global\evenfootline={\hfil}
\global\oddfootline={\hfil}
\global\evenheadline={\line{\thischapter\hfil\folio}}
\global\oddheadline={\line{\thischapter\hfil\folio}}
\global\let\contentsalignmacro = \chappager
}

\ifx\today\undefined
\def\today{%
  \number\day\space
  \ifcase\month
  \or\putwordMJan\or\putwordMFeb\or\putwordMMar\or\putwordMApr
  \or\putwordMMay\or\putwordMJun\or\putwordMJul\or\putwordMAug
  \or\putwordMSep\or\putwordMOct\or\putwordMNov\or\putwordMDec
  \fi
  \space\number\year}
\fi

\def\thistitle{\putwordNoTitle}

\message{tables,}

\newdimen\tableindent \tableindent=.8in
\newdimen\itemindent  \itemindent=.3in
\newdimen\itemmargin  \itemmargin=.1in

\newdimen\itemmax


\newif\ifitemxneedsnegativevskip

\def\itemxpar{\par\ifitemxneedsnegativevskip\nobreak\vskip-\parskip\nobreak\fi}

\def\internalBitem{\smallbreak \parsearg\itemzzz}
\def\internalBitemx{\itemxpar \parsearg\itemzzz}

\def\itemzzz #1{\begingroup %
  \advance\hsize by -\rightskip
  \advance\hsize by -\tableindent
  \setbox0=\hbox{\itemindicate{#1}}%
  \itemindex{#1}%
  \nobreak 
  %
  \ifdim \wd0>\itemmax
    %
    \begingroup
      \advance\leftskip by-\tableindent
      \advance\hsize by\tableindent
      \advance\rightskip by0pt plus1fil
      \leavevmode\unhbox0\par
    \endgroup
    %
    \nobreak \vskip-\parskip
    %
    %
    \penalty 10001
    \endgroup
    \itemxneedsnegativevskipfalse
  \else
    \noindent
    \nobreak\kern-\tableindent
    \dimen0 = \itemmax  \advance\dimen0 by \itemmargin \advance\dimen0 by -\wd0
    \unhbox0
    \nobreak\kern\dimen0
    \endgroup
    \itemxneedsnegativevskiptrue
  \fi
}

\def\item{\errmessage{@item while not in a list environment}}
\def\itemx{\errmessage{@itemx while not in a list environment}}

\envdef\table{%
  \let\itemindex\gobble
  \tablecheck{table}%
}
\envdef\ftable{%
  \def\itemindex ##1{\doind {fn}{\code{##1}}}%
  \tablecheck{ftable}%
}
\envdef\vtable{%
  \def\itemindex ##1{\doind {vr}{\code{##1}}}%
  \tablecheck{vtable}%
}
\def\tablecheck#1{%
  \ifnum \the\catcode`\^^M=\active
    \endgroup
    \errmessage{This command won't work in this context; perhaps the problem is
      that we are \inenvironment\thisenv}%
    \def\next{\doignore{#1}}%
  \else
    \let\next\tablex
  \fi
  \next
}
\def\tablex#1{%
  \def\itemindicate{#1}%
  \parsearg\tabley
}
\def\tabley#1{%
  {%
    \makevalueexpandable
    \edef\temp{\noexpand\tablez #1\space\space\space}%
    \expandafter
  }\temp \endtablez
}
\def\tablez #1 #2 #3 #4\endtablez{%
  \aboveenvbreak
  \ifnum 0#1>0 \advance \leftskip by #1\mil \fi
  \ifnum 0#2>0 \tableindent=#2\mil \fi
  \ifnum 0#3>0 \advance \rightskip by #3\mil \fi
  \itemmax=\tableindent
  \advance \itemmax by -\itemmargin
  \advance \leftskip by \tableindent
  \exdentamount=\tableindent
  \parindent = 0pt
  \parskip = \smallskipamount
  \ifdim \parskip=0pt \parskip=2pt \fi
  \let\item = \internalBitem
  \let\itemx = \internalBitemx
}


\newcount \itemno

\envdef\itemize{\parsearg\doitemize}

\def\doitemize#1{%
  \aboveenvbreak
  \itemmax=\itemindent
  \advance\itemmax by -\itemmargin
  \advance\leftskip by \itemindent
  \exdentamount=\itemindent
  \parindent=0pt
  \parskip=\smallskipamount
  \ifdim\parskip=0pt \parskip=2pt \fi
  \def\itemcontents{#1}%
  \ifx\itemcontents\empty\def\itemcontents{\bullet}\fi
  \let\item=\itemizeitem
}

%
\def\itemizeitem{%
  \advance\itemno by 1  
  {\let\par=\endgraf \smallbreak}
  {%
   \ifnum\lastpenalty<10000 \parskip=0in \fi
   \noindent
   \hbox to 0pt{\hss \itemcontents \kern\itemmargin}%
   \vadjust{\penalty 1200}}
  \flushcr
}

%
\def\splitoff#1#2\endmark{\def\first{#1}\def\rest{#2}}%

%
\envparseargdef\enumerate{\enumeratey #1  \endenumeratey}
\def\enumeratey #1 #2\endenumeratey{%
  \def\thearg{#1}%
  \ifx\thearg\empty \def\thearg{1}\fi
  %
  \expandafter\splitoff\thearg\endmark
  \ifx\rest\empty
    %
    %
    \ifnum\lccode\expandafter`\thearg=0\relax
      \numericenumerate 
    \else
      \ifnum\lccode\expandafter`\thearg=\expandafter`\thearg\relax
        \lowercaseenumerate 
      \else
        \uppercaseenumerate 
      \fi
    \fi
  \else
    \numericenumerate
  \fi
}

%
\def\numericenumerate{%
  \itemno = \thearg
  \startenumeration{\the\itemno}%
}

\def\lowercaseenumerate{%
  \itemno = \expandafter`\thearg
  \startenumeration{%
    \ifnum\itemno=0
      \errmessage{No more lowercase letters in @enumerate; get a bigger
                  alphabet}%
    \fi
    \char\lccode\itemno
  }%
}

\def\uppercaseenumerate{%
  \itemno = \expandafter`\thearg
  \startenumeration{%
    \ifnum\itemno=0
      \errmessage{No more uppercase letters in @enumerate; get a bigger
                  alphabet}
    \fi
    \char\uccode\itemno
  }%
}

%
\def\startenumeration#1{%
  \advance\itemno by -1
  \doitemize{#1.}\flushcr
}

\newskip\multitableparskip
\newskip\multitableparindent
\newdimen\multitablecolspace
\newskip\multitablelinespace
\multitableparskip=0pt
\multitableparindent=6pt
\multitablecolspace=12pt
\multitablelinespace=0pt

%
\let\endsetuptable\relax
\def\xendsetuptable{\endsetuptable}
\let\columnfractions\relax
\def\xcolumnfractions{\columnfractions}
\newif\ifsetpercent

%
\def\pickupwholefraction#1 {%
  \global\advance\colcount by 1
  \expandafter\xdef\csname col\the\colcount\endcsname{#1\hsize}%
  \setuptable
}

\newcount\colcount
\def\setuptable#1{%
  \def\firstarg{#1}%
  \ifx\firstarg\xendsetuptable
    \let\go = \relax
  \else
    \ifx\firstarg\xcolumnfractions
      \global\setpercenttrue
    \else
      \ifsetpercent
         \let\go\pickupwholefraction
      \else
         \global\advance\colcount by 1
         \setbox0=\hbox{#1\unskip\space}
         \expandafter\xdef\csname col\the\colcount\endcsname{\the\wd0}%
      \fi
    \fi
    \ifx\go\pickupwholefraction
      \def\go{\pickupwholefraction#1}%
    \else
      \let\go = \setuptable
    \fi%
  \fi
  \go
}

%
\def\headitem{\checkenv\multitable \crcr \global\everytab={\bf}\the\everytab}%
%
\def\tab{\checkenv\multitable &\the\everytab}%

%
\newtoks\everytab  
\envdef\multitable{%
  \vskip\parskip
  \startsavinginserts
  %
  \def\item{\crcr}%
  \tolerance=9500
  \hbadness=9500
  \setmultitablespacing
  \parskip=\multitableparskip
  \parindent=\multitableparindent
  \overfullrule=0pt
  \global\colcount=0
  \everycr = {%
    \noalign{%
      \global\everytab={}%
      \global\colcount=0 
      \checkinserts
    }%
  }%
  \parsearg\domultitable
}
\def\domultitable#1{%
  \setuptable#1 \endsetuptable
  %
  \halign\bgroup &%
    \global\advance\colcount by 1
    \multistrut
    \vtop{%
      \hsize=\expandafter\csname col\the\colcount\endcsname
      %
      %
      %
      %
      \rightskip=0pt
      \ifnum\colcount=1
	\advance\hsize by\leftskip
      \else
	\ifsetpercent \else
	  \advance\hsize by \multitablecolspace
	\fi
      \leftskip=\multitablecolspace
      \fi
      \noindent\ignorespaces##\unskip\multistrut
    }\cr
}
\def\Emultitable{%
  \crcr
  \egroup 
  \global\setpercentfalse
}

\def\setmultitablespacing{%
  \def\multistrut{\strut}
  %
\ifdim\multitablelinespace=0pt
\setbox0=\vbox{X}\global\multitablelinespace=\the\baselineskip
\global\advance\multitablelinespace by-\ht0
\fi
\ifdim\multitableparskip>\multitablelinespace
\global\multitableparskip=\multitablelinespace
\global\advance\multitableparskip-7pt 
\fi%
\ifdim\multitableparskip=0pt
\global\multitableparskip=\multitablelinespace
\global\advance\multitableparskip-7pt 
\fi}

\message{conditionals,}

%
\def\makecond#1{%
  \expandafter\let\csname #1\endcsname = \relax
  \expandafter\let\csname iscond.#1\endcsname = 1
}
\makecond{iftex}
\makecond{ifnotdocbook}
\makecond{ifnothtml}
\makecond{ifnotinfo}
\makecond{ifnotplaintext}
\makecond{ifnotxml}

%

%
\newcount\doignorecount

\def\doignore#1{\begingroup
  \catcode`\@ = \other
  \catcode`\{ = \other
  \catcode`\} = \other
  %
  \spaceisspace
  %
  \doignorecount = 0
  %
  \dodoignore{#1}%
}

{ \catcode`_=11 
  \obeylines %
  \gdef\dodoignore#1{%
    %
    \long\def\doignoretext##1^^M@end #1{\doignoretextyyy##1^^M@#1\_STOP_}%
    \long\def\doignoretextyyy##1^^M@#1##2\_STOP_{\doignoreyyy{##2}\_STOP_}%
    %
    \obeylines %
    \doignoretext ^^M%
  }%
}

\def\doignoreyyy#1{%
  \def\temp{#1}%
  \ifx\temp\empty			
    \let\next\doignoretextzzz
  \else					
    \advance\doignorecount by 1
    \let\next\doignoretextyyy		
  \fi
  \next #1
}

%
\def\doignoretextzzz#1{%
  \ifnum\doignorecount = 0	
    \let\next\enddoignore
  \else				
    \advance\doignorecount by -1
    \let\next\doignoretext      
  \fi
  \next
}

\def\enddoignore{\endgroup\ignorespaces}

%
%
\parseargdef\set{\setyyy#1 \endsetyyy}
\def\setyyy#1 #2\endsetyyy{%
  {%
    \makevalueexpandable
    \def\temp{#2}%
    \edef\next{\gdef\makecsname{SET#1}}%
    \ifx\temp\empty
      \next{}%
    \else
      \setzzz#2\endsetzzz
    \fi
  }%
}
\def\setzzz#1 \endsetzzz{\next{#1}}

%
\parseargdef\clear{%
  {%
    \makevalueexpandable
    \global\expandafter\let\csname SET#1\endcsname=\relax
  }%
}

\def\value{\begingroup\makevalueexpandable\valuexxx}
\def\valuexxx#1{\expandablevalue{#1}\endgroup}
{
  \catcode`\- = \active \catcode`\_ = \active
  \gdef\makevalueexpandable{%
    \let\value = \expandablevalue
    \catcode`\-=\other \catcode`\_=\other
    \let-\realdash \let_\normalunderscore
  }
}

%
\def\expandablevalue#1{%
  \expandafter\ifx\csname SET#1\endcsname\relax
    {[No value for ``#1'']}%
    \message{Variable `#1', used in @value, is not set.}%
  \else
    \csname SET#1\endcsname
  \fi
}

%
%
\makecond{ifset}
\def\ifset{\parsearg{\doifset{\let\next=\ifsetfail}}}
\def\doifset#1#2{%
  {%
    \makevalueexpandable
    \let\next=\empty
    \expandafter\ifx\csname SET#2\endcsname\relax
      #1
    \fi
    \expandafter
  }\next
}
\def\ifsetfail{\doignore{ifset}}

%
%
\makecond{ifclear}
\def\ifclear{\parsearg{\doifset{\else \let\next=\ifclearfail}}}
\def\ifclearfail{\doignore{ifclear}}



\message{indexing,}

\edef\newwrite{\makecsname{ptexnewwrite}}

%
\def\newindex#1{%
  \iflinks
    \expandafter\newwrite \csname#1indfile\endcsname
    \openout \csname#1indfile\endcsname \jobname.#1 
  \fi
  \expandafter\xdef\csname#1index\endcsname{
    \noexpand\doindex{#1}}
}

%

%

%
\def\newcodeindex#1{%
  \iflinks
    \expandafter\newwrite \csname#1indfile\endcsname
    \openout \csname#1indfile\endcsname \jobname.#1
  \fi
  \expandafter\xdef\csname#1index\endcsname{%
    \noexpand\docodeindex{#1}}%
}

%
%
\def\synindex#1 #2 {\dosynindex\doindex{#1}{#2}}
\def\syncodeindex#1 #2 {\dosynindex\docodeindex{#1}{#2}}

\def\dosynindex#1#2#3{%
  \expandafter \ifx\csname donesynindex#2\endcsname \undefined
    \expandafter\closeout\csname#2indfile\endcsname
    \expandafter\let\csname\donesynindex#2\endcsname = 1
  \fi
  \expandafter\let\expandafter\temp\expandafter=\csname#3indfile\endcsname
  \expandafter\let\csname#2indfile\endcsname=\temp
  \expandafter\xdef\csname#2index\endcsname{\noexpand#1{#3}}%
}




\def\doindex#1{\edef\indexname{#1}\parsearg\singleindexer}
\def\singleindexer #1{\doind{\indexname}{#1}}

\def\docodeindex#1{\edef\indexname{#1}\parsearg\singlecodeindexer}
\def\singlecodeindexer #1{\doind{\indexname}{\code{#1}}}

%
\def\indexdummies{%
  \def\@{@}
  \def\ {\realbackslash\space }%
  \let\{ = \mylbrace
  \let\} = \myrbrace
  %
  %
  %
  %
  \def\definedummyword##1{%
    \expandafter\def\csname ##1\endcsname{\realbackslash ##1\space}%
  }%
  \def\definedummyletter##1{%
    \expandafter\def\csname ##1\endcsname{\realbackslash ##1}%
  }%
  \let\definedummyaccent\definedummyletter
  %
  \commondummies
}

%
\def\atdummies{%
  \def\@{@@}%
  \def\ {@ }%
  \let\{ = \lbraceatcmd
  \let\} = \rbraceatcmd
  %
  \def\definedummyword##1{%
    \expandafter\def\csname ##1\endcsname{@##1\space}%
  }%
  \def\definedummyletter##1{%
    \expandafter\def\csname ##1\endcsname{@##1}%
  }%
  \let\definedummyaccent\definedummyletter
  %
  \commondummies
}

%
\def\commondummies{%
  \normalturnoffactive
  \commondummiesnofonts
  \definedummyletter{_}%
  %
  \definedummyword{AA}%
  \definedummyword{AE}%
  \definedummyword{L}%
  \definedummyword{OE}%
  \definedummyword{O}%
  \definedummyword{aa}%
  \definedummyword{ae}%
  \definedummyword{l}%
  \definedummyword{oe}%
  \definedummyword{o}%
  \definedummyword{ss}%
  \definedummyword{exclamdown}%
  \definedummyword{questiondown}%
  \definedummyword{ordf}%
  \definedummyword{ordm}%
  %
  \definedummyword{bf}%
  \definedummyword{gtr}%
  \definedummyword{hat}%
  \definedummyword{less}%
  \definedummyword{sf}%
  \definedummyword{sl}%
  \definedummyword{tclose}%
  \definedummyword{tt}%
  \definedummyword{LaTeX}%
  \definedummyword{TeX}%
  %
  \definedummyword{bullet}%
  \definedummyword{comma}%
  \definedummyword{copyright}%
  \definedummyword{registeredsymbol}%
  \definedummyword{dots}%
  \definedummyword{enddots}%
  \definedummyword{equiv}%
  \definedummyword{error}%
  \definedummyword{euro}%
  \definedummyword{expansion}%
  \definedummyword{minus}%
  \definedummyword{pounds}%
  \definedummyword{point}%
  \definedummyword{print}%
  \definedummyword{result}%
  %
  \makevalueexpandable
  %
  \unsepspaces
  %
  \turnoffmacros
}

%
{
  \catcode`\~=\other
  \gdef\commondummiesnofonts{%
    \definedummyletter{!}%
    \definedummyaccent{"}%
    \definedummyaccent{'}%
    \definedummyletter{*}%
    \definedummyaccent{,}%
    \definedummyletter{.}%
    \definedummyletter{/}%
    \definedummyletter{:}%
    \definedummyaccent{=}%
    \definedummyletter{?}%
    \definedummyaccent{^}%
    \definedummyaccent{`}%
    \definedummyaccent{~}%
    \definedummyword{u}%
    \definedummyword{v}%
    \definedummyword{H}%
    \definedummyword{dotaccent}%
    \definedummyword{ringaccent}%
    \definedummyword{tieaccent}%
    \definedummyword{ubaraccent}%
    \definedummyword{udotaccent}%
    \definedummyword{dotless}%
    %
    \definedummyword{b}%
    \definedummyword{i}%
    \definedummyword{r}%
    \definedummyword{sc}%
    \definedummyword{t}%
    %
    \definedummyword{acronym}%
    \definedummyword{cite}%
    \definedummyword{code}%
    \definedummyword{command}%
    \definedummyword{dfn}%
    \definedummyword{emph}%
    \definedummyword{env}%
    \definedummyword{file}%
    \definedummyword{kbd}%
    \definedummyword{key}%
    \definedummyword{math}%
    \definedummyword{option}%
    \definedummyword{samp}%
    \definedummyword{strong}%
    \definedummyword{tie}%
    \definedummyword{uref}%
    \definedummyword{url}%
    \definedummyword{var}%
    \definedummyword{verb}%
    \definedummyword{w}%
  }
}

%
\def\indexnofonts{%
  \def\definedummyaccent##1{%
    \expandafter\let\csname ##1\endcsname\asis
  }%
  \def\definedummyletter##1{%
    \expandafter\def\csname ##1\endcsname{}%
  }%
  \let\definedummyword\definedummyaccent
  \commondummiesnofonts
  %
  %
  \def\ { }%
  \def\@{@}%
  \def\_{\normalunderscore}%
  %
  \def\AA{AA}%
  \def\AE{AE}%
  \def\L{L}%
  \def\OE{OE}%
  \def\O{O}%
  \def\aa{aa}%
  \def\ae{ae}%
  \def\l{l}%
  \def\oe{oe}%
  \def\o{o}%
  \def\ss{ss}%
  \def\exclamdown{!}%
  \def\questiondown{?}%
  \def\ordf{a}%
  \def\ordm{o}%
  \def\LaTeX{LaTeX}%
  \def\TeX{TeX}%
  %
  \def\bullet{bullet}%
  \def\comma{,}%
  \def\copyright{copyright}%
  \def\registeredsymbol{R}%
  \def\dots{...}%
  \def\enddots{...}%
  \def\equiv{==}%
  \def\error{error}%
  \def\euro{euro}%
  \def\expansion{==>}%
  \def\minus{-}%
  \def\pounds{pounds}%
  \def\point{.}%
  \def\print{-|}%
  \def\result{=>}%
  %
  \emptyusermacros
}

\let\indexbackslash=0  
\let\SETmarginindex=\relax 

\def\doind#1#2{\dosubind{#1}{#2}{}}

%
\def\dosubind#1#2#3{%
  \iflinks
  {%
    \toks0 = {#2}%
    \def\thirdarg{#3}%
    \ifx\thirdarg\empty \else
      \toks0 = \expandafter{\the\toks0 \space #3}%
    \fi
    \edef\writeto{\csname#1indfile\endcsname}%
    \ifvmode
      \dosubindsanitize
    \else
      \dosubindwrite
    \fi
  }%
  \fi
}

%
\def\dosubindwrite{%
  \ifx\SETmarginindex\relax\else
    \insert\margin{\hbox{\vrule height8pt depth3pt width0pt \the\toks0}}%
  \fi
  %
  \indexdummies 
  \escapechar=`\\
  \def\backslashcurfont{\indexbackslash}
  %
  {\indexnofonts
   \edef\temp{\the\toks0}
   \xdef\indexsorttmp{\temp}%
  }%
  %
  \edef\temp{%
    \write\writeto{%
      \string\entry{\indexsorttmp}{\noexpand\folio}{\the\toks0}}%
  }%
  \temp
}

%
%
%
%
%
%
\edef\zeroskipmacro{\expandafter\the\csname z@skip\endcsname}
%
%
\def\dosubindsanitize{%
  \skip0 = \lastskip
  \edef\lastskipmacro{\the\lastskip}%
  \count255 = \lastpenalty
  %
  \ifx\lastskipmacro\zeroskipmacro
  \else
    \vskip-\skip0
  \fi
  \dosubindwrite
  \ifx\lastskipmacro\zeroskipmacro
    %
    \ifnum\count255>9999 \penalty\count255 \fi
  \else
    \nobreak\vskip\skip0
  \fi
}



\def\cindexsub {\begingroup\obeylines\cindexsub}
{\obeylines %
\gdef\cindexsub "#1" #2^^M{\endgroup %
\dosubind{cp}{#2}{#1}}}


%
\parseargdef\printindex{\begingroup
  \dobreak \chapheadingskip{10000}%
  \smallfonts \rm
  \tolerance = 9500
  \everypar = {}
  %
  \catcode`\@ = 11
  \openin 1 \jobname.#1s
  \ifeof 1
    \putwordIndexNonexistent
  \else
    %
    \read 1 to \temp
    \ifeof 1
      \putwordIndexIsEmpty
    \else
      \def\indexbackslash{\backslashcurfont}%
      \catcode`\\ = 0
      \escapechar = `\\
      \begindoublecolumns
      \input \jobname.#1s
      \enddoublecolumns
    \fi
  \fi
  \closein 1
\endgroup}


\def\initial#1{{%
  \let\tentt=\sectt \let\tt=\sectt \let\sf=\sectt
  %
  \removelastskip
  %
  \nobreak
  \vskip 0pt plus 3\baselineskip
  \penalty 0
  \vskip 0pt plus -3\baselineskip
  %
  %
  \vskip 1.67\baselineskip plus .5\baselineskip
  \leftline{\secbf #1}%
  \nobreak
  \vskip .33\baselineskip plus .1\baselineskip
}}

%
%
\def\entry{%
  \begingroup
    %
    \par
    %
    \parfillskip = 0in
    %
    \parskip = 0in
    %
    \finalhyphendemerits = 0
    %
    %
    \hangindent = 2em
    %
    \rightskip = 0pt plus1fil
    %
    \vskip 0pt plus1pt
    %
    \afterassignment\doentry
    \let\temp =
}
\def\doentry{%
    \bgroup 
      \noindent
      \aftergroup\finishentry
}
\def\finishentry#1{%
    %
    \def\tempa{{\rm }}%
    \def\tempb{#1}%
    \edef\tempc{\tempa}%
    \edef\tempd{\tempb}%
    \ifx\tempc\tempd
      \ %
    \else
      %
      \hfil\penalty50
      \null\nobreak\indexdotfill 
      %
      \ifpdf
	\pdfgettoks#1.%
	\ \the\toksA
      \else
	\ #1%
      \fi
    \fi
    \par
  \endgroup
}

\def\indexdotfill{\cleaders
  \hbox{$\mathsurround=0pt \mkern1.5mu ${\it .}$ \mkern1.5mu$}\hskip 1em plus 1fill}

\newskip\secondaryindent \secondaryindent=0.5cm
\def\secondary#1#2{{%
  \parfillskip=0in
  \parskip=0in
  \hangindent=1in
  \hangafter=1
  \noindent\hskip\secondaryindent\hbox{#1}\indexdotfill
  \ifpdf
    \pdfgettoks#2.\ \the\toksA 
  \else
    #2
  \fi
  \par
}}

\catcode`\@=11

\newbox\partialpage
\newdimen\doublecolumnhsize

\def\begindoublecolumns{\begingroup 
  \output = {%
    %
    \ifvoid\partialpage \else
      \onepageout{\pagecontents\partialpage}%
    \fi
    \global\setbox\partialpage = \vbox{%
      \unvbox\PAGE
      \kern-\topskip \kern\baselineskip
    }%
  }%
  \eject 
  %
  \output = {\doublecolumnout}%
  %
  %
  %
  %
  \doublecolumnhsize = \hsize
    \advance\doublecolumnhsize by -.04154\hsize
    \divide\doublecolumnhsize by 2
  \hsize = \doublecolumnhsize
  %
  \vsize = 2\vsize
}

%
\def\doublecolumnout{%
  \splittopskip=\topskip \splitmaxdepth=\maxdepth
  \dimen@ = \vsize
  \divide\dimen@ by 2
  \advance\dimen@ by -\ht\partialpage
  %
  \setbox0=\vsplit255 to\dimen@ \setbox2=\vsplit255 to\dimen@
  \onepageout\pagesofar
  \unvbox255
  \penalty\outputpenalty
}
%
\def\pagesofar{%
  \unvbox\partialpage
  \hsize = \doublecolumnhsize
  \wd0=\hsize \wd2=\hsize
  \hbox to\pagewidth{\box0\hfil\box2}%
}
%
\def\enddoublecolumns{%
  \output = {%
    \balancecolumns
    %
    \global\output = {\onepageout{\pagecontents\PAGE}}%
  }%
  \eject
  \endgroup 
  %
  \pagegoal = \vsize
}
%
\def\balancecolumns{%
  \setbox0 = \vbox{\unvbox255}
  \dimen@ = \ht0
  \advance\dimen@ by \topskip
  \advance\dimen@ by-\baselineskip
  \divide\dimen@ by 2 
  \splittopskip = \topskip
  {%
    \vbadness = 10000
    \loop
      \global\setbox3 = \copy0
      \global\setbox1 = \vsplit3 to \dimen@
    \ifdim\ht3>\dimen@
      \global\advance\dimen@ by 1pt
    \repeat
  }%
  \setbox0=\vbox to\dimen@{\unvbox1}%
  \setbox2=\vbox to\dimen@{\unvbox3}%
  \pagesofar
}
\catcode`\@ = \other

\message{sectioning,}

\newcount\unnumberedno \unnumberedno = 10000
\newcount\chapno
\newcount\secno        \secno=0
\newcount\subsecno     \subsecno=0
\newcount\subsubsecno  \subsubsecno=0

\newcount\appendixno  \appendixno = `\@
%
%
\def\appendixletter{%
  \ifnum\appendixno=`A A%
  \else\ifnum\appendixno=`B B%
  \else\ifnum\appendixno=`C C%
  \else\ifnum\appendixno=`D D%
  \else\ifnum\appendixno=`E E%
  \else\ifnum\appendixno=`F F%
  \else\ifnum\appendixno=`G G%
  \else\ifnum\appendixno=`H H%
  \else\ifnum\appendixno=`I I%
  \else\ifnum\appendixno=`J J%
  \else\ifnum\appendixno=`K K%
  \else\ifnum\appendixno=`L L%
  \else\ifnum\appendixno=`M M%
  \else\ifnum\appendixno=`N N%
  \else\ifnum\appendixno=`O O%
  \else\ifnum\appendixno=`P P%
  \else\ifnum\appendixno=`Q Q%
  \else\ifnum\appendixno=`R R%
  \else\ifnum\appendixno=`S S%
  \else\ifnum\appendixno=`T T%
  \else\ifnum\appendixno=`U U%
  \else\ifnum\appendixno=`V V%
  \else\ifnum\appendixno=`W W%
  \else\ifnum\appendixno=`X X%
  \else\ifnum\appendixno=`Y Y%
  \else\ifnum\appendixno=`Z Z%
  \else\char\the\appendixno
  \fi\fi\fi\fi\fi\fi\fi\fi\fi\fi\fi\fi\fi
  \fi\fi\fi\fi\fi\fi\fi\fi\fi\fi\fi\fi\fi}

\def\thischapter{}
\def\thissection{}

\newcount\absseclevel 
\newcount\secbase\secbase=0 

\def\raisesections{\global\advance\secbase by -1}

\def\lowersections{\global\advance\secbase by 1}

\chardef\maxseclevel = 3
%
\chardef\unmlevel = \maxseclevel
%
\def\chapheadtype{N}

\def\genhead#1#2#3{%
  \absseclevel=#2
  \advance\absseclevel by \secbase
  \ifnum \absseclevel < 0
    \absseclevel = 0
  \else
    \ifnum \absseclevel > 3
      \absseclevel = 3
    \fi
  \fi
  \def\headtype{#1}%
  \if \headtype U%
    \ifnum \absseclevel < \unmlevel
      \chardef\unmlevel = \absseclevel
    \fi
  \else
    \ifnum \absseclevel = 0
      \edef\chapheadtype{\headtype}%
    \else
      \if \headtype A\if \chapheadtype N%
	\errmessage{@appendix... within a non-appendix chapter}%
      \fi\fi
    \fi
    \ifnum \absseclevel > \unmlevel
      \def\headtype{U}%
    \else
      \chardef\unmlevel = 3
    \fi
  \fi
  \if \headtype U%
    \ifcase\absseclevel
	\unnumberedzzz{#3}%
    \or \unnumberedseczzz{#3}%
    \or \unnumberedsubseczzz{#3}%
    \or \unnumberedsubsubseczzz{#3}%
    \fi
  \else
    \if \headtype A%
      \ifcase\absseclevel
	  \appendixzzz{#3}%
      \or \appendixsectionzzz{#3}%
      \or \appendixsubseczzz{#3}%
      \or \appendixsubsubseczzz{#3}%
      \fi
    \else
      \ifcase\absseclevel
	  \chapterzzz{#3}%
      \or \seczzz{#3}%
      \or \numberedsubseczzz{#3}%
      \or \numberedsubsubseczzz{#3}%
      \fi
    \fi
  \fi
  \suppressfirstparagraphindent
}

\def\numhead{\genhead N}
\def\apphead{\genhead A}
\def\unnmhead{\genhead U}

%
\let\chaplevelprefix = \empty
\outer\parseargdef\chapter{\numhead0{#1}} 
\def\chapterzzz#1{%
  \global\secno=0 \global\subsecno=0 \global\subsubsecno=0
    \global\advance\chapno by 1
  %
  \gdef\chaplevelprefix{\the\chapno.}%
  \resetallfloatnos
  \message{\putwordChapter\space \the\chapno}%
  %
  \chapmacro{#1}{Ynumbered}{\the\chapno}%
  %
  \global\let\section = \numberedsec
  \global\let\subsection = \numberedsubsec
  \global\let\subsubsection = \numberedsubsubsec
}

\outer\parseargdef\appendix{\apphead0{#1}} 
\def\appendixzzz#1{%
  \global\secno=0 \global\subsecno=0 \global\subsubsecno=0
    \global\advance\appendixno by 1
  \gdef\chaplevelprefix{\appendixletter.}%
  \resetallfloatnos
  \def\appendixnum{\putwordAppendix\space \appendixletter}%
  \message{\appendixnum}%
  \chapmacro{#1}{Yappendix}{\appendixletter}%
  \global\let\section = \appendixsec
  \global\let\subsection = \appendixsubsec
  \global\let\subsubsection = \appendixsubsubsec
}

\outer\parseargdef\unnumbered{\unnmhead0{#1}} 
\def\unnumberedzzz#1{%
  \global\secno=0 \global\subsecno=0 \global\subsubsecno=0
    \global\advance\unnumberedno by 1
  %
  \global\let\chaplevelprefix = \empty
  \resetallfloatnos
  %
  %
  \toks0 = {#1}%
  \message{(\the\toks0)}%
  \chapmacro{#1}{Ynothing}{\the\unnumberedno}%
  \global\let\section = \unnumberedsec
  \global\let\subsection = \unnumberedsubsec
  \global\let\subsubsection = \unnumberedsubsubsec
}

\outer\parseargdef\centerchap{%
  \let\centerparametersmaybe = \centerparameters
  \unnmhead0{#1}%
  \let\centerparametersmaybe = \relax
}


\outer\parseargdef\numberedsec{\numhead1{#1}} 
\def\seczzz#1{%
  \global\subsecno=0 \global\subsubsecno=0  \global\advance\secno by 1
  \sectionheading{#1}{sec}{Ynumbered}{\the\chapno.\the\secno}%
}

\outer\parseargdef\appendixsection{\apphead1{#1}} 
\def\appendixsectionzzz#1{%
  \global\subsecno=0 \global\subsubsecno=0  \global\advance\secno by 1
  \sectionheading{#1}{sec}{Yappendix}{\appendixletter.\the\secno}%
}
\let\appendixsec\appendixsection

\outer\parseargdef\unnumberedsec{\unnmhead1{#1}} 
\def\unnumberedseczzz#1{%
  \global\subsecno=0 \global\subsubsecno=0  \global\advance\secno by 1
  \sectionheading{#1}{sec}{Ynothing}{\the\unnumberedno.\the\secno}%
}

\outer\parseargdef\numberedsubsec{\numhead2{#1}} 
\def\numberedsubseczzz#1{%
  \global\subsubsecno=0  \global\advance\subsecno by 1
  \sectionheading{#1}{subsec}{Ynumbered}{\the\chapno.\the\secno.\the\subsecno}%
}

\outer\parseargdef\appendixsubsec{\apphead2{#1}} 
\def\appendixsubseczzz#1{%
  \global\subsubsecno=0  \global\advance\subsecno by 1
  \sectionheading{#1}{subsec}{Yappendix}%
                 {\appendixletter.\the\secno.\the\subsecno}%
}

\outer\parseargdef\unnumberedsubsec{\unnmhead2{#1}} 
\def\unnumberedsubseczzz#1{%
  \global\subsubsecno=0  \global\advance\subsecno by 1
  \sectionheading{#1}{subsec}{Ynothing}%
                 {\the\unnumberedno.\the\secno.\the\subsecno}%
}

\outer\parseargdef\numberedsubsubsec{\numhead3{#1}} 
\def\numberedsubsubseczzz#1{%
  \global\advance\subsubsecno by 1
  \sectionheading{#1}{subsubsec}{Ynumbered}%
                 {\the\chapno.\the\secno.\the\subsecno.\the\subsubsecno}%
}

\outer\parseargdef\appendixsubsubsec{\apphead3{#1}} 
\def\appendixsubsubseczzz#1{%
  \global\advance\subsubsecno by 1
  \sectionheading{#1}{subsubsec}{Yappendix}%
                 {\appendixletter.\the\secno.\the\subsecno.\the\subsubsecno}%
}

\outer\parseargdef\unnumberedsubsubsec{\unnmhead3{#1}} 
\def\unnumberedsubsubseczzz#1{%
  \global\advance\subsubsecno by 1
  \sectionheading{#1}{subsubsec}{Ynothing}%
                 {\the\unnumberedno.\the\secno.\the\subsecno.\the\subsubsecno}%
}

\let\section = \numberedsec
\let\subsection = \numberedsubsec
\let\subsubsection = \numberedsubsubsec



\def\majorheading{%
  {\advance\chapheadingskip by 10pt \chapbreak }%
  \parsearg\chapheadingzzz
}

\def\chapheadingzzz#1{%
  {\chapfonts \vbox{\hyphenpenalty=10000\tolerance=5000
                    \parindent=0pt\raggedright
                    \rm #1\hfill}}%
  \bigskip \par\penalty 200\relax
  \suppressfirstparagraphindent
}

\parseargdef\heading{\sectionheading{#1}{sec}{Yomitfromtoc}{}
  \suppressfirstparagraphindent}
\parseargdef\subheading{\sectionheading{#1}{subsec}{Yomitfromtoc}{}
  \suppressfirstparagraphindent}
\parseargdef\subsubheading{\sectionheading{#1}{subsubsec}{Yomitfromtoc}{}
  \suppressfirstparagraphindent}


\def\dobreak#1#2{\par\ifdim\lastskip<#1\removelastskip\penalty#2\vskip#1\fi}


\newskip\chapheadingskip

\def\chapbreak{\dobreak \chapheadingskip {-4000}}
\def\chappager{\par\vfill\supereject}
\def\chapoddpage{\chappager \ifodd\pageno \else \hbox to 0pt{} \chappager\fi}

\def\setchapternewpage #1 {\csname CHAPPAG#1\endcsname}

\def\CHAPPAGoff{%
\global\let\contentsalignmacro = \chappager
\global\let\pchapsepmacro=\chapbreak
\global\let\pagealignmacro=\chappager}

\def\CHAPPAGon{%
\global\let\contentsalignmacro = \chappager
\global\let\pchapsepmacro=\chappager
\global\let\pagealignmacro=\chappager
\global\def\HEADINGSon{\HEADINGSsingle}}

\def\CHAPPAGodd{%
\global\let\contentsalignmacro = \chapoddpage
\global\let\pchapsepmacro=\chapoddpage
\global\let\pagealignmacro=\chapoddpage
\global\def\HEADINGSon{\HEADINGSdouble}}

\CHAPPAGon

%
%
\def\Ynothingkeyword{Ynothing}
\def\Yomitfromtockeyword{Yomitfromtoc}
\def\Yappendixkeyword{Yappendix}
\def\chapmacro#1#2#3{%
  \pchapsepmacro
  {%
    \chapfonts \rm
    %
    \gdef\thissection{#1}%
    \gdef\thischaptername{#1}%
    %
    \def\temptype{#2}%
    \ifx\temptype\Ynothingkeyword
      \setbox0 = \hbox{}%
      \def\toctype{unnchap}%
      \gdef\thischapter{#1}%
    \else\ifx\temptype\Yomitfromtockeyword
      \setbox0 = \hbox{}
      \def\toctype{omit}%
      \gdef\thischapter{}%
    \else\ifx\temptype\Yappendixkeyword
      \setbox0 = \hbox{\putwordAppendix{} #3\enspace}%
      \def\toctype{app}%
      %
      \xdef\thischapter{\putwordAppendix{} \appendixletter:
                        \noexpand\thischaptername}%
    \else
      \setbox0 = \hbox{#3\enspace}%
      \def\toctype{numchap}%
      \xdef\thischapter{\putwordChapter{} \the\chapno:
                        \noexpand\thischaptername}%
    \fi\fi\fi
    %
    \writetocentry{\toctype}{#1}{#3}%
    %
    \donoderef{#2}%
    %
    \vbox{\hyphenpenalty=10000 \tolerance=5000 \parindent=0pt \raggedright
          \hangindent=\wd0 \centerparametersmaybe
          \unhbox0 #1\par}%
  }%
  \nobreak\bigskip 
  \nobreak
}

\let\centerparametersmaybe = \relax
\def\centerparameters{%
  \advance\rightskip by 3\rightskip
  \leftskip = \rightskip
  \parfillskip = 0pt
}

%
\def\setchapterstyle #1 {\csname CHAPF#1\endcsname}

\def\chfopen #1#2{\chapoddpage {\chapfonts
\vbox to 3in{\vfil \hbox to\hsize{\hfil #2} \hbox to\hsize{\hfil #1} \vfil}}%
\par\penalty 5000 %
}
\def\centerchfopen #1{%
\chapoddpage {\chapfonts \vbox{\hyphenpenalty=10000\tolerance=5000
                       \parindent=0pt
                       \hfill {\rm #1}\hfill}}\bigskip \par\nobreak
}
\def\CHAPFopen{%
  \global\let\chapmacro=\chfopen
  \global\let\centerchapmacro=\centerchfopen}

%
\newskip\secheadingskip

\newskip\subsecheadingskip


%
%
\def\sectionheading#1#2#3#4{%
  {%
    \csname #2fonts\endcsname \rm
    %
    \csname #2headingbreak\endcsname
    %
    \def\sectionlevel{#2}%
    \def\temptype{#3}%
    \ifx\temptype\Ynothingkeyword
      \setbox0 = \hbox{}%
      \def\toctype{unn}%
      \gdef\thissection{#1}%
    \else\ifx\temptype\Yomitfromtockeyword
      \setbox0 = \hbox{}%
      \def\toctype{omit}%
      \let\sectionlevel=\empty
    \else\ifx\temptype\Yappendixkeyword
      \setbox0 = \hbox{#4\enspace}%
      \def\toctype{app}%
      \gdef\thissection{#1}%
    \else
      \setbox0 = \hbox{#4\enspace}%
      \def\toctype{num}%
      \gdef\thissection{#1}%
    \fi\fi\fi
    %
    \writetocentry{\toctype\sectionlevel}{#1}{#4}%
    %
    \donoderef{#3}%
    %
    \vbox{\hyphenpenalty=10000 \tolerance=5000 \parindent=0pt \raggedright
          \hangindent=\wd0  
          \unhbox0 #1}%
  }%
  \kern .5 \csname #2headingskip\endcsname
  %
  \nobreak
  %
  \vskip-\parskip
  %
  %
  \penalty 10001
}

\message{toc,}
\newwrite\tocfile

%
%
%
\newif\iftocfileopened
\def\omitkeyword{omit}%
\def\writetocentry#1#2#3{%
  \edef\writetoctype{#1}%
  \ifx\writetoctype\omitkeyword \else
    \iftocfileopened\else
      \immediate\openout\tocfile = \jobname.toc
      \global\tocfileopenedtrue
    \fi
    \iflinks
      \toks0 = {#2}%
      \toks2 = \expandafter{\lastnode}%
      \edef\temp{\write\tocfile{\realbackslash #1entry{\the\toks0}{#3}%
                               {\the\toks2}{\noexpand\folio}}}%
      \temp
    \fi
  \fi
  %
  \ifpdf \global\pdfmakepagedesttrue \fi
}

\newskip\contentsrightmargin \contentsrightmargin=1in
\newcount\savepageno
\newcount\lastnegativepageno \lastnegativepageno = -1

%
\def\startcontents#1{%
  \contentsalignmacro
  \immediate\closeout\tocfile
  %
  \def\thischapter{}%
  \chapmacro{#1}{Yomitfromtoc}{}%
  \savepageno = \pageno
  \begingroup                  
    \catcode`\\=0  \catcode`\{=1  \catcode`\}=2  \catcode`\@=11
    \raggedbottom             
    \advance\hsize by -\contentsrightmargin 
    %
    \ifnum \pageno>0 \global\pageno = \lastnegativepageno \fi
}

\def\contents{%
  \startcontents{\putwordTOC}%
    \openin 1 \jobname.toc
    \ifeof 1 \else
      \input \jobname.toc
    \fi
    \vfill \eject
    \contentsalignmacro 
    \ifeof 1 \else
      \pdfmakeoutlines
    \fi
    \closein 1
  \endgroup
  \lastnegativepageno = \pageno
  \global\pageno = \savepageno
}

\def\summarycontents{%
  \startcontents{\putwordShortTOC}%
    \let\numchapentry = \shortchapentry
    \let\appentry = \shortchapentry
    \let\unnchapentry = \shortunnchapentry
    \secfonts
    \let\rm=\shortcontrm \let\bf=\shortcontbf
    \let\sl=\shortcontsl \let\tt=\shortconttt
    \rm
    \hyphenpenalty = 10000
    \advance\baselineskip by 1pt 
    \def\numsecentry##1##2##3##4{}
    \let\appsecentry = \numsecentry
    \let\unnsecentry = \numsecentry
    \let\numsubsecentry = \numsecentry
    \let\appsubsecentry = \numsecentry
    \let\unnsubsecentry = \numsecentry
    \let\numsubsubsecentry = \numsecentry
    \let\appsubsubsecentry = \numsecentry
    \let\unnsubsubsecentry = \numsecentry
    \openin 1 \jobname.toc
    \ifeof 1 \else
      \input \jobname.toc
    \fi
    \closein 1
    \vfill \eject
    \contentsalignmacro 
  \endgroup
  \lastnegativepageno = \pageno
  \global\pageno = \savepageno
}
\let\shortcontents = \summarycontents

%
\def\shortchaplabel#1{%
  %
  \hbox to 1em{#1\hss}%
}


\def\numchapentry#1#2#3#4{\dochapentry{#2\labelspace#1}{#4}}
%
\def\shortchapentry#1#2#3#4{%
  \tocentry{\shortchaplabel{#2}\labelspace #1}{\doshortpageno\bgroup#4\egroup}%
}

%
\def\appendixbox#1{%
  \setbox0 = \hbox{\putwordAppendix{} M}%
  \hbox to \wd0{\putwordAppendix{} #1\hss}}
\def\appentry#1#2#3#4{\dochapentry{\appendixbox{#2}\labelspace#1}{#4}}

\def\unnchapentry#1#2#3#4{\dochapentry{#1}{#4}}
\def\shortunnchapentry#1#2#3#4{\tocentry{#1}{\doshortpageno\bgroup#4\egroup}}

\def\numsecentry#1#2#3#4{\dosecentry{#2\labelspace#1}{#4}}
\let\appsecentry=\numsecentry
\def\unnsecentry#1#2#3#4{\dosecentry{#1}{#4}}

\def\numsubsecentry#1#2#3#4{\dosubsecentry{#2\labelspace#1}{#4}}
\let\appsubsecentry=\numsubsecentry
\def\unnsubsecentry#1#2#3#4{\dosubsecentry{#1}{#4}}

\def\numsubsubsecentry#1#2#3#4{\dosubsubsecentry{#2\labelspace#1}{#4}}
\let\appsubsubsecentry=\numsubsubsecentry
\def\unnsubsubsecentry#1#2#3#4{\dosubsubsecentry{#1}{#4}}

\newdimen\tocindent \tocindent = 15pt

%
\def\dochapentry#1#2{%
   \penalty-300 \vskip1\baselineskip plus.33\baselineskip minus.25\baselineskip
   \begingroup
     \chapentryfonts
     \tocentry{#1}{\dopageno\bgroup#2\egroup}%
   \endgroup
   \nobreak\vskip .25\baselineskip plus.1\baselineskip
}

\def\dosecentry#1#2{\begingroup
  \secentryfonts \leftskip=\tocindent
  \tocentry{#1}{\dopageno\bgroup#2\egroup}%
\endgroup}

\def\dosubsecentry#1#2{\begingroup
  \subsecentryfonts \leftskip=2\tocindent
  \tocentry{#1}{\dopageno\bgroup#2\egroup}%
\endgroup}

\def\dosubsubsecentry#1#2{\begingroup
  \subsubsecentryfonts \leftskip=3\tocindent
  \tocentry{#1}{\dopageno\bgroup#2\egroup}%
\endgroup}

\let\tocentry = \entry

\def\labelspace{\hskip1em \relax}

\def\dopageno#1{{\rm #1}}
\def\doshortpageno#1{{\rm #1}}

\def\chapentryfonts{\secfonts \rm}
\def\secentryfonts{\textfonts}
\def\subsecentryfonts{\textfonts}
\def\subsubsecentryfonts{\textfonts}

\message{environments,}

%
%
\def\point{$\star$}
\def\result{\leavevmode\raise.15ex\hbox to 1em{\hfil$\Rightarrow$\hfil}}
\def\expansion{\leavevmode\raise.1ex\hbox to 1em{\hfil$\mapsto$\hfil}}
\def\print{\leavevmode\lower.1ex\hbox to 1em{\hfil$\dashv$\hfil}}
\def\equiv{\leavevmode\lower.1ex\hbox to 1em{\hfil$\ptexequiv$\hfil}}

%
\newbox\errorbox
{\tentt \global\dimen0 = 3em}
\dimen2 = .55pt 
\setbox0 = \hbox{\kern-.75pt \tensf error\kern-1.5pt}
\setbox\errorbox=\hbox to \dimen0{\hfil
   \hsize = \dimen0 \advance\hsize by -5.8pt 
   \advance\hsize by -2\dimen2 
   \vbox{%
      \hrule height\dimen2
      \hbox{\vrule width\dimen2 \kern3pt          
         \vtop{\kern2.4pt \box0 \kern2.4pt}
         \kern3pt\vrule width\dimen2}
      \hrule height\dimen2}
    \hfil}
\def\error{\leavevmode\lower.7ex\copy\errorbox}


\envdef\tex{%
  \catcode `\\=0 \catcode `\{=1 \catcode `\}=2
  \catcode `\$=3 \catcode `\&=4 \catcode `\#=6
  \catcode `\^=7 \catcode `\_=8 \catcode `\~=\active \let~=\tie
  \catcode `\%=14
  \catcode `\+=\other
  \catcode `\"=\other
  \catcode `\|=\other
  \catcode `\<=\other
  \catcode `\>=\other
  \escapechar=`\\
  \let\b=\ptexb
  \let\bullet=\ptexbullet
  \let\c=\ptexc
  \let\,=\ptexcomma
  \let\.=\ptexdot
  \let\dots=\ptexdots
  \let\equiv=\ptexequiv
  \let\!=\ptexexclam
  \let\i=\ptexi
  \let\indent=\ptexindent
  \let\noindent=\ptexnoindent
  \let\{=\ptexlbrace
  \let\+=\tabalign
  \let\}=\ptexrbrace
  \let\/=\ptexslash
  \let\*=\ptexstar
  \let\t=\ptext
  \def\endldots{\mathinner{\ldots\ldots\ldots\ldots}}%
  \def\enddots{\relax\ifmmode\endldots\else$\mathsurround=0pt \endldots\,$\fi}%
  \def\@{@}%
}


\newskip\lispnarrowing \lispnarrowing=0.4in

\def\lisppar{\null\endgraf}

\newskip\envskipamount \envskipamount = 0pt

%
\def\aboveenvbreak{{%
  \ifnum \lastpenalty=10000 \else
    \advance\envskipamount by \parskip
    \endgraf
    \ifdim\lastskip<\envskipamount
      \removelastskip
      \ifnum\lastpenalty<10000 \penalty-50 \fi
      \vskip\envskipamount
    \fi
  \fi
}}

\let\afterenvbreak = \aboveenvbreak

\let\nonarrowing=\relax

\font\circle=lcircle10
\newdimen\circthick
\newdimen\cartouter\newdimen\cartinner
\newskip\normbskip\newskip\normpskip\newskip\normlskip
\circthick=\fontdimen8\circle
\def\ctl{{\circle\char'013\hskip -6pt}}
\def\ctr{{\hskip 6pt\circle\char'010}}
\def\cbl{{\circle\char'012\hskip -6pt}}
\def\cbr{{\hskip 6pt\circle\char'011}}
\def\carttop{\hbox to \cartouter{\hskip\lskip
        \ctl\leaders\hrule height\circthick\hfil\ctr
        \hskip\rskip}}
\def\cartbot{\hbox to \cartouter{\hskip\lskip
        \cbl\leaders\hrule height\circthick\hfil\cbr
        \hskip\rskip}}
\newskip\lskip\newskip\rskip

\envdef\cartouche{%
  \ifhmode\par\fi  
  \startsavinginserts
  \lskip=\leftskip \rskip=\rightskip
  \leftskip=0pt\rightskip=0pt 
  \cartinner=\hsize \advance\cartinner by-\lskip
  \advance\cartinner by-\rskip
  \cartouter=\hsize
  \advance\cartouter by 18.4pt	
  \normbskip=\baselineskip \normpskip=\parskip \normlskip=\lineskip
  \let\nonarrowing=\comment
  \vbox\bgroup
      \baselineskip=0pt\parskip=0pt\lineskip=0pt
      \carttop
      \hbox\bgroup
	  \hskip\lskip
	  \vrule\kern3pt
	  \vbox\bgroup
	      \kern3pt
	      \hsize=\cartinner
	      \baselineskip=\normbskip
	      \lineskip=\normlskip
	      \parskip=\normpskip
	      \vskip -\parskip
	      \comment 
}
\def\Ecartouche{%
              \ifhmode\par\fi
	      \kern3pt
	  \egroup
	  \kern3pt\vrule
	  \hskip\rskip
      \egroup
      \cartbot
  \egroup
  \checkinserts
}

\def\nonfillstart{%
  \aboveenvbreak
  \hfuzz = 12pt 
  \sepspaces 
  \let\par = \lisppar 
  \obeylines 
  \parskip = 0pt
  \parindent = 0pt
  \emergencystretch = 0pt 
  \ifx\nonarrowing\relax
    \advance \leftskip by \lispnarrowing
    \exdentamount=\lispnarrowing
  \fi
  \let\exdent=\nofillexdent
}

%
\def\smallword{small}
\def\nosmallword{nosmall}
\let\SETdispenvsize\relax
\def\setnormaldispenv{%
  \ifx\SETdispenvsize\smallword
    \smallexamplefonts \rm
  \fi
}
\def\setsmalldispenv{%
  \ifx\SETdispenvsize\nosmallword
  \else
    \smallexamplefonts \rm
  \fi
}

\def\makedispenv #1#2{
  \expandafter\envdef\csname#1\endcsname {\setnormaldispenv #2}
  \expandafter\envdef\csname small#1\endcsname {\setsmalldispenv #2}
  \expandafter\let\csname E#1\endcsname \afterenvbreak
  \expandafter\let\csname Esmall#1\endcsname \afterenvbreak
}

\def\maketwodispenvs #1#2#3{
  \makedispenv{#1}{#3}
  \makedispenv{#2}{#3}
}

%
%
\maketwodispenvs {lisp}{example}{%
  \nonfillstart
  \tt
  \let\kbdfont = \kbdexamplefont 
  \gobble       
}

%
\makedispenv {display}{%
  \nonfillstart
  \gobble
}

%
\makedispenv{format}{%
  \let\nonarrowing = t%
  \nonfillstart
  \gobble
}

\envdef\flushleft{%
  \let\nonarrowing = t%
  \nonfillstart
  \gobble
}

%
\envdef\flushright{%
  \let\nonarrowing = t%
  \nonfillstart
  \advance\leftskip by 0pt plus 1fill
  \gobble
}

%
\envdef\quotation{%
  {\parskip=0pt \aboveenvbreak}
  \parindent=0pt
  %
  \ifx\nonarrowing\relax
    \advance\leftskip by \lispnarrowing
    \advance\rightskip by \lispnarrowing
    \exdentamount = \lispnarrowing
    \let\nonarrowing = \relax
  \fi
  \parsearg\quotationlabel
}

%
\def\Equotation{%
  \par
  \ifx\quotationauthor\undefined\else
    \leftline{\kern 2\leftskip \sl ---\quotationauthor}%
  \fi
  {\parskip=0pt \afterenvbreak}%
}

\def\quotationlabel#1{%
  \def\temp{#1}%
  \ifx\temp\empty \else
    {\bf #1: }%
  \fi
}

%
%
\def\dospecials{%
  \do\ \do\\\do\{\do\}\do\$\do\&%
  \do\#\do\^\do\^^K\do\_\do\^^A\do\%\do\~%
  \do\<\do\>\do\|\do\@\do+\do\"%
}
%
\def\uncatcodespecials{%
  \def\do##1{\catcode`##1=\other}\dospecials}
%
\begingroup
  \catcode`\`=\active\gdef`{\relax\lq}
\endgroup
%
%
\begingroup
  \catcode`\^^I=\active
  
\endgroup
%

%
\newdimen\tabw \setbox0=\hbox{\tt\space} \tabw=8\wd0 
\def\starttabbox{\setbox0=\hbox\bgroup}
\begingroup
  \catcode`\^^I=\active
  \gdef\tabexpand{%
    \catcode`\^^I=\active
    \def^^I{\leavevmode\egroup
      \dimen0=\wd0 
      \divide\dimen0 by\tabw
      \multiply\dimen0 by\tabw 
      \advance\dimen0 by\tabw  
      \wd0=\dimen0 \box0 \starttabbox
    }%
  }
\endgroup
\def\setupverbatim{%
  \nonfillstart
  \advance\leftskip by -\defbodyindent
  \tt
  \def\par{\leavevmode\egroup\box0\endgraf}%
  \catcode`\`=\active
  \tabexpand
  \obeylines \uncatcodespecials \sepspaces
  \everypar{\starttabbox}%
}

%
%
\begingroup
  \catcode`[=1\catcode`]=2\catcode`\{=\other\catcode`\}=\other
  \gdef\doverb{#1[\def\next##1#1}[##1\endgroup]\next]
\endgroup
%

%
%
%
%
%
%
\begingroup
  \catcode`\ =\active
  \obeylines %
  \xdef\doverbatim#1^^M#2@end verbatim{#2\noexpand\end\gobble verbatim}%
\endgroup
\envdef\verbatim{%
    \setupverbatim\doverbatim
}

%

%

%
%

\def\docopying#1@end copying{\endgroup\def\copyingtext{#1}}

\message{defuns,}

\newskip\defbodyindent \defbodyindent=.4in
\newskip\defargsindent \defargsindent=50pt
\newskip\deflastargmargin \deflastargmargin=18pt

\def\startdefun{%
  \ifnum\lastpenalty<10000
    \medbreak
  \else
    %
    \ifnum\lastpenalty=10002 \penalty2000 \fi
    %
    \medskip  
  \fi
  \parindent=0in
  \advance\leftskip by \defbodyindent
  \exdentamount=\defbodyindent
}

\def\dodefunx#1{%
  \checkenv#1%
  %
  \ifnum\lastpenalty=10002 \penalty3000 \fi
  %
  \expandafter\gobbledefun#1%
}
\def\gobbledefun#1\startdefun{}

%
\def\printdefunline#1#2{%
  \begingroup
    #1#2 \endheader
    \interlinepenalty = 10000
    \advance\rightskip by 0pt plus 1fil
    \endgraf
    \nobreak\vskip -\parskip
    \penalty 10002  
    \checkparencounts
  \endgroup
}

\def\Edefun{\endgraf\medbreak}

%
\def\makedefun#1{%
  \expandafter\let\csname E#1\endcsname = \Edefun
  \edef\temp{\noexpand\domakedefun
    \makecsname{#1}\makecsname{#1x}\makecsname{#1header}}%
  \temp
}

%
%
\def\domakedefun#1#2#3{%
  \envdef#1{%
    \startdefun
    \parseargusing\activeparens{\printdefunline#3}%
  }%
  \def#2{\dodefunx#1}%
  \def#3%
}


\makedefun{deffn}{\deffngeneral{}}

\makedefun{defop}#1 {\defopon{#1\ \putwordon}}

\def\defopon#1#2 {\deffngeneral{\putwordon\ \code{#2}}{#1\ \code{#2}} }

%
\def\deffngeneral#1#2 #3 #4\endheader{%
  \dosubind{fn}{\code{#3}}{#1}%
  \defname{#2}{}{#3}\magicamp\defunargs{#4\unskip}%
}


\makedefun{deftypefn}{\deftypefngeneral{}}

\makedefun{deftypeop}#1 {\deftypeopon{#1\ \putwordon}}

\def\deftypeopon#1#2 {\deftypefngeneral{\putwordon\ \code{#2}}{#1\ \code{#2}} }

%
\def\deftypefngeneral#1#2 #3 #4 #5\endheader{%
  \dosubind{fn}{\code{#4}}{#1}%
  \defname{#2}{#3}{#4}\defunargs{#5\unskip}%
}


\makedefun{deftypevr}{\deftypecvgeneral{}}

\makedefun{deftypecv}#1 {\deftypecvof{#1\ \putwordof}}

\def\deftypecvof#1#2 {\deftypecvgeneral{\putwordof\ \code{#2}}{#1\ \code{#2}} }

%
\def\deftypecvgeneral#1#2 #3 #4 #5\endheader{%
  \dosubind{vr}{\code{#4}}{#1}%
  \defname{#2}{#3}{#4}\defunargs{#5\unskip}%
}


\makedefun{defvr}#1 {\deftypevrheader{#1} {} }

\makedefun{defcv}#1 {\defcvof{#1\ \putwordof}}

\def\defcvof#1#2 {\deftypecvof{#1}#2 {} }

\makedefun{deftp}#1 #2 #3\endheader{%
  \doind{tp}{\code{#2}}%
  \defname{#1}{}{#2}\defunargs{#3\unskip}%
}

\makedefun{defun}{\deffnheader{\putwordDeffunc} }
\makedefun{defmac}{\deffnheader{\putwordDefmac} }
\makedefun{defspec}{\deffnheader{\putwordDefspec} }
\makedefun{deftypefun}{\deftypefnheader{\putwordDeffunc} }
\makedefun{defvar}{\defvrheader{\putwordDefvar} }
\makedefun{defopt}{\defvrheader{\putwordDefopt} }
\makedefun{deftypevar}{\deftypevrheader{\putwordDefvar} }
\makedefun{defmethod}{\defopon\putwordMethodon}
\makedefun{deftypemethod}{\deftypeopon\putwordMethodon}
\makedefun{defivar}{\defcvof\putwordInstanceVariableof}
\makedefun{deftypeivar}{\deftypecvof\putwordInstanceVariableof}

%
%
\def\defname#1#2#3{%
  \advance\leftskip by -\defbodyindent
  %
  \def\temp{#1}%
  \setbox0=\hbox{\kern\deflastargmargin \ifx\temp\empty\else [\rm\temp]\fi}
  %
  \dimen0=\hsize  \advance\dimen0 by -\wd0  \advance\dimen0 by \rightskip
  \dimen2=\hsize  \advance\dimen2 by -\defargsindent
  \parshape 2 0in \dimen0 \defargsindent \dimen2
  %
  \noindent
  \hbox to 0pt{%
    \hfil\box0 \kern-\hsize
    \kern\leftskip
  }%
  %
  \tolerance=10000 \hbadness=10000
  \exdentamount=\defbodyindent
  {%
    \df \tt
    \def\temp{#2}
    \ifx\temp\empty\else \tclose{\temp} \fi
    #3
  }%
  {\rm\enskip}
  \boldbrax
}

%
\def\defunargs#1{%
  \df \sl \hyphenchar\font=0
  %
  \let\var=\ttslanted
  #1%
  \sl\hyphenchar\font=45
}

%
\def\activeparens{%
  \catcode`\(=\active \catcode`\)=\active
  \catcode`\[=\active \catcode`\]=\active
  \catcode`\&=\active
}

\let\lparen = ( \let\rparen = )

{
  \activeparens
  \global\let(=\lparen \global\let)=\rparen
  \global\let[=\lbrack \global\let]=\rbrack
  \global\let& = \&

  \gdef\boldbrax{\let(=\opnr\let)=\clnr\let[=\lbrb\let]=\rbrb}
  \gdef\magicamp{\let&=\amprm}
}

\newcount\parencount

\newif\ifampseen
\def\amprm#1 {\ampseentrue{\bf\&#1 }}

\def\parenfont{%
  \ifampseen
    \ifnum \parencount=1 \rm \fi
  \else
    \sf
  \fi
}
\def\infirstlevel#1{%
  \ifampseen
    \ifnum\parencount=1
      #1%
    \fi
  \fi
}
\def\bfafterword#1 {#1 \bf}

\def\opnr{%
  \global\advance\parencount by 1
  {\parenfont(}%
  \infirstlevel \bfafterword
}
\def\clnr{%
  {\parenfont)}%
  \infirstlevel \sl
  \global\advance\parencount by -1
}

\newcount\brackcount
\def\lbrb{%
  \global\advance\brackcount by 1
  {\bf[}%
}
\def\rbrb{%
  {\bf]}%
  \global\advance\brackcount by -1
}

\def\checkparencounts{%
  \ifnum\parencount=0 \else \badparencount \fi
  \ifnum\brackcount=0 \else \badbrackcount \fi
}
\def\badparencount{%
  \errmessage{Unbalanced parentheses in @def}%
  \global\parencount=0
}
\def\badbrackcount{%
  \errmessage{Unbalanced square braces in @def}%
  \global\brackcount=0
}

\message{macros,}

\ifx\eTeXversion\undefined
  \newwrite\macscribble
  \def\scantokens#1{%
    \toks0={#1}%
    \immediate\openout\macscribble=\jobname.tmp
    \immediate\write\macscribble{\the\toks0}%
    \immediate\closeout\macscribble
    \input \jobname.tmp
  }
\fi

\def\scanmacro#1{%
  \begingroup
    \newlinechar`\^^M
    \let\xeatspaces\eatspaces
    \catcode`\@=0 \catcode`\\=\active \escapechar=`\@
    \spaceisspace
    %
    %
    \scantokens{#1 }%
  \endgroup
}

\def\scanexp#1{%
  \edef\temp{\noexpand\scanmacro{#1}}%
  \temp
}

\newcount\paramno   
\newtoks\macname    
\newif\ifrecursive  
\def\macrolist{}    

%
\def\cslet#1#2{%
  \expandafter\let
  \csname#1\expandafter\endcsname
  \csname#2\endcsname
}

{\catcode`\@=11
\gdef\eatspaces #1{\expandafter\trim@\expandafter{#1 }}
\gdef\trim@ #1{\trim@@ @#1 @ #1 @ @@}
\gdef\trim@@ #1@ #2@ #3@@{\trim@@@\empty #2 @}
\def\unbrace#1{#1}
\unbrace{\gdef\trim@@@ #1 } #2@{#1}
}

{\catcode`\^^M=\other \catcode`\Q=3%
\gdef\eatcr #1{\eatcra #1Q^^MQ}%
\gdef\eatcra#1^^MQ{\eatcrb#1Q}%
\gdef\eatcrb#1Q#2Q{#1}%
}



\def\scanctxt{%
  \catcode`\"=\other
  \catcode`\+=\other
  \catcode`\<=\other
  \catcode`\>=\other
  \catcode`\@=\other
  \catcode`\^=\other
  \catcode`\_=\other
  \catcode`\|=\other
  \catcode`\~=\other
}

\def\scanargctxt{%
  \scanctxt
  \catcode`\\=\other
  \catcode`\^^M=\other
}

\def\macrobodyctxt{%
  \scanctxt
  \catcode`\{=\other
  \catcode`\}=\other
  \catcode`\^^M=\other
  \usembodybackslash
}

\def\macroargctxt{%
  \scanctxt
  \catcode`\\=\other
}


{\catcode`@=0 @catcode`@\=@active
 @gdef@usembodybackslash{@let\=@mbodybackslash}
 @gdef@mbodybackslash#1\{@csname macarg.#1@endcsname}
}
\expandafter\def\csname macarg.\endcsname{\realbackslash}

\def\macroxxx#1{%
  \getargs{#1}
  \ifx\argl\empty       
     \paramno=0%
  \else
     \expandafter\parsemargdef \argl;%
  \fi
  \if1\csname ismacro.\the\macname\endcsname
     \message{Warning: redefining \the\macname}%
  \else
     \expandafter\ifx\csname \the\macname\endcsname \relax
     \else \errmessage{Macro name \the\macname\space already defined}\fi
     \global\cslet{macsave.\the\macname}{\the\macname}%
     \global\expandafter\let\csname ismacro.\the\macname\endcsname=1%
     \toks0 = \expandafter{\macrolist\do}%
     \xdef\macrolist{\the\toks0
       \expandafter\noexpand\csname\the\macname\endcsname}%
  \fi
  \begingroup \macrobodyctxt
  \ifrecursive \expandafter\parsermacbody
  \else \expandafter\parsemacbody
  \fi}

\parseargdef\unmacro{%
  \if1\csname ismacro.#1\endcsname
    \global\cslet{#1}{macsave.#1}%
    \global\expandafter\let \csname ismacro.#1\endcsname=0%
    \begingroup
      \expandafter\let\csname#1\endcsname \relax
      \let\do\unmacrodo
      \xdef\macrolist{\macrolist}%
    \endgroup
  \else
    \errmessage{Macro #1 not defined}%
  \fi
}

%
\def\unmacrodo#1{%
  \ifx#1\relax
  \else
    \noexpand\do \noexpand #1%
  \fi
}

\def\getargs#1{\getargsxxx#1{}}
\def\getargsxxx#1#{\getmacname #1 \relax\getmacargs}
\def\getmacname #1 #2\relax{\macname={#1}}
\def\getmacargs#1{\def\argl{#1}}


%

\def\parsemargdef#1;{\paramno=0\def\paramlist{}%
        \let\hash\relax\let\xeatspaces\relax\parsemargdefxxx#1,;,}
\def\parsemargdefxxx#1,{%
  \if#1;\let\next=\relax
  \else \let\next=\parsemargdefxxx
    \advance\paramno by 1%
    \expandafter\edef\csname macarg.\eatspaces{#1}\endcsname
        {\xeatspaces{\hash\the\paramno}}%
    \edef\paramlist{\paramlist\hash\the\paramno,}%
  \fi\next}


\long\def\parsemacbody#1@end macro%
{\xdef\temp{\eatcr{#1}}\endgroup\defmacro}%
\long\def\parsermacbody#1@end rmacro%
{\xdef\temp{\eatcr{#1}}\endgroup\defmacro}%

\def\defmacro{%
  \let\hash=##
  \ifrecursive
    \ifcase\paramno
      \expandafter\xdef\csname\the\macname\endcsname{%
        \noexpand\scanmacro{\temp}}%
    \or 
      \expandafter\xdef\csname\the\macname\endcsname{%
         \bgroup\noexpand\macroargctxt
         \noexpand\braceorline
         \expandafter\noexpand\csname\the\macname xxx\endcsname}%
      \expandafter\xdef\csname\the\macname xxx\endcsname##1{%
         \egroup\noexpand\scanmacro{\temp}}%
    \else 
      \expandafter\xdef\csname\the\macname\endcsname{%
         \bgroup\noexpand\macroargctxt
         \noexpand\csname\the\macname xx\endcsname}%
      \expandafter\xdef\csname\the\macname xx\endcsname##1{%
          \expandafter\noexpand\csname\the\macname xxx\endcsname ##1,}%
      \expandafter\expandafter
      \expandafter\xdef
      \expandafter\expandafter
        \csname\the\macname xxx\endcsname
          \paramlist{\egroup\noexpand\scanmacro{\temp}}%
    \fi
  \else
    \ifcase\paramno
      \expandafter\xdef\csname\the\macname\endcsname{%
        \noexpand\norecurse{\the\macname}%
        \noexpand\scanmacro{\temp}\egroup}%
    \or 
      \expandafter\xdef\csname\the\macname\endcsname{%
         \bgroup\noexpand\macroargctxt
         \noexpand\braceorline
         \expandafter\noexpand\csname\the\macname xxx\endcsname}%
      \expandafter\xdef\csname\the\macname xxx\endcsname##1{%
        \egroup
        \noexpand\norecurse{\the\macname}%
        \noexpand\scanmacro{\temp}\egroup}%
    \else 
      \expandafter\xdef\csname\the\macname\endcsname{%
         \bgroup\noexpand\macroargctxt
         \expandafter\noexpand\csname\the\macname xx\endcsname}%
      \expandafter\xdef\csname\the\macname xx\endcsname##1{%
          \expandafter\noexpand\csname\the\macname xxx\endcsname ##1,}%
      \expandafter\expandafter
      \expandafter\xdef
      \expandafter\expandafter
      \csname\the\macname xxx\endcsname
      \paramlist{%
          \egroup
          \noexpand\norecurse{\the\macname}%
          \noexpand\scanmacro{\temp}\egroup}%
    \fi
  \fi}

\def\norecurse#1{\bgroup\cslet{#1}{macsave.#1}}

\def\braceorline#1{\let\next=#1\futurelet\nchar\braceorlinexxx}
\def\braceorlinexxx{%
  \ifx\nchar\bgroup\else
    \expandafter\parsearg
  \fi \next}

\def\turnoffmacros{\begingroup \def\do##1{\let\noexpand##1=\relax}%
  \edef\next{\macrolist}\expandafter\endgroup\next}

%
%
\def\emptyusermacros{\begingroup
  \def\do##1{\let\noexpand##1=\noexpand\asis}%
  \edef\next{\macrolist}\expandafter\endgroup\next}


\def\aliasyyy #1=#2\relax{%
  {%
    \expandafter\let\obeyedspace=\empty
    \xdef\next{\global\let\makecsname{#1}=\makecsname{#2}}%
  }%
  \next
}

\message{cross references,}

\newwrite\auxfile

\newif\ifhavexrefs    
\newif\ifwarnedxrefs  


\def\inforefzzz #1,#2,#3,#4**{\putwordSee{} \putwordInfo{} \putwordfile{} \file{\ignorespaces #3{}},
  node \samp{\ignorespaces#1{}}}

%
\parseargdef\node{\checkenv{}\donode #1 ,\finishnodeparse}
%
\def\donode#1 ,#2\finishnodeparse{\dodonode #1,\finishnodeparse}
\def\dodonode#1,#2\finishnodeparse{\gdef\lastnode{#1}}

\let\lastnode=\empty

%
\def\donoderef#1{%
  \ifx\lastnode\empty\else
    \setref{\lastnode}{#1}%
    \global\let\lastnode=\empty
  \fi
}

%
\newcount\savesfregister
\def\savesf{\relax \ifhmode \savesfregister=\spacefactor \fi}
\def\restoresf{\relax \ifhmode \spacefactor=\savesfregister \fi}

%
%
\def\setref#1#2{%
  \pdfmkdest{#1}%
  \iflinks
    {%
      \atdummies  
      \turnoffactive
      \otherbackslash
      \edef\writexrdef##1##2{%
	\write\auxfile{@xrdef{#1-
	  ##1}{##2}}
      }%
      \toks0 = \expandafter{\thissection}%
      \immediate \writexrdef{title}{\the\toks0 }%
      \immediate \writexrdef{snt}{\csname #2\endcsname}
      \writexrdef{pg}{\folio}
    }%
  \fi
}

%

\def\ref#1{\xrefX[#1,,,,,,,]}
\def\xrefX[#1,#2,#3,#4,#5,#6]{\begingroup
  \unsepspaces
  \def\printedmanual{\ignorespaces #5}%
  \def\printedrefname{\ignorespaces #3}%
  \setbox1=\hbox{\printedmanual\unskip}%
  \setbox0=\hbox{\printedrefname\unskip}%
  \ifdim \wd0 = 0pt
    \expandafter\ifx\csname SETxref-automatic-section-title\endcsname\relax
      \def\printedrefname{\ignorespaces #1}%
    \else
      \ifdim \wd1 > 0pt
        \def\printedrefname{\ignorespaces #1}%
      \else
        \ifhavexrefs
          \def\printedrefname{\refx{#1-title}{}}%
        \else
          \def\printedrefname{\ignorespaces #1}%
        \fi%
      \fi
    \fi
  \fi
  %
  \ifpdf
    \leavevmode
    \getfilename{#4}%
    {\turnoffactive \otherbackslash
     \ifnum\filenamelength>0
       \startlink attr{/Border [0 0 0]}%
         goto file{\the\filename.pdf} name{#1}%
     \else
       \startlink attr{/Border [0 0 0]}%
         goto name{\pdfmkpgn{#1}}%
     \fi
    }%
    \linkcolor
  \fi
  %
  {%
    \indexnofonts
    \turnoffactive
    \otherbackslash
    \expandafter\global\expandafter\let\expandafter\Xthisreftitle
      \csname XR#1-title\endcsname
  }%
  \iffloat\Xthisreftitle
    \ifdim\wd0 = 0pt
      \refx{#1-snt}%
    \else
      \printedrefname
    \fi
    %
    \ifdim \wd1 > 0pt
      \space \putwordin{} \cite{\printedmanual}%
    \fi
  \else
    %
    \ifdim \wd1 > 0pt
      \putwordsection{} ``\printedrefname'' \putwordin{} \cite{\printedmanual}%
    \else
      {\turnoffactive \otherbackslash
       \setbox2 = \hbox{\ignorespaces \refx{#1-snt}{}}%
       \ifdim \wd2 > 0pt \refx{#1-snt}\space\fi
      }%
      \xrefprintnodename\printedrefname
      %
      ,\space
      %
      \turnoffactive \otherbackslash \putwordpage\tie\refx{#1-pg}{}%
    \fi
  \fi
  \endlink
\endgroup}

%
\def\xrefprintnodename#1{[#1]}

%

\def\Ynumbered{%
  \ifnum\secno=0
    \putwordChapter@tie \the\chapno
  \else \ifnum\subsecno=0
    \putwordSection@tie \the\chapno.\the\secno
  \else \ifnum\subsubsecno=0
    \putwordSection@tie \the\chapno.\the\secno.\the\subsecno
  \else
    \putwordSection@tie \the\chapno.\the\secno.\the\subsecno.\the\subsubsecno
  \fi\fi\fi
}
\def\Yappendix{%
  \ifnum\secno=0
     \putwordAppendix@tie @char\the\appendixno{}%
  \else \ifnum\subsecno=0
     \putwordSection@tie @char\the\appendixno.\the\secno
  \else \ifnum\subsubsecno=0
    \putwordSection@tie @char\the\appendixno.\the\secno.\the\subsecno
  \else
    \putwordSection@tie
      @char\the\appendixno.\the\secno.\the\subsecno.\the\subsubsecno
  \fi\fi\fi
}

%
\def\refx#1#2{%
  {%
    \indexnofonts
    \otherbackslash
    \expandafter\global\expandafter\let\expandafter\thisrefX
      \csname XR#1\endcsname
  }%
  \ifx\thisrefX\relax
    \angleleft un\-de\-fined\angleright
    \iflinks
      \ifhavexrefs
        \message{\linenumber Undefined cross reference `#1'.}%
      \else
        \ifwarnedxrefs\else
          \global\warnedxrefstrue
          \message{Cross reference values unknown; you must run TeX again.}%
        \fi
      \fi
    \fi
  \else
    \thisrefX
  \fi
  #2
}

%
\def\xrdef#1#2{%
  \expandafter\gdef\csname XR#1\endcsname{#2}
  %
  \expandafter\iffloat\csname XR#1\endcsname
    \expandafter\let\expandafter\floatlist
      \csname floatlist\iffloattype\endcsname
    %
    \expandafter\ifx\floatlist\relax
      \toks0 = {\do}
    \else
      \toks0 = \expandafter{\floatlist\do}%
    \fi
    %
    \expandafter\xdef\csname floatlist\iffloattype\endcsname{\the\toks0{#1}}%
  \fi
}

%
\def\tryauxfile{%
  \openin 1 \jobname.aux
  \ifeof 1 \else
    \readauxfile
    \global\havexrefstrue
  \fi
  \closein 1
}

\def\readauxfile{\begingroup
  \catcode`\^^@=\other
  \catcode`\^^A=\other
  \catcode`\^^B=\other
  \catcode`\^^C=\other
  \catcode`\^^D=\other
  \catcode`\^^E=\other
  \catcode`\^^F=\other
  \catcode`\^^G=\other
  \catcode`\^^H=\other
  \catcode`\^^K=\other
  \catcode`\^^L=\other
  \catcode`\^^N=\other
  \catcode`\^^P=\other
  \catcode`\^^Q=\other
  \catcode`\^^R=\other
  \catcode`\^^S=\other
  \catcode`\^^T=\other
  \catcode`\^^U=\other
  \catcode`\^^V=\other
  \catcode`\^^W=\other
  \catcode`\^^X=\other
  \catcode`\^^Z=\other
  \catcode`\^^[=\other
  \catcode`\^^\=\other
  \catcode`\^^]=\other
  \catcode`\^^^=\other
  \catcode`\^^_=\other
  %
  %
  \catcode`\^=\other
  %
  \catcode`\~=\other
  \catcode`\[=\other
  \catcode`\]=\other
  \catcode`\"=\other
  \catcode`\_=\other
  \catcode`\|=\other
  \catcode`\<=\other
  \catcode`\>=\other
  \catcode`\$=\other
  \catcode`\#=\other
  \catcode`\&=\other
  \catcode`\%=\other
  \catcode`+=\other 
  %
  \catcode`\\=\other
  %
  {%
    \count 1=128
    \def\loop{%
      \catcode\count 1=\other
      \advance\count 1 by 1
      \ifnum \count 1<256 \loop \fi
    }%
  }%
  %
  \catcode`\{=1
  \catcode`\}=2
  \catcode`\@=0
  \input \jobname.aux
\endgroup}

\message{insertions,}

\newcount \footnoteno

\def\supereject{\par\penalty -20000\footnoteno =0 }


{\catcode `\@=11
%
\gdef\footnote{%
  \let\indent=\ptexindent
  \let\noindent=\ptexnoindent
  \global\advance\footnoteno by \@ne
  \edef\thisfootno{$^{\the\footnoteno}$}%
  %
  \let\@sf\empty
  \ifhmode\edef\@sf{\spacefactor\the\spacefactor}\ptexslash\fi
  %
  \unskip
  \thisfootno\@sf
  \dofootnote
}%

%
%
\gdef\dofootnote{%
  \insert\footins\bgroup
  \hsize=\pagewidth
  \interlinepenalty\interfootnotelinepenalty
  \splittopskip\ht\strutbox 
  \splitmaxdepth\dp\strutbox
  \floatingpenalty\@MM
  \leftskip\z@skip
  \rightskip\z@skip
  \spaceskip\z@skip
  \xspaceskip\z@skip
  \parindent\defaultparindent
  \smallfonts \rm
  %
  \let\noindent = \relax
  %
  \everypar = {\hang}%
  \textindent{\thisfootno}%
  %
  \footstrut
  \futurelet\next\fo@t
}
}


%
\def\startsavinginserts{%
  \ifx \insert\ptexinsert
    \let\insert\saveinsert
  \else
    \let\checkinserts\relax
  \fi
}

%
\def\saveinsert#1{%
  \edef\next{\noexpand\savetobox \makeSAVEname#1}%
  \afterassignment\next
  \let\temp =
}
\def\makeSAVEname#1{\makecsname{SAVE\expandafter\gobble\string#1}}
\def\savetobox#1{\global\setbox#1 = \vbox\bgroup \unvbox#1}

\def\checksaveins#1{\ifvoid#1\else \placesaveins#1\fi}

\def\placesaveins#1{%
  \ptexinsert \csname\expandafter\gobblesave\string#1\endcsname
    {\box#1}%
}

{
  \def\dospecials{\do S\do A\do V\do E} \uncatcodespecials  
  \gdef\gobblesave @SAVE{}
}

\def\newsaveins #1{%
  \edef\next{\noexpand\newsaveinsX \makeSAVEname#1}%
  \next
}
\def\newsaveinsX #1{%
  \csname newbox\endcsname #1%
  \expandafter\def\expandafter\checkinserts\expandafter{\checkinserts
    \checksaveins #1}%
}

\let\checkinserts\empty
\newsaveins\footins
\newsaveins\margin

%
\openin 1 = epsf.tex
\ifeof 1 \else
  %
  \input epsf.tex
\fi
\closein 1
%
\newif\ifwarnednoepsf
\newhelp\noepsfhelp{epsf.tex must be installed for images to
  work.  It is also included in the Texinfo distribution, or you can get
  it from ftp://tug.org/tex/epsf.tex.}
\def\image#1{%
  \ifx\epsfbox\undefined
    \ifwarnednoepsf \else
      \errhelp = \noepsfhelp
      \errmessage{epsf.tex not found, images will be ignored}%
      \global\warnednoepsftrue
    \fi
  \else
    \imagexxx #1,,,,,\finish
  \fi
}
%
\newif\ifimagevmode
\def\imagexxx#1,#2,#3,#4,#5,#6\finish{\begingroup
  \catcode`\^^M = 5     
  \normalturnoffactive  
  \ifvmode
    \imagevmodetrue
    \nobreak\bigskip
    \nobreak\vskip\parskip
    \nobreak
    \line\bgroup\hss
  \fi
  %
  \ifpdf
    \dopdfimage{#1}{#2}{#3}%
  \else
    \setbox0 = \hbox{\ignorespaces #2}\ifdim\wd0 > 0pt \epsfxsize=#2\relax \fi
    \setbox0 = \hbox{\ignorespaces #3}\ifdim\wd0 > 0pt \epsfysize=#3\relax \fi
    \epsfbox{#1.eps}%
  \fi
  \ifimagevmode \hss \egroup \bigbreak \fi  
\endgroup}

%
\envparseargdef\float{\eatcommaspace\eatcommaspace\dofloat#1, , ,\finish}

\def\eatcommaspace#1, {#1,}

%
%
%
\let\resetallfloatnos=\empty
\def\dofloat#1,#2,#3,#4\finish{%
  \let\thiscaption=\empty
  \let\thisshortcaption=\empty
  %
  %
  %
  \startsavinginserts
  %
  \par
  \vtop\bgroup
    \def\floattype{#1}%
    \def\floatlabel{#2}%
    \def\floatloc{#3}
    \ifx\floattype\empty
      \let\safefloattype=\empty
    \else
      {%
        \indexnofonts
        \turnoffactive
        \xdef\safefloattype{\floattype}%
      }%
    \fi
    %
    \ifx\floatlabel\empty \else
      %
      \expandafter\getfloatno\csname\safefloattype floatno\endcsname
      \global\advance\floatno by 1
      {%
        %
        \edef\thissection{\floatmagic=\safefloattype}%
        \setref{\floatlabel}{Yfloat}%
      }%
    \fi
    %
    \vskip\parskip
    %
    \restorefirstparagraphindent
}

%
\def\Efloat{%
    \let\floatident = \empty
    %
    \ifx\floattype\empty \else \def\floatident{\floattype}\fi
    %
    \ifx\floatlabel\empty \else
      \ifx\floattype\empty \else 
        \appendtomacro\floatident{\tie}%
      \fi
      \appendtomacro\floatident{\chaplevelprefix\the\floatno}%
    \fi
    %
    \let\captionline = \floatident
    \ifx\thiscaption\empty \else
      \ifx\floatident\empty \else
	\appendtomacro\captionline{: }
      \fi
      %
      \appendtomacro\captionline{\scanexp\thiscaption}%
    \fi
    %
    \ifx\captionline\empty \else
      \vskip.5\parskip
      \captionline
      %
      \vskip\parskip
    \fi
    %
    \ifx\floatlabel\empty \else
      {%
        \atdummies \turnoffactive \otherbackslash
	\scanexp{%
	  \xdef\noexpand\gtemp{%
	    \ifx\thisshortcaption\empty
	      \thiscaption
	    \else
	      \thisshortcaption
	    \fi
	  }%
	}%
        \immediate\write\auxfile{@xrdef{\floatlabel-lof}{\floatident
	  \ifx\gtemp\empty \else : \gtemp \fi}}%
      }%
    \fi
  \egroup  
  %
  %
  %
  \checkinserts
}

%
\def\appendtomacro#1#2{%
  \expandafter\def\expandafter#1\expandafter{#1#2}%
}

%
\def\caption{\docaption\thiscaption}

\def\docaption{\checkenv\float \bgroup\scanargctxt\defcaption}
\def\defcaption#1#2{\egroup \def#1{#2}}

\def\getfloatno#1{%
  \ifx#1\relax
      \csname newcount\endcsname #1%
      %
      \expandafter\gdef\expandafter\resetallfloatnos
        \expandafter{\resetallfloatnos #1=0 }%
  \fi
  \let\floatno#1%
}

%
\def\Yfloat{\floattype@tie \chaplevelprefix\the\floatno}%

\def\floatmagic{!!float!!}

%
\def\iffloat#1{\expandafter\doiffloat#1==\finish}
%
%
\def\doiffloat#1=#2=#3\finish{%
  \def\temp{#1}%
  \def\iffloattype{#2}%
  \ifx\temp\floatmagic
}

%
\parseargdef\listoffloats{%
  \def\floattype{#1}
  {%
    \indexnofonts
    \turnoffactive
    \xdef\safefloattype{\floattype}%
  }%
  %
  \expandafter\ifx\csname floatlist\safefloattype\endcsname \relax
    \ifhavexrefs
      \message{\linenumber No `\safefloattype' floats to list.}%
    \fi
  \else
    \begingroup
      \leftskip=\tocindent  
      \let\do=\listoffloatsdo
      \csname floatlist\safefloattype\endcsname
    \endgroup
  \fi
}

%
%
\def\listoffloatsdo#1{\listoffloatsdoentry#1\finish}
\def\listoffloatsdoentry#1-title\finish{{%
  \toksA = \expandafter{\csname XR#1-lof\endcsname}%
  %
  \edef\writeentry{\noexpand\entry{\the\toksA}{\csname XR#1-pg\endcsname}}%
  \writeentry
}}

\message{localization,}

%
\parseargdef\documentlanguage{%
  \tex 
    \openin 1 txi-#1.tex
    \ifeof 1
      \errhelp = \nolanghelp
      \errmessage{Cannot read language file txi-#1.tex}%
    \else
      \input txi-#1.tex
    \fi
    \closein 1
  \endgroup
}
\newhelp\nolanghelp{The given language definition file cannot be found or
is empty.  Maybe you need to install it?  In the current directory
should work if nowhere else does.}


%
\newdimen\defaultparindent \defaultparindent = 15pt

\chapheadingskip = 15pt plus 4pt minus 2pt
\secheadingskip = 12pt plus 3pt minus 2pt
\subsecheadingskip = 9pt plus 2pt minus 2pt

\vbadness = 10000

\hbadness = 2000

\widowpenalty=10000
\clubpenalty=10000

%
\def\setemergencystretch{%
  \ifx\emergencystretch\thisisundefined
    \def\emergencystretch{\dimen0}%
  \else
    \emergencystretch = .15\hsize
  \fi
}

%
%
\def\internalpagesizes#1#2#3#4#5#6#7#8{%
  \voffset = #3\relax
  \topskip = #6\relax
  \splittopskip = \topskip
  \vsize = #1\relax
  \advance\vsize by \topskip
  \outervsize = \vsize
  \advance\outervsize by 2\topandbottommargin
  \pageheight = \vsize
  \hsize = #2\relax
  \outerhsize = \hsize
  \advance\outerhsize by 0.5in
  \pagewidth = \hsize
  \normaloffset = #4\relax
  \bindingoffset = #5\relax
  \ifpdf
    \pdfpageheight #7\relax
    \pdfpagewidth #8\relax
  \fi
  \setleading{\textleading}
  \parindent = \defaultparindent
  \setemergencystretch
}

\def\letterpaper{{\globaldefs = 1
  \parskip = 3pt plus 2pt minus 1pt
  \textleading = 13.2pt
  %
  \internalpagesizes{46\baselineskip}{6in}%
                    {\voffset}{.25in}%
                    {\bindingoffset}{36pt}%
                    {11in}{8.5in}%
}}






%
\parseargdef\pagesizes{\pagesizesyyy #1,,\finish}
\def\pagesizesyyy#1,#2,#3\finish{{%
  \setbox0 = \hbox{\ignorespaces #2}\ifdim\wd0 > 0pt \hsize=#2\relax \fi
  \globaldefs = 1
  \parskip = 3pt plus 2pt minus 1pt
  \setleading{\textleading}%
  \dimen0 = #1
  \advance\dimen0 by \voffset
  \dimen2 = \hsize
  \advance\dimen2 by \normaloffset
  \internalpagesizes{#1}{\hsize}%
                    {\voffset}{\normaloffset}%
                    {\bindingoffset}{44pt}%
                    {\dimen0}{\dimen2}%
}}

%
\letterpaper

\message{and turning on texinfo input format.}

\catcode`\"=\other
\catcode`\~=\other
\catcode`\^=\other
\catcode`\_=\other
\catcode`\|=\other
\catcode`\<=\other
\catcode`\>=\other
\catcode`\+=\other
\catcode`\$=\other

\def\normalunderscore{_}

\def\normaldollar{$}

%
%
\def\ifusingtt#1#2{\ifdim \fontdimen3\font=0pt #1\else #2\fi}

\def\ifusingit#1#2{\ifdim \fontdimen1\font>0pt #1\else #2\fi}


\catcode`\"=\active
\def\activedoublequote{{\tt\char34}}
\let"=\activedoublequote
\catcode`\~=\active
\def~{{\tt\char126}}
\chardef\hat=`\^
\catcode`\^=\active
\def^{{\tt \hat}}

\catcode`\_=\active
\def_{\ifusingtt\normalunderscore\_}
\def\_{\leavevmode \kern.07em \vbox{\hrule width.3em height.1ex}\kern .07em }

\catcode`\|=\active
\def|{{\tt\char124}}
\chardef \less=`\<
\catcode`\<=\active
\def<{{\tt \less}}
\chardef \gtr=`\>
\catcode`\>=\active
\def>{{\tt \gtr}}
\catcode`\+=\active
\def+{{\tt \char 43}}
\catcode`\$=\active
\def${\ifusingit{{\sl\$}}\normaldollar}


\catcode`\@=0

\global\chardef\backslashcurfont=`\\

{\catcode`\\=\active
 @gdef@rawbackslash{@let\=@backslashcurfont}
 @gdef@otherbackslash{@let\=@realbackslash}
}

{\catcode`\\=\other @gdef@realbackslash{\}}


\catcode`\\=\active

@def@turnoffactive{%
  @let"=@normaldoublequote
  @let\=@realbackslash
  @let~=@normaltilde
  @let^=@normalcaret
  @let_=@normalunderscore
  @let|=@normalverticalbar
  @let<=@normalless
  @let>=@normalgreater
  @let+=@normalplus
  @let$=@normaldollar 
  @unsepspaces
}

%
@def@normalturnoffactive{@turnoffactive @let\=@normalbackslash}

@otherifyactive

%
@gdef@eatinput input texinfo{@fixbackslash}
@global@let\ = @eatinput

%
@gdef@fixbackslash{%
  @ifx\@eatinput @let\ = @normalbackslash @fi
  @catcode`+=@active
  @catcode`@_=@active
}

@escapechar = `@@

@catcode`@& = @other
@catcode`@# = @other
@catcode`@

@c Local variables:
@c eval: (add-hook 'write-file-hooks 'time-stamp)
@c page-delimiter: "^\\\\message"
@c time-stamp-start: "def\\\\texinfoversion{"
@c time-stamp-format: "
@c time-stamp-end: "}"
@c End:

@c vim:sw=2:

@ignore
   arch-tag: e1b36e32-c96e-4135-a41a-0b2efa2ea115
@end ignore

@setfilename gracos.info
@settitle GRACOS

@include version.atexi
@setchapternewpage odd

@macro mathenv{param}
@ifhtml
\param\
@end ifhtml
@iftex
@math{\param\}
@end iftex
@end macro

@macro mathenvd{paramtex,paramhtml}
@iftex
@math{\paramtex\}
@end iftex
@ifhtml
\paramhtml\
@end ifhtml
@end macro

@c For a final copy, take out the rectangles
@c that mark overfull boxes (in case you have decided
@c that the text looks ok even though it passes the margin).
@finalout

@ifinfo
This is a GRACOS Users Reference.

@include copyright.texi
@end ifinfo

@titlepage
@sp 10
@comment The title is printed in a large font.
@title GRACOS Users Reference
@subtitle For version @value{VERSION}, @value{UPDATED}
@author Alexander Shirokov

@c The following two commands start the copyright page.
@page
@vskip 0pt plus 1filll

@include copyright.texi

@end titlepage

@c Output the table of contents at the beginning.
@contents

@node top
@ifhtml
@top GRACOS
@end ifhtml

@exampleindent 1

@c 
@c @alias cvar = strong
@c @alias cfun = strong
@c @alias fopt = strong
@c @alias fopv = strong
@c 

@node Introduction
@chapter Introduction

This reference documents version @value{VERSION} of @dfn{GRACOS}, the
@emph{GRA}vitational @emph{COS}mology suite, or a collection of
software for cosmological N-body type simulations and data
analysis. The primary component of @emph{GRACOS} is the
@dfn{@command{gracos} code}, -- a parallel cosmological N-body code.  

The @command{gracos} executable uses a set of optimization techniques
developed over many years and first then completely documented on
March 2004 in the @uref{http://library.mit.edu, @emph{MIT Ph.D. Thesis
}} by @emph{Alexander Shirokov}, and
@command{astro-ph/0505087} (in the
reduced form). The most up-to-date version of @command{gracos} can be
found at @uref{http://www.gracos.org}.

Cosmological N-body type work generally employs highly computationally
challenging and versatile applications.  
@c
However it is not just the solving the computational algorithm that is
challenging, but also bringing it in a usable form to @emph{you}, the end user. 
@c
@c
It is generally considered very complicated to run an N-body
simulation, especially for large problems, because many, if not all,
of the current N-body code distributions come without many commonly
used options, such as ability to generate the initial conditions given
a set of cosmological parameters, especially when heavy amounts of
data are involved; file input-output format is still an outstanding
issue. This situation gets more complicated after one includes common
data analysis tasks, such as the group finder, imaging or power
spectrum estimation.

@c
@dfn{@command{gracos}} comes with a built-in command-line interface that
makes it easy to manipulate with the parallel application, raising the
level of the communication between the researcher and the
implementation, and thus substantially simplifying all the engineering
tasks associated with the scientific problems and helping document the
simulations.
@c
New commands can be introduced easily (once you know how to do it).
@c
@emph{GRACOS} distribution provides a rather extensive the
@command{gracos-examples} directory containing the scripts, that can be
readily used upon the completion of the installation procedure. These
scripts are to be used for testing, as templates for large N-body
simulations and for doing some common parallel cosmological simulation
tasks.
@c
@c
Command line interface extends @command{gracos} development horizons to a
new level.  

@c
This reference is mostly error-free, because a substantial fraction of
its content is automatically generated in a one time procedure that
runs @emph{GRACOS} to generate most plots, images and tables
presented throughout this reference (@command{doc/src/README} shows how
to run the procedure). In particular, the procedure executes many
scripts later described, following the guidelines in this reference.

In addition to the main executable, @emph{GRACOS} installation
provides the C-libraries and the associated C-header files that are used
for @command{gracos} but are also suitable for more generic applications.

@node Reference to gracos Executable
@chapter Reference to @command{gracos} Executable 

This chapter documents @command{gracos} executable, - our main product.

N-body simulations tend to be massive; however before any
computationally large problem is submitted to a cluster it is more
than useful to test the same problem for a reduced problem size. The
script examples presented thoughout this reference are just those
small test cases. In order to extend them to larger problems just
increase the number of processes, the gridsize and make sure no part
of the script is intrinsically unscalable.  This chapter assumes the
successful completion of the installation procedure,
@xref{Installation Procedure}.

@node Introduction to Interface
@section Introduction to Interface

Sophisticated interface is used to extend @command{gracos} capabilities
and make it easier to document the simulations. This section provides
necessary introduction.

@node Running gracos
@subsection Running @command{gracos}

@command{gracos} can be run both in parallel and serially; typing
@command{gracos --help} or @command{gracos -?} yields a useful help
message.
@noindent
@example
@include gracos-help.atexi
@end example
@noindent
The user interacts with @command{gracos} by passing the commands within
the @command{gracos} scripting session. The commands can be entered from
standard input (option @command{-t} or the default), a file (option
@command{-i}), or passed in the command line (option @command{-e}).  The
commands are executed using the @emph{Tcl} embedded interpreter, See
@ref{Commands} for more detail.

To enter the commands through the standard input, simply type
@command{gracos} or, for a parallel run on @var{Nproc} logical
processes type

@smallexample
@var{SHELL$} mpirun -np @var{Nproc} @command{gracos}
@end smallexample
@noindent
You may use any integer number @var{Nproc} of processes greater than
zero on any machine, however it is best to accomodate the agreement
between the number @var{Nproc} of logical processes and the number of
physical processes and cores available on the system.

The @command{gracos
a new command from the standard input in the newly opened
@emph{@command{gracos} scripting session}.  As can be seen from the
following example by a user familiar with @emph{Tcl}, indeed
@command{gracos} uses the facilities provided by the @emph{Tcl} computing
language:

@smallexample
@include prompt.atexi
@end smallexample
@noindent
The input is distinguished from the output by the presence or absence
of the prompts (``@command{gracos
for incomplete commands). In the above example, @command{$i} refers to a
@emph{Tcl} variable, and @command{foreach} is a @emph{Tcl} built-in
command. As soon as the complete command is entered within the
scripting session, its copies are sent to the remaining MPI processes
and the command is processed with the embedded @emph{Tcl} interpreter
individually on each process, See @ref{Commands} for more
information.

@c brief introduction to Tcl
In order to master @command{gracos}, the user should learn some basics
of @emph{Tcl}. Fortunately, @emph{Tcl} is usually found easy to learn.
A number of resources are available online; we refer the users to the
@uref{http://aspn.activestate.com/ASPN/docs/ActiveTcl/8.4/tcl/tcl_contents.htm,
@emph{Tcl} reference}, the @uref{http://www.tcl.tk/man/tcl8.4,
@emph{Tcl} manual} and the
@uref{http://www.tcl.tk/man/tcl8.5/tutorial/tcltutorial.html,
@emph{Tcl} tutorial} for version 8.5.
@c 
In @command{gracos} we avoid the use of any advanced @emph{Tcl} feature,
when possible.
@c
For those readers whose access to the above online resources is
restricted, let us just say that the @emph{Tcl} built-in commands
@command{set} and @command{puts} are used to manage the @emph{Tcl} variables
(setting a variable and displaying its value); the square brackets '['
and ']' are used for an in-line expansion of a @emph{Tcl} expression;
the curly brackets '@{' and '@}' as well and the double quotation
marks are used for quoting in slightly different contexts (mainly: the
expansion of subexpressions is allowed in double quoted expressions
but is not allowed in the curly bracket quotations); the @command{expr}
@emph{Tcl} built-in command is used for arithmetic expression
evaluation, the @command{exec} is for external shell expression
evaluation, and the @command{eval} is for the evaluation of the arguments
passed to the command in the current scripting session.  Finally
@command{cd} can be used to change the current working directory path.

In addition to the more generic commands intrinsic to the
@emph{Tcl}, @command{gracos} has a number of commands that are only
specific to the @command{gracos} executable; their implementation forms
the trunk of the @command{gracos} C-code.

As soon as @command{gracos} executable is ran, a number of useful files
appear in the working directory.  File @command{gracos_history} contains
the current listing of all the executed commands; files @command{so-1}
@dots{} @command{so-}@var{(Nproc-1)} show the standard output on
non-server processes. The standard output on process zero appears on
the screen, it is not redirected unless the @command{--f} command line
option is used.

To execute the same commands as in the example above by reading them
from a file, instead of standard input, we create a temporary file
with the @command{mktemp} linux command, fill it with the (identical)
commands to be executed and pass the filename with the @command{-i}
option to @command{gracos}

@smallexample
@var{SHELL$} TmpFile=`mktemp tmp.XXXXXX` 
@var{SHELL$} cat >  $TmpFile
puts "Hello"
foreach i @{ 1 2 3 4 @} @{
   puts $i
@}
@var{SHELL$} mpirun -np 4 @command{gracos} @command{-i} $TmpFile
gracos
Hello
gracos
   puts $i
@}

1
2
3
4
@var{SHELL$}
@end smallexample

Now, instead of having to type the commands in through the shell
session, we can place them into a new file, make the file executable
and run.  Shown below is the copy of the working script, also provided
at @command{gracos-examples/hello.sh}
@example
@include gracos/gracos-examples/hello.sh.c
@end example
@noindent
Typing 
@smallexample
chmod +x ./hello.sh
./hello.sh
@end smallexample
@noindent
yields the expected result.  The important detail to notice in the
last example above is using the backslash character (\) necessary to
avoid the conflict between the interpretations of the '@command{$}'
character, which is used for variable dereferencing by both @var{Tcl}
and @var{bash}. If you do not use the backslash character in the
example above the variable @command{$i} is interpreted as a
@command{bash} shell variable and, according to the bash syntax, be
expanded to empty value because no such @command{bash} variable is
defined in the script. Because of the use of backslash character, this
variable is interpreted as a @emph{Tcl} variable which is set in the
preceeding @command{foreach} @emph{Tcl} statement. One just has to be
careful using backslashes to avoid the similar misinterpretations.

@node gracos-examples
@subsection @command{gracos-examples}

Each case considered in this reference is presented as a @command{gracos}
script ready to be executed in your Unix-flavored shell, any time
following the @emph{GRACOS} installation procedure.
@c
These scripts are organized within the so called @dfn{@command{gracos-examples}
directory} whose archieve is one of the products of @emph{GRACOS}
installation.  In order to make a local copy of the directory, simply
type:

@smallexample
@var{SHELL$} cp `gracos-config info pkgdatadir`/gracos-examples.tar.gz .
@var{SHELL$} tar zxf gracos-examples.tar.gz
@end smallexample
@noindent 
in your Unix flavored shell. The procedure creates the
@command{gracos-examples} directory which contains all the example
scripts presented in this reference. For example, the @command{hello.sh}
script presented in @ref{Running gracos} is included as
@command{gracos-examples/hello.sh}; you may execute it any time following
the installation procedure, @xref{Installation Procedure}.

The user may become confused by the large number of files
automatically generated as the examples are ran; type
@command{gracos-examples -r} to clean the @command{gracos-examples}
directory and return to the initial pristine state.

@node Command Line Options
@subsection    Command Line Options

The full list of command line options is provided in @ref{Running gracos}.

If you use a batch system for job submission we recommend using option
@command{-f}. Job submission systems spool extra information into their
standard output file, making it difficult to extract the standard
output of process zero from the standard output of the job, the latter
produced only at the end of the run also. Do not use this option for
interactive sessions as you will not be able to see the prompt.

Regarding the option @command{-t}, note that one of the previously
released versions of LAM MPI are known to show an erroneous behavior
on standard input: the interactive session would terminate as soon as
it is started without prompting the user. Even though the bug was
corrected in the subsequent versions of LAM, it may be a good idea to
always use a non-interactive session where possible (copy your input
to a file and pass the using the @command{-i} option to @command{gracos}).

Option @command{-v} controls the verbosity associated with the
@command{tverb} internal @command{gracos} bitwise variable. Use this option
to set ``on'' the memory allocation verbosity when debugging memory
leaks. Using @command{-v}@command{-1} sets verbosity to the maximum
(all bits are ``on'' for @command{-1} in its bitwise representation).

@c
@node Commands
@subsection Commands

@command{gracos} interacts with the user via its own extendable command
line-based language

The complete list of @command{gracos} commands is given in @ref{Reference to gracos Executable}.

@command{gracos} uses the @var{Tcl} (@emph{Tool Command Language}) as the
@emph{embedded interpreter}; which means that any command processed
within the @command{gracos} scripting session is parsed and executed by
the @emph{Tcl} (version 8.4) interpreter. This interpreter is extended
within @command{gracos} to include new @command{gracos}-specific commands
and procedures. @emph{Tcl} is currently chosen as the interpreter due
to its ease of use, portability, light weight, and the open source
license.
@c default Tcl session
Typing @command{tclsh} in your Unix-type prompt brings you to the default
@emph{Tcl} scripting session (in the same way as typing @command{python}
for example brings you to the @emph{Python} scripting session). 
@c
We refer the users who are new to @emph{Tcl} to the online
@uref{http://aspn.activestate.com/ASPN/docs/ActiveTcl/8.4/tcl/tcl_contents.htm,
@emph{Tcl} reference}, @uref{http://www.tcl.tk/man/tcl8.4,
@emph{Tcl} manual} and 
@uref{http://www.tcl.tk/man/tcl8.5/tutorial/tcltutorial.html,
@emph{Tcl} (version 8.5) tutorial}.
@c
The book @emph{Tcl and Tk Toolkit 1994}, by @emph{John
K. Ousterhout} is also recommended for its coverage of embedding;
however it covers a much earlier version of @emph{Tcl}.
@c difference of gracos from the Tcl scripting session
The @dfn{@command{gracos} interpreter} differs from the default
@emph{Tcl} interpreter exactly by the presence of the specific
@var{Tcl} variables, procedures defined in the @command{init.tcl} startup
file, and the new commands linked to the internal @command{gracos}
@emph{C}-functions using the @command{Tcl_CreateCommand} function.
File @command{init.tcl} is sourced each time @command{gracos} is
started. This file contains definitions for custom @command{gracos}
@emph{Tcl} procedures and variables.
@c parallel runs
@emph{Tcl} commands always originate on process zero for parallel runs
with MPI. As soon as a complete @emph{Tcl} command is entered on
process zero its copies are immediately sent to the remaining MPI
processes. The identical commands are then synchronously interpreted
on all the remaining MPI processes.

@node Variables
@subsection Variables
@c Distinction of bound variables from the Tcl variables
In addition to the usual @emph{Tcl} variables, @command{gracos} operates
with the so called @command{gracos} variables (those variables are bound to the
pointers inside the C-code).

The complete list of the @command{gracos} variables is given in @ref{Reference to gracos Executable}.

@command{gracos} variables are operated by @command{varset}, @command{varunset},
@command{varputs}, @command{vars}, and @command{varchmod} gracos commands,
@xref{gracos Commands}.  In addition, they are automatically set or
updated within the source code, at any point in the source code where
such setting is rational.

@command{gracos} variables are linked to the internal source code
variables so that whenever a value of a @command{gracos} variable is
changed in the script the corresponding variable within the source
code is also immediately changed.  @command{gracos} variables are not
associated with @emph{Tcl} variables in any way and are referred to
in this reference as simply @dfn{variables}.  The list of all gracos
variables for this version of @command{gracos} is given in
@ref{gracos Variables}.

@c @command{gracos} variable Access method
Commands, @command{varset}, @command{varunset}, @command{varputs},
@command{varchmod} and @command{vars} are used to manage the @command{gracos} variables
from within the @command{gracos} scripting session.  C-macros
@command{varset}, @command{varget} are used to manage @command{gracos} variables
within the C-code. 
A @command{gracos} variable should be carefully distinguished from a regular Tcl
variable; the easiest way to do it is by keeping in mind that the
'@command{$}' character is always used for referring to a @emph{Tcl}
variable, but is never used for a @command{gracos} variable.

@c 
Intentionally, no new @command{gracos} variables can be introduced in the
runtime.  Each variable is assigned a unique key and a type which do
not change in the runtime, See @ref{gracos Variables} for the
complete list of @command{gracos} variables with their keys and types; however,
as we already mentioned, a given variable can be either set or unset,
that is - its status may change in the runtime.

@c 
@command{gracos} variables provide us with an important method of error control
intrinsic to the @emph{gracos} scripts. In particular, any attempt to
access a variable whose current status is unset results in termination
with an error message, regardless of the value of the memory space
associated with the variable. The termination behavior can be avoided
if necessary, by setting the @command{unvar} mode of @command{tetol} bitwise
variable, See @ref{Bitwise Variables}.

The list of all the currently set @command{gracos} variables is displayed
by typing @command{vars} within the @emph{gracos} scripting session (use
the @command{vars} @command{-a} for the complete list, including
uninitialized variables). The first three columns in the output of
this command are each one character wide. The first column says if the
variable is set (@command{' '}) or unset (@command{'?'}); the second
column says if the variable is currently writable (@command{' '}) or
not (@command{'r'}); the third column says if the variable is an array
(@command{'A'}) or a scalar variable (@command{' '}).  The last two
columns of @command{vars} output show respectively the variable name and
its value or values (for the array variables).

Command @command{varchmod} can be used to switch write protection status
of a variable and is useful for such purposes, as keeping the
verbosity constant while loading a datafile which may or may not
contain verbosity assignments. The syntax is @code{varchmod @{+w|-w@}
@var{KEY}} where @code{@var{KEY}} is the name of the @command{gracos}
variable. The usage of other @var{var}commands is explained in the
next subsection.

In the following example, we illustrate the use of the @command{gracos}
variables versus usual @emph{Tcl} variables.  We first display the
list of the @command{gracos} variables using @command{vars} command twice:
with and without the @command{-a} flag. Then we set a @command{gracos}
variable @command{nstep} to a specific legal value using the
@command{varset} command; then we set a @emph{Tcl} variable to a
different value choosing the same variable name. Then we display their
values by using @command{varputs} for @command{gracos} variable and
@command{puts} for @emph{Tcl} variable. We observe that the variables
have kept their originally assigned values and are therefore indeed
distinct.
@example
@include gracos/gracos-examples/interface/vars.sh.c
@end example

@node Bitwise Variables
@subsection Bitwise Variables

@dfn{Bitwise variables} are technically just regular @command{gracos} 4
byte integer variables; they are used to control miscellaneous bitwise
(``yes'' or ``no''s) modes.  The @command{varset} and @command{varputs}
commands applicable to the @command{gracos} variables in general can be
used to manage values of bitwise variables as well.  Each bitwise
variables can currently be assigned no more than 32 modes.

Type @command{bitvar -vars}  to see the list of all bitwise variables;
@command{bitvars} is a @emph{Tcl} procedure defined in
@command{init.tcl}, @xref{Commands}.

In @emph{bitwise string} representation bitwise variable is shown as a string
of zeros and ones.

The in @emph{integer representation} of bitwise variables is more compact
than bitwise string representation and is convenient in certain
situations.  Setting a bitwise variable to @command{-1} enables all
the modes because all bits of the @command{-1} integer are
``yes''. Setting the integer to zero disables all the modes at once.
It is not convenient however to operate with integer values if one
wants to set some specific bit of a bitwise variable.

@emph{Token representation} uses tokens to specify values of each
mode. The user conveniently operates with tokens, combining them with
the @command{OR} ('|'), @command{AND} ('&') and @command{NOT} ('~')
bitwise operators. Section @ref{gracos Variables} lists all tokens with
brief description for each bitwise variable.

The @command{bitvar} procedure defined in @command{init.tcl} can be used to
manipulate bitwise variables in their token representations.  The
highly commented example below illustrates different ways of
manilulating bitwise variables on the example of @command{tverb} bitwise
variable. 

@example
@include gracos/gracos-examples/interface/bitvars.sh.c
@end example

@node Runtime Administration
@subsection Runtime Administration

@dfn{Runtime administration}  technique is the
perfect tool for doing things on the fly (as the code runs).

The user requests runtime administration by creating specifically
named @dfn{runtime administration file}(s) within the working
directory either before or during the run.  The runtime administration
files are read by @command{gracos} as soon as the control reaches any of
the @dfn{runtime administration checkpoints} within the source code
(functions @command{admin_checkpoint)}) with a unique label
@command{admin_label} passed as an argument, See @ref{Runtime Administration} for
their complete list.

Runtime administration can be invoked explicitly, using the
@command{admin} command. There are a few kinds of runtime administration
described in the following text.

There are two important points to keep in mind
@itemize @bullet
@item @emph{Avoid partial updates of the runtime administration files}
If you edit runtime administration file during the run, casual saving
an incomplete version of file may lead to errors if the checkpoint is
reached at the time.
@item @emph{Avoid possibility of infinite recursion loop.}
If your runtime administration method invokes one of the commands that
themselves contain a runtime administration checkpoint the infinite
recursion loop is avoided by restricting the scope of administraction
to particular checkpoints; See @ref{Example: Tabulated Transfer Function} for example.
@end itemize

The runtime administration files do not need to have the executable
file permissions.

The following is the list of all runtime administration checkpoint
ids. 

@table @code
@item integ
Invoked each timestep of the particle integration run, @xref{integ: Running an N-body simulation}.
@item integ_pre
Invoked at the beginning in the @command{integ} command, just before the
main timestepping loop of the integration of particle trajectories
starts.
@item image
Invoked just after the image data is projected to form a numerical
matrix-plate but before it is converted to produce a @file{gif} file,
during the @command{image} command, @xref{image: Particle Data Imaging}.  This
administration checkpoint can be used to tune the dynamic range
covered by the image to the brightness of the dimmest and the
brightest spots on the image plate.
@item admin
Invoked when command @command{admin} is executed in the interpreter.
@item env
Invoked when command @command{env} is executed in the interpreter.
@end table

@node One Time Administration
@subsubsection One Time Administration

File @command{inc.tcl}, if found at the runtime administration
checkpoint, is interpreted as an input @command{gracos} script and is
then moved to @command{inc.tcl~}, unles the new @command{admin_keep}
variable is set to @command{1} while processing that script.

@emph{Tcl} variable @code{admin_label}, temporarily defined as the
label of the current administration checkpoint, may be used to
restrict the scope of one time administration to a particular
administration checkpoint in order to avoid infinite recursion. An
example of one time administration with such restricted scope of
administration is presented in @ref{Example: Tabulated Transfer Function}.

As the most simple example, try executing the following
@emph{bash} shell command within the working directory during any of
the @emph{GRACOS} runs to see the run stop:

@smallexample
@var{SHELL$} echo end > @command{inc.tcl}
@end smallexample
@noindent
The newly created file @command{inc.tcl} contains just a single
@command{gracos} command @command{end} to request an end of the gracos
session as soon as possible. One should not worry about infinite
recursion in this example because, as we know from @ref{Runtime Administration} the
command @command{end} does not itself contain another runtime
administration checkpoints.

@node Repeatable Administration
@subsubsection Repeatable Administration

The @command{gracos} variable @command{admin_path}, if set, defines the name
of the @dfn{repeatable administration file}, which gets processed at
each administration checkpoint without getting deleted. If the
repeatable administration file is defined but does not exist the code
shall terminate with the error message.
@c
In contrast to the one time administration file, the path of the
repeatable administration file is not fixed and the file is not
removed after getting executed, thus getting processed each
administration checkpoint for as long as it is defined and the run
is continued.

The content of the repeatable administration file is processed
according to its filename extension.

@c Tcl
A file whose extension is @command{.tcl} is interpreted in the same way
as described in Section @ref{One Time Administration}, except that the
file is not replaced after the execution.

@c Python
A file whose extension is @command{.py} is processed under @emph{Python}
interpreter. The details of the intrinsic procedure are as follows.
@c the header
First an automatically generated header @emph{Python} script is
processed in which a number of variables are defined.
Type

@smallexample
gracos -e env
@end smallexample
@noindent
to see what this @emph{Python}s header roughly looks like (the real
header will differ only by the runtime generated values in the
assignments). 
First, all the @emph{GRACOS} variables are set under
their original @emph{GRACOS} names (floating point numbers may lose
significant bits in this assignment); those @emph{GRACOS} variables
that are unset are initialized to the @emph{Python}s @code{None}.
@c
Next, an unrestricted choice of variables are defined by convenience,
according to their setting within the source code.
@c
The @code{admin_label} is also defined and can be used to restrict the
scope of the runtime administration to a particular checkpoint.
@c the body
The content of the repeatable administration file is then processed in
the same @emph{Python} session.
@c the output
Should the user define the @command{admin_out} @emph{Python} variable,
the content of this variable is sources as an input @emph{Tcl} script
within the @emph{GRACOS} scripting session immediately following this
checkpoint instance.

Note that @emph{Python} interpreter is invoked by using the
@command{system} C-function call, which is known to have portability
issues on some rare systems (the @emph{Infiniband} networking systems
is the only such system, as far as we currently know).  You must have
configured @emph{GRACOS} without @command{--disable-system-calls}
@command{configure} option for the @emph{Python} runtime
administration to work; See Sections @ref{Installation Procedure}, and
@ref{ibtest: System Call Portability Test} for more details.

@node Scheduled Administration
@subsubsection Scheduled Administration

Scheduled administration method is useful for scheduling certain tasks
to be executed at a pre-specified time.

File @file{sched.txt}, if present in the working directory, must be
the text file containing the integer (of may be floating point number)
representing the number of seconds since 1970-01-01 00:00:00 UTC at
which to schedule executing a command.  the standard of output of
C-command @command{time(NULL)} of shell command @command{date +

The command, which is the content of variable @command{cmd_sched}, will
be executed as soon as control reaches one of the scheduled
administration checkpoints and time exceeds the number written in file
@file{sched.txt}. The content of the variable is subject to the same
substitution rules as those applicable to the output control variables
described in @ref{Checkpoint Control Variables}.

Below is the exanmple of scheduled administration
@example
@include gracos/gracos-examples/problems/sched.sh.c
@end example

@node Primary Output Files
@subsection    Primary Output Files

A run with @command{gracos} executable produces a number of @dfn{primary
output files} within the working directory. These files contain useful
information about the run and the used executable. 

@itemize @bullet
@item File @command{gracos_history}: 
The list of commands submitted to @command{gracos} interpreter so far.
This file is is created on process zero and is updated during the run
as soon as a new command starts being interpreted. If you look at this
file during the runtime, the line at the bottom indicates the
currently processed command or procedure.

@item Files @command{so-@var{Rank}}:
The standard output of process whose MPI rank is @var{Rank} (an
integer). The files are created, one for each process, and are updated
as the run progresses. The amount of data written into these files
depends on the settings of the verbosities, controlled by
the bitwise variables @command{tverb} and @command{sverb}
@xref{Bitwise Variables}.

@item File @command{gracos_info}:
Structured document containing the version- and run-specific
information. This file is updated only a few times during the run and
is created on process zero only.  

The file is structured as an XML document, and is thus easily
parseable by any of the standard XML parsers, such as @command{xml.dom}
in @emph{Python}.

The first XML node @emph{configure} contains the configuration and
version information on the executable. The second node @emph{run}
contains information specific to each particular run: the time at
which run was started in two formats (@emph{time} and
@emph{date}), the number of MPI processes associated with the run
(@emph{nproc}), and the process specific information for each
process (@emph{processes}): the @emph{rank}, the
@emph{hostname} of the hosting node as the current process, the
process ID on hosting node, and the @dfn{multiplicity} - the total
number of processes assigned to the same hosting node as the current
process.  The greater than one value of the multiplicity indicates
that the resources of the node are shared with the other processes 
associated with the same run.

Shown below is the example of this
file, when created by the run on four processes in the @command{hello.sh}
example in @ref{Running gracos}

@ifhtml
@smallexample
@include prompt_info.atexi
@end smallexample
@end ifhtml

@iftex
@smallexample
@include prompt_info_short.atexi
@end smallexample
@end iftex

@end itemize
@noindent

@node Error Behavior
@subsection Error Behavior

If @command{gracos} halts before having reached the @code{return}
statement of @code{main} @emph{C}-function it does so abnormally.
These failures may be due to a variety of reasons. Excepting the batch
job resource limit and 

@table @var
@item Internal test failures
Internal test failures always trigger a call to one of the C-macros
@code{stop} or @code{stopm(@var{"ERROR MESSAGE"})} defined in
@command{zero.h}.  The macros produce message(s) similar to these

@smallexample
Exit (gracos.c:278) rank=@var{rank}

@r{or simply}

ERROR MESSAGE
Exit (gracos.c:278) rank=@var{rank}
@end smallexample
in both the standard output and error outputs of the process(es) where
the internal test failure(s) occurred. These messages show the
location of error in the source code and the rank of the process that
triggered the error message. The @emph{ERROR MESSAGE}, if present,
usually provides some hints on the nature of the problem; e.g.  that a
particular variable is used but not initialized.

@item Warnings

Rare warnings produce messages similar to those above but do not
result in program termination.

@item Memory leaks
A memory leak always indicates the presence of a bug, and resolving it
is a very high priority.  In @command{GRACOS} all allocated memory is
carefully tracked to ensure against memory leaks. If it does occur,
the following message appears at the end of standard output and
error outputs of proces zero:

@smallexample
!!!!!!!!!!!!!!!!!!!!!!!!!!!!!!!!!!!!!!!!!!!!!!!!!!!!!!!!!!!!
WARNING: memory leak on at least one process:
cntsum=4  memsum=20
@end smallexample
@noindent

if you observe a memory leak message please report it to @command{gracos}
authors.

@end table

@node File Input and Output with dataman
@section File Input and Output with @command{dataman}

@emph{GRACOS} supports a number of formats for primary (e.g. suitable
for a whole restart of the simulation) file input-output; they all are
implemented with the @command{dataman} command managing high level file
input-output. The data is converted when needed from the the units
particular to the datafile format into the units used by
@command{gracos}. Sometimes, as in case considered in Section @ref{Example: Santa Barbara Cluster Project}
such unit conversion requires same @command{gracos} variables to be set;
if so, not having them set will result in an error message indicating
which of the required parameters are missing.

A data file format can be @emph{static} or @emph{dynamic}.  The static
datafile format allows only a predefined number and type of the
simulation parameters be stored; while the dynamic format allows more
freedom.  In addition, a data file format can be @emph{distributed}
and @emph{serial}.  In the distributed File I/O format, more than one
process are allowed to write data on disks.  For the serial datafile
format the whole file I/O procedure is always performed on one process
only. In the parallel runs the processes exchange messages with the
process using MPI.

Below is the full description of the options; Section @ref{File Input and Output with dataman}
lists all the available datafile formats for @command{dataman}.

@table @code

@item @command{dataman} @var{OPTIONS....}
Load or save the data from or into a datafile(s).
@end table

The following table lists the available @var{OPTIONS}.
@table @code

@item @command{-mode}=@var{string}
Mandatory flag used to specify the required action. The possible
values of the string argument are @command{save},@command{load}, or
@command{delete}, indicating file output, input or deletion.

@item @command{-fmtid}=@var{string}
This is a mandatory flag. The data formats used for file I/O are
identified by their @dfn{Format ID}, the argument of @command{-fmtid}
specifies the @var{Format ID} (See below for the list of all the
supported formats). In the absence of this option, the code tries to
determine the format using the extension of the path given by the
@command{-path} flag argument. If both methods fail the run stops with an
error message, since the machine does not know which format to use for
file I/O.

@item @command{-path}=@var{string}
This is a mandatory flag; its argument provides the longest path
prefix that is a prefix of all the paths used for I/O in the current
instance of file input-output. The final file pathname will be
appended when necessary: for example, in parallel I/O the path is
appended with MPI ID of the process performing the I/O; in case when
many file paths are used on one process a proper extension used to
distinguish the path is added; See also the @command{-suf} and
@command{-prefixes} options.

@item @command{-suf}=@var{string}
Optional flag, used to append a string to the end of the file path.

@item @command{-prefixes}=@var{string}
When local disks are cross-mounted, this option can be used to modify
the leading parts of @command{-path}.  The argument of @command{-prefixes} is
a text file, readable on process zero, and is the list of the prefixes
in order of their process ids. The leading portion of the @command{-path}
argument is appended individually for each process whose process id is
@var{rank} with the content of the row number @var{rank} in this
file. If the file has less than @var{rank} rows the path is reset to
@var{NULL} for this process. The file should end with a newline.

@item @command{-spec}=@var{string}
The argument provides the format specification string and is used in
conjunction with the format ids for which such format specification is
applicable; see the description for each format below for documentation.

@item @command{-masseq}
If masses of all particles are the same it is not efficient to store
them in a datafile. If masses of particles are not stored in a
datafile used for particle data input, this option should be used to
initialize them to the same value.

@item @command{-sort}
Sort particles with respect to their particle ids before writing them
to disk. This option is applicable only for data output
(@var{-fmtid=save}) using serial data file formats and is silently
ignored when used in all other cases.

@end table

@node Physical Units
@subsection Physical Units

The particle data are normally expressed within @command{gracos} source
code in so called @dfn{gracos code units}: the particle positions are
comoving, in the units of @command{dx}; the velocities are proper, in the
units of (@command{h0}*@command{dx}/@command{a}), @xref{gracos Variables} for the 
definitions of @command{h0}, @command{dx}, and @command{a}.  The units of mass
are chosen so that the total mass of all the particles in the
simulation box is normalized to equal the number of PM density/force
mesh grid points @command{nx}*@command{ny}*@command{nz}.  For the case of
equivalent particles, one particle per each PM-mesh cell, each
particle has a unit weight in code units.

@node binpart: The Simplest Format
@subsection @command{binpart}: The Simplest Format 
In this dynamic serial format (@command{binpart} signifies the @var{Format
ID}) the particle data are stored in a single file in the most simple
possible way, controlled by the user. No header is used, and the data
is read or written in a simple sequence of records, one for each
particle in the simulation volume.

The user must provide the definition for the layout of data in the
record using the @dfn{format specification} option @command{-spec}
to the @command{dataman} command. The syntax for the @command{-spec} argument
string is a comma-separated list of tokens given below
@table @code
@table @code

@item @command{x}, @command{y} , or @command{z}
The x,y, or z- components of the comoving particles coordinates, in Megaparsec.

@item @command{vx}, @command{vy}, or @command{vz}
The x,y, or z- components of the proper peculiar velocity, in km/s.

@item @command{vxc}, @command{vyc}, or @command{vzc}
The x,y, or z- components of the comoving peculiar velocity, in km/s.

@item @command{m}
The mass of the particle in the solar masses.
@item @command{id}
The particle id, currently a 4 byte integer.

@item @command{xs}, @command{ys}, @command{zs}
@item @command{vxs}, @command{vys}, or @command{vzs}
@item @command{ms}
@item @command{gxs}, @command{gys}, or @command{gzs}
The coordinates, velocities, masses and gravitational accelerations,
in the @emph{GRACOS} code units, @xref{Physical Units}.
@end table
@end table

If the @command{id} specifier is not supplied on input, the particle ids
are assigned sequentially in order of their appearance in the file,
starting from 1.

The @command{binpart} format is already used widely due to its
simplicity. For example
@uref{http://t8web.lanl.gov/people/heitmann/arxiv/sb.html, The Santa
Barbara Cluster project} has adopted a variation of this format to
publish their particle data. Section @ref{Example: Santa Barbara Cluster Project} shows how, using the @command{dataman} to load the particle data in
their format directly into @command{gracos} and start N-body simulation
in the same session.

This format does not have any built-in header for holding any
variables. Some additional parameters are generally required if one
uses this format for input, e.g. for the domain decomposition. One has
to manually supply the missing information with @command{varset} before
the @command{dataman} command can be called successfully. If a required
variable is unset the run terminates with an error message indicating
the name of the required variable.  The @command{npartall} variable is
set automatically based on the input file size and the format
specification.

@node udf: Unified Datafile Format
@subsection  @command{udf}: Unified Datafile Format
This dynamic distributed format (@var{Format ID}: @emph{udf}) is the
perfect choice for easy simulation restart and runtime backups
automatically stores all the necessary @command{gracos} variables and
data, being also very fast due to the usage of local disk space (some
networking systems however do not have local disk spaces). The @emph{udf} format
is however somewhat frequently changed between different @command{gracos}
versions and is hard to be interpreted by other software. However, if
the data is written in this format one can always restart the
simulation and convert the data to other easily interpretable formats
such as @emph{binpart}, @xref{binpart: The Simplest Format}.

The @emph{udf} format is customizable and self-describing. It allows
the option for the arbitrary number of the output files not exceeding
the number of running MPI processes.  Each process writes at most one
output file.  This format is portable between the machines of the same
CPU-endian types. The datafiles are currently designed to store all
the code variables, particle data, and the refinement numbers of the
chaining mesh cell.

The @command{-spec} argument is used for file output only.  The argument
of @command{-spec} option is a string: a comma-separated list of tokens
with their subarguments, following the below syntax

@table @code
@table @code
@item @command{eq}
Request equal sized output files. The cost is slightly slower
completion due to the necessary interprocessor communications.  This
option is useful for use in clusters with local disc spaces on each
node; it ensures that the disc space gets used up uniformly.

@item @command{nfio:}@var{int32}
Integer (the default value is @var{Nproc}); option results in the
specified number of the output files. If option @command{-eq} is given,
these files will be equal sized. The actual data is not moved while
their copies are moved between the processes in the failure tolerant
way so as to accommodate the specified number of the output
files. This option should be used if one wants to reduce the number of
output files without compressing the data to the bottom processes. if
@command{nfio} is less than @command{nshr} (below) then the option @command{-eq}
is automatically set.  If @command{nshr} is less than @command{nfio} then
@command{nfio}=@command{nshr} is set automatically.

@item @command{nshr:}@var{int32}
Integer, less than or equal to @var{Nproc}; the option allows user the
ability to later restart the simulation on fewer number of processes
than the running number of processes (@var{Nproc}).  Repartitioning of
domain decomposition is performed so that the whole domain gets
divided evenly (weighed by the workload per cell) in the result
between the @var{nshr} (as opposed to @var{Nproc} as it normally is)
bottom processes before data is saved into the datafiles. The
simulation can be restarted in a separate @command{gracos} session using
a number of processes (that is equal or higher than @code{nshr}). Do
not use this option if you do not plan to restart the same simulation
on fewer number of processes.

@end table

@end table

@node tpm: Tree-Particle-Mesh format
@subsection @command{tpm}:  Tree-Particle-Mesh format
The static serial data format (@var{Format ID}: @command{tpm}) adopted by
the TPM code version 1.2, See
@url{http://www.astro.princeton.edu/~bode/TPM} is composed of 6 binary
files with the same basename and extensions @file{.px}, @file{.py},
@file{.pz} and @file{.vx}, @file{.vy}, @file{.vz}. They contain
identical information in the header and the corresponding component of
position or velocities of the particles in their main bodies.

@node p3mdat: Fortran p3m Datafile Format
@subsection @command{p3mdat}: Fortran p3m Datafile Format 
The static serial format (@var{Format ID}: @command{p3mdat}) used by the
p3m serial @emph{Fortran 77} codes. The data is saved onto one file
consisting of the header and the main body. The number of particles
that can be saved is limited in the format to the maximum numbed held
by an integer (@command{INT_MAX} as defined in the standard
@file{limits.h} @emph{C}-header). The error message will appear and
the run will stop if this limit is exceeded. 

Particle IDs are not stored. However one may use particle sorting
procedure to store particle IDs implicitly. Pass @command{-sort} option to
@command{dataman} and particles will be written into the disk in the
order of increasing particle IDs.  While loading from a @command{p3mdat}
datafile, particle ids are automatically set to the sequential number
(starting from one) in which the particle appears in the
datafile. In this way particle IDs are easily recovered.

The masses of particles are also not directly stored. The
@command{-masseq} option must be used to initialize the masses to the equal
value.  The @command{p3mdat} format is supported by the Serial
@emph{Fortran 77} codes that we are using for tests and verification
against the older @command{p3m2_93} codes.

The following table lists the data items stored in a p3m.dat header in
the order of their appearance in case.
@table @code

@item @var{long unsigned}
@emph{Fortran 77} conventional mark for the start of a binary File I/O record.

@item npartall
Same as @command{npartall} from @ref{gracos Variables}, but here
constrained to 32 bits. The number of particles can not
exceed @command{INT_MAX} in this format.
@item @command{nx}, @command{ny}, @command{nz}
      @xref{gracos Variables}.
@item @command{dx}, @command{epsilon}
      Expressed in the comoving Megaparsecs. 
@item @command{a}, @command{omegam}, @command{omegav}, @command{h0}, @command{dtsin}, @command{etat}, and @command{nstep}
      @xref{gracos Variables}.
@item ekin@var{, } egrav@var{, and } egint
      Same as in Section @ref{gracos Variables}, but expressed in the
physical units; the conversion factor is given in Section @ref{Physical Units}.

@item @var{long unsigned}
End of the @emph{Fortran 77}  record.
@end table

@node mwhite: Martin White format
@subsection @command{mwhite}: Martin White format

This format has at some point been adopted by @emph{Martin White} for
his codes.  Its original description follows
@quotation
The phase space data
files have to be in a specific format. First there is a 4-byte
integer, which should be 1.  Then an integer 20. Then a header which
contains: number of DM particles, number of SPH particles (0 in our
case), number of star particles (0), scale-factor of the output,
spline softening length in units of the box size.  The first 3 entries
are 4 byte ints, the last 2 are 4 byte floats. If you know a Plummer
softening, the spline softening is a factor 2.8 larger.  Then all of
the particle positions, 3x 4-byte floats per particle, in units of the
box length.  So all coordinates should run [0,1) and the order is
x0,y0,z0,x1,y1,z1, etc. After the positions are the velocities
vx0,vy0,vz0,vx1,vy1,vz1,...  These are stored in units of the Hubble
velocity across the box.  If you have them in km/s then divide by
a.H.Lbox.  Finally the particle ID number for each particle, as a
4-byte integer.  And that's it.
@end quotation

This format allows the following specification with @command{-spec}
option, whose argument should be a comma-separated list of tokens
given in table below
@table @code
@item @command{nfh:}@var{int32}
Allows one to use @command{nfh} files instead of just one (the default);
the files are still written/read on process zero. Files written with
some value of the argument of @command{nfh} should be read with the same
argument to @command{nfh}.
@end table

@node Example: Datafile Manipulations
@subsection  Example: Datafile Manipulations

N-body problems generally require storing many kinds of data on a disk
for various purposes such as to be able to restart the simulation from
a midpoint, or perform data analysis on particles in a separate run.
New datafile formats are routinely devised in the N-body community in
order to fit particular problems.

@command{gracos} offers a file interface that supports a few data
formats, @xref{File Input and Output with dataman}. The interface is
implemented through the @command{dataman} command, allowing the user to
load, save, or delete the data stored in dataliles, or make a
conversion between the supported formats. The list of the supported
file formats will be updated based on the need.

Most of the examples in this reference use @command{dataman} in one way or
another. Any of the supported datafile formats can be used, for
example, to start a full N-body simulation with @command{gracos} in the
same session.  The example in @ref{Example: Santa Barbara Cluster Project}
shows how to load data files using the format adopted by the Santa
Barbara Cluster Project.

In the densely commented example below we perform a few conversion
between various supported formats listed in @ref{File Input and Output with dataman}.

@example
@include gracos/gracos-examples/interface/fio.sh.c
@end example

@node Example: Santa Barbara Cluster Project
@subsection    Example: Santa Barbara Cluster Project

In our next example, we start the simulation with the Santa Barbara
Cluster Project (SBCP) initial conditions. These particle data initial
conditions must first be downloaded and unarchieved using the following
Unix-flavored shell commands

@smallexample
@var{SHELL$ }wget http://t8web.lanl.gov/cosmic/ic_sb128.tar
@var{SHELL$ }tar xvf ic_sb128.tar
particles_ic_sb128
input_sb
@end smallexample
@noindent
The data in the unpacked data files is written using the following
format, as quoted from the @emph{SBCP} webpage:
@quotation
@dots{} There are two files in the archives: particles_name, which
has the following content: x[Mpc], v_x[km/s], y[Mpc], v_y[km/s],
z[Mpc], v_z[km/s], particle mass[M_sol], particle tag. This file is
written in single precision and binary format.  @dots{}
All data are given in comoving units  @dots{}
@end quotation
@noindent
Note that the velocities are expressed in comoving km/s, which is unusual.

In order to load the datafiles into @command{gracos}, we use the above
description to set the @command{-spec} flag appropriately in the example
below, @xref{binpart: The Simplest Format}.

The example below is a script that should be executed in the same
directory as the above download procedure.  The important thing to
notice here is that the particle datafile alone does not completely
specify the simulation parameters in this case. Additional simulation
parameters must be set in order for @emph{GRACOS} to read the
particle datafile; this is because part of the necessary information
is supplied in the text file @file{input_sb}, some more information
(the initial redshift or the starting expansion factor @command{a}, etc)
are provided in the @emph{SBCP} web page; additionally,
@emph{GRACOS} uses its own units for data and domain decomposition
which depend on a few more simulation parameters such as
@command{epsilon} and @command{etat}. It is easy to find the list of the
necessary parameters since skipping any of the settings inevitably
results in a crash with the appropriate error message indicating which
of the parameters is missing. Another important thing is the use of
the @command{-spec} option to @command{dataman}. The format specification is
made consistent with the description quoted above. 

Next to loading the particle data from the SBCP initial conditions
data, the script executable below generates the power spectrum, and
the images of the density distribution simulation box as commented in
the script. According to the setting for @command{img_slabs} code
variable, four images are produced in slabs along the line of sight,
See @ref{image: Particle Data Imaging} for more examples and explanations on
the image parameters. 

We encourage the users to execute the below script (after finishing
the above procedure for downloading the SBPC data) to reproduce the
results.

@example
@include gracos/gracos-examples/problems/sbarb.sh.c
@end example

The image below shows the power spectrum measurement for matter
distribution in the particle data volume. The image data are generated
as the result of the @command{pmpower} call in the above example; the
data are written into file @command{pmpower.dat} and the image itself is
generated by typing the same @command{pmpower} (in @emph{MATLAB}) in the
same directory, @xref{pmpower: Power Spectrum Estimation}. The lower
half of error bars os longer than the upper one due to the
non-linearity of logarythm while we use linear values power spectra
for variance estimators.

@ifhtml @c .gif file
@sp 2
@image{gracos/gracos-examples/problems/sbarb/pmpower-sbarb,,,,jpg}
@sp 2
@end ifhtml
@iftex @c .eps file
@image{gracos/gracos-examples/problems/sbarb/pmpower-sbarb}
@end iftex

The following four images appear as the result of the @command{image}
call in the above example; the number of images is defined by the
@command{img_slabs} variable setting the number of image slab along the
line of sight.

@ifhtml @c .gif file
@sp 2
@image{gracos/gracos-examples/problems/sbarb/sbarb-0,,,,gif}
@image{gracos/gracos-examples/problems/sbarb/sbarb-1,,,,gif}
@image{gracos/gracos-examples/problems/sbarb/sbarb-2,,,,gif}
@image{gracos/gracos-examples/problems/sbarb/sbarb-3,,,,gif}
@sp 2
@end ifhtml
@iftex @c .eps file
@image{gracos/gracos-examples/problems/sbarb/sbarb}
@end iftex

@node integ: Running an N-body simulation
@section @command{integ}: Running an N-body simulation

The @command{integ} command starts the N-body simulation on the currently
loaded particle data. The acceleration mode is controlled by the
@command{acc} bitwise variable whose default value is ``full
acceleration'': @command{(@command{PM} | @command{PP} | @command{FPM} |
@command{FPP})}, @xref{Bitwise Variables}.

Timesteps are normally computed using a particular time stepping
criterion during an N-body simulation; the positions and velocities of
the particles are made synchronized at the certain points in time.
The output expansion factors specified using the @command{-aouts} option do
not generally coinside with the expansion factors corresponding to the
particle data synchronization points; when, during the integration
procedure, the next output expansion factor falls within the next
timestep, the value of this timestep is normally truncated to match
the output expansion factor to the roundoff error; See the
@command{disable_timestep_truncation} sub-option.

In addition to variables needed for domain decomposition, variable
@command{etat} must be set before calling @command{integ}. A reasonable
value for this variable is 0.05; setting this variable to lower values
can be used to increase time resolution of the simulation.

The following two options are available
@table @code

@item @command{-aouts}

Mandatory opton, a string containing the comma separated list of
output specifiers each formatted as @emph{A[:B]}, where @emph{A} is
either the floating point number setting the output expansion factor
at which to produce data output or the word @command{INF} to indicate
unlimited integration of particle trajectories. Optional column
separated modifier @emph{B} (the default value is @command{r}) may have
the following mnemonic values

@table @code
@item @command{r}
@emph{Restart output}, using the expansion of variable @command{out_restart}, @xref{Checkpoint Control Variables}.
@item @command{a}
@emph{Analysis output}, using the expansion of variable @command{out_analysis}, @xref{Checkpoint Control Variables}.
@item @command{b}
@emph{Backup output}, using the expansion of variable @command{out_backup}, @xref{Checkpoint Control Variables}.
@item @command{n}
No output
@end table

For example, command @command{integ -aout=0.1:a,1:ra} indicates that
the analysis output will be executed at expansion factor 0.1; both the
restart and analysis outputs will be executed at expansion factor 1.

The user may define actions performed at the restart, backup and
analysis outputs by setting the variables @command{out_restart},
@command{out_backup} and @command{out_analysis}, @xref{Checkpoint Control Variables}.

@item @command{-mode}
Optional argument. The default value is
``@command{-mode}=@command{scheme}:@command{DKD}''

@table @code
@item @command{scheme}:@var{string}
The @command{scheme} specifies the time integration scheme used for
particles. The allowed values are @var{@command{KDK}} for Kick-Drift-Kick
integration scheme and @var{@command{DKD}} for Drift-Kick-Drift
integration scheme; the default is @var{@command{DKD}}.
@item @command{reset_econ}
Reset the energy conservation measure
(@command{egrav}+@command{ekin}-@command{egint}) to zero at the beginning of
the run, by modifying @command{egint}.  The status of energy conservation
is normally something that accumulates from the initial conditions
produced by the particle initial conditions generator. The use of this
option leads to resetting the energy conservation history to zero, so
that any accumulated value during the integration run will have
resulted from energy computation occurred during the current run as
opposed to the cumulative value.

@item @command{disable_timestep_truncation}
By default, this option is disabled. The use of this option disables
timestep truncation needed to match the output expansion factors set
by the @command{-aouts} flag. For each output expansion factor the data is
saved at the end of the current timestep without timestep
truncation. This leads to preserving the natural, non-truncated values
for all time steps in the simulation at the inconvenience of having
slightly @emph{higher} output expansion factors than those specified
by the user.

@item @command{break}:@var{nsteps}
This option should not be normally used.  The mandatory positive
integer argument @command{nsteps} of this option gives number of
time steps for the simulation, after which the integration loop is
exited even if the aimed expansion factors specified by the
@command{aout} are not reached.

@item @command{mult}
Use multiple timestepping scheme (this option is not currently available).
@c The time steps are computed specifically for each particle in the
@c simulation volume. All the individual particle time steps match at the
@c certain points at which the whole simulation box is synchronized and
@c is suitable for the output.  @command{-aout} flag.

@end table
@end table

@node Checkpoint Control Variables
@subsection Checkpoint Control Variables

The expanded values of @dfn{output control variables} 
@table @code
@item 
out_restart
@item 
out_analyze
@item 
out_backup
@end table
yield the command to be executed at a checkpoint, @xref{integ: Running an N-body simulation}.

In addition to usual Tcl expansion, the following pattern expansion
takes place
@table @code
@item @@MODE@@
Expands to @command{-mode=save}, this pattern should be used in the @command{dataman} commands.
@item @@OUTID@@
In case of @command{integ}, expands to the id of the current output
specifier given in @command{-aouts} flag, @xref{integ: Running an N-body simulation}.

In all other cases expands to ``outid''.
@item @@LABEL@@
expands to ``r'', ``a'', ``b'', or ``n'' for restart,
analysis,  backup or no outputs, respectively.
@end table

@node checkpoint Command
@subsection @command{checkpoint} Command

If you wish to checkpoint at an arbitrary time, independently of the outputs
specified by the @command{-aouts} argument to @command{integ}, use  the @command{checkpoint}
command; Its usage is as follows
@table @code

@table @code
@item @command{checkpoint} -labels=@var{LABELS}
where @var{LABELS} is a string containing any combination of ``r'', ``a'', and ``b'', @xref{Checkpoint Control Variables}.
@end table
@end table

@node Automatic Backup Option
@subsection Automatic Backup Option

Variable @command{nchkpt} (See @ref{gracos Variables}), if set, indicates the
backup period, in seconds. For large integration times it may it may
be necessary to leverage the system failure risk with this option. The
timer is started at the beginning of a run with @command{gracos}. Once
@command{nchkpt} seconds are passed, variable @command{out_backup} is
expanded and its content is executed. No backups are performed if
@command{nchkpt} variable is unset.

Given extremely large integration times, many instances of the backup
are executed. To avoid unnecessary file accumulation, old backups are
removed at each new backup output, at this point the
@command{@@MODE@@} pattern is expanded into @command{-mode=delete} and
@command{@@OUTID@@} expands to the value @emph{OUTID}-1.

@node image: Particle Data Imaging
@section @command{image}: Particle Data Imaging

@command{image} is a @command{gracos} command implementing the imaging of
particles with variaus choices for the projection volume, direction
and other imaging parameters.  The maximum dynamic range of 256 colors
is currently allowed for imaging. The color 0 in the palette
corresponds to the lowest threshold value below which no contrast can
be discerned from the image. Similarly, the color 255 corresponds to
the highest threshold in the plotted value.

A number of code variables must be set before this command is called.
One can avoid having to set them all manually by using the @command{-set-defaults}
flag:

@smallexample
gracos
@end smallexample
to generate a reasonable guess to all the required variables from the
current data, See Section @ref{Example: Particle Data Imaging} or
@ref{Example: Tabulated Transfer Function} for the examples of the use of this command.

The following is the complete list of the required variables
@table @code
@item  @command{img_cmap}      
An integer within the range 0-4, specifying the colormap to be used
for images. We found the colormaps 1 and 4 to be most useful.  The
colormap 4 produces black and white images.

@item  @command{img_dim}
An integer 1,2 or 3, stands for the direction (@var{x},@var{y} or
@var{z}) of the axis used for the projection of the simulation box.
The plane of the generated image will span the remaining two
dimensions as shown in the following figure.

@float
@ifhtml 
@sp 2
@image{img_dim,,,,gif}
@sp 2
@end ifhtml
@iftex
@image{img_dim}
@end iftex
@end float

@item  @command{img_cx}, @command{img_cy}, @command{img_cz}
Specify the center of the main image box within the simulation volume.

@item  @command{img_bx}, @command{img_by}, @command{img_bz}
Specify the size of the main image box.

@item  @command{img_slabs}     
An integer greater than zero, setting the number of images in the
output of the @command{image} command. If this number is 1 (the default),
then the content of the whole main image box, whose dimensions are
given by (@command{img_cx}, @command{img_cy}, @command{img_cz}), and
(@command{img_bx}, @command{img_by}, @command{img_bz}) is projected into the
image plane. If this number is greater than one, then the main image
box is split into the same number of slabs of equal thickness and the
output images are the projections for each individual slab in the
sequence.

@item  @command{img_pixw}
Specify the output image width, in pixels. The height is found
automatically so as to observe isotropic scaling in all directions
within the image plane.

@item  @command{img_rsmoo}
Image smoothing length.

@item  @command{img_rbl}, @command{img_rbr}
The two floating point variables set the contrast. 

@end table

@node Example: Particle Data Imaging
@subsection    Example: Particle Data Imaging

In the example below, we first load the particle data from a sample
datafile (supplied along with the source code distribution) using
@command{dataman}, then we produce the images of the distribution of
particles in the simulation box using different colormaps, projection
dimensions, image sharpness and image dimensions.

@example
@include gracos/gracos-examples/problems/imaging.sh.c
@end example

The following runtime administration file must be placed into the same
directory under the name @command{imaging-admin.py}, @xref{Runtime Administration}.
@example
@include gracos/gracos-examples/problems/imaging-admin.py.c
@end example

The four images below have been produced by running the above
script. They are originated in the left-to-right, top to bottom order
as files @command{im1.gif}, @command{im2.gif}, @command{im3.gif}, and
@command{im4.gif}.

@ifhtml @c .gif file
@sp 2
@image{gracos/gracos-examples/problems/imaging/im1,,,,gif}
@image{gracos/gracos-examples/problems/imaging/im2,,,,gif}
@image{gracos/gracos-examples/problems/imaging/im3,,,,gif}
@image{gracos/gracos-examples/problems/imaging/im4,,,,gif}
@sp 2
@end ifhtml

@iftex @c .eps file
@image{gracos/gracos-examples/problems/imaging/images}
@end iftex

Note that the third image is particularly overexposed. This is
because, the default value (@command{10.}) for the @command{img_rbr} code
variable, giving the upper threshold for the brightness range sampled
by the available colormap, is too low for this overdense halo; so that
most pixels on the image fall above the threshold (this default value
is set by the @command{image --set-defaults} in the above
script). In order to tune the threshold to the brightness distribution
of the pixels we have to use runtime administration technique
described in @ref{Repeatable Administration} because it is generally
not possible to determine the correct value for the upper brightness
threshold before the actual data is projected onto the image plate. The
instance of runtime administration is invoked for the last image, at the @command{image} 
checkpoint in
which the upper brightness threshold is set exactly to the value of
the brightness threshold @command{Out_cmlup} exceeded by a low
fraction (0.003) of the total number of pixels in the image. This is an 
example where runtime administration technique is useful.

@node pmpower: Power Spectrum Estimation
@section @command{pmpower}: Power Spectrum Estimation

Power spectrum is a very frequently used statistics of density field
with many analytical fits and expressions given in the literature and
is directly related to two point correlation function.

Typing in @emph{gracos} prompt

@smallexample
pmpower [--path=@var{FILE} ] [--nbins=@var{NBINS}]
@end smallexample
@noindent
yields the binned power spectrum of the currently loaded particle data
written in the text file @code{@var{FILE}} (the default is
@command{'pmpower.dat'}); @code{@var{NBINS}} is the number of bins used
within the range of the wavevectors of the main Brillouin zone.

The data can be visualized by typing the same command in
@emph{MATLAB}:

@smallexample
>> pmpower [(@var{FILE})]
@end smallexample
@noindent
using the same default name for @code{@var{FILE}}.

In @ref{Example: Santa Barbara Cluster Project} and @ref{Example: Tabulated Transfer Function} we
show an example of thus produced image of the power spectrum.

The internal procedure for computing the power spectrum is as follows. 
First, particle mass is interpolated on the grid using the TSC density
assignment scheme; then the Fourier transform is performed. The value
of the power spectrum is estimated for each grid point in the fourier
space as proportional to the square of the Fourier transform of the
density at the same gridpoint. The binned values of the power spectrum
are found by averaging of these values for each wave vector bin, up to
the Nyquist frequency.  The rms deviation (the error bars) of the
binned values are found by dividing the rms deviation of the
measurements by the square root of the number of measurements for each
bin (the Central Limit Theorem).

@node Initial Condition Generators
@section Initial Condition Generators

In general, any data simulation or data analysis task is preceded by
the specification of the current state of the system. In our case, we
specify the data such as the particle positions, velocities etc;
followed by the parallel domain decomposition.  

This section shows the examples of running a large simulation with
@emph{GRACOS} for various choices of the initial conditions; these
examples are presented in the form of readily executable scripts.
Cosmological initial conditions are currently generated using the
Zeldovich approximation with four methods of specifying the initial
power spectrum of density perturbations. This section lists all the
currently available methods of automatically generating the initial
conditions; both rectangular and glass initial particle grid can be
used.

Once the initial conditions are generated: the data can be analyzed or
evolved, and saved into the datafiles using one of the methods
described in Section @ref{File Input and Output with dataman}. Once saved into
datafiles, the data can be loaded evolved and analyzed again in a
separate @emph{GRACOS} scripting session.  In general, to start a
cosmological simulation, the user is only required to provide the
cosmological and other simulation parameters. @emph{GRACOS} starts
with those parameters in the input and finishes with the complete
realization of the required data, usually the particle data at the
given expansion factors.

@node grafic: the Initial Conditions Generator
@subsection @command{grafic}: the Initial Conditions Generator

This command implements an extension of the @command{grafic1} initial
conditions generator originally presented as an @emph{Fortran77} code
by @emph{E. Bertschinger}, @xref{Serial Fortran N-body Codes}.

Section @ref{sec_cgf} provides us
with the test scripts that should be used to verify the compatibility of
the result of the @command{gracos} command with the output of original
@emph{Fortran} codes.

The following @command{gracos} variables must be set before this command
is called

@table @code
@item @command{nx},@command{ny},@command{nz},@command{dx}
Mandatory
@item @command{epsilon}
Unless the option @command{-nodom} is used
@item @command{omegam}, @command{omegav} and @command{h0}
Unless the option @command{-lingerdat} is used
@item @command{a}
If this variable is set, it defines the starting expansion factor of
the universe. Either set this variable or use the @command{-sigstart}
option described below; error message terminates the run when
attempting to use both ways.
@end table

@vskip 10pt

@noindent
The @command{grafic} command accepts the following options

@vskip 10pt

@table @code

@item @command{-icase}=@var{int32}
Select the type of initial power spectrum (matter transfer function T(k)).
The Integer takes the following values:
@table @code
@item 0
Use tabulated values for the power spectrum. See the @command{-table} option.
@item 1
Use the transfer command tabulated in a @file{linger.dat} file.
See the @command{-lingerdat} option.
Set the @command{omegam}, @command{omegav} and @command{h0}
code variables from the @file{linger.dat} file header.
@item 2 
Use the BBKS analytical approximation for the T(k).
@item 3
Use T(k) = 1 (the scale-free case).
@end table

@item @command{-pnorm}=@var{float32}
Mandatory option. A floating point number, sets the desired normalization at a = 1:

@table @code 
@item for @command{icase} = 0
Mandatory option. @command{pnorm} equals @command{-sigma_8} at the present epoch (only
negative @command{pnorm} is cureently allowed).

@item for @command{icase} = 1,2
@table @code 
@item if @command{pnorm} > 0
@command{pnorm} equals Qrms-ps/micro-K
@item if @command{pnorm} < 0
@command{pnorm} equals @command{-sigma_8} at the present epoch.
@end table

@item for @command{icase} = 3
@table @code 
@item if @command{pnorm} > 0
@command{pnorm} equals k^3 * P(k), where P(k) is an unfiltered value,
e.g. with the Gaussian or Hanning filters, of the initial power
spectrum, and k = pi/@command{dx} (or k = pi/@var{rsmoo}, if the
@command{-rsmoo} option (down the list) is set and @var{rsmoo} >
@command{dx}).
@item if @command{pnorm} < 0
@command{pnorm} equals @command{-sigma_8} at the present epoch.
@end table
@end table

@item @command{-lingerdat}
Provide the name of the lingerdat file (for @command{-icase} value 1 only).

@item @command{-enable-baryons}
@emph{GRACOS} currently does not have the capability for gas N-body
simulations. By using this option, however, one generates the velocities
for baryonic matter, just as they would be generated with the original
@emph{grafic1} code, @xref{Serial Fortran N-body Codes}; this data,
however is not used for CDM or any other particle
initializations. Using this option in conjunction with
@command{-ic-vel-files} allows one to save baryonic data into files and
analyze it.

@c @item @command{-offvelb}=@var{offvelb}
@item @command{-offvelc}=@var{int32}
@item @command{-offvelb}=@var{int32}
Integer parameters for offsetting baryon and CDM velocity fields by
0.5 grid spacing. The argument should be +/-1 or 0; the default value
is +1, just as the default in the @emph{grafic1} code. The origin of
the baryon velocity field will be
@command{offvelb}*(0.5,0.5,0.5)*@command{dx} and similarly for @command{offvelc}.

@item @command{-sigstart}=@var{float32}
This option is mandatory unless the @command{a} code variable is set
before calling @command{grafic}.

@var{sigstart} sets the initial linear density fluctuation amplitude
on the ultimate sub grid scale (or the approximately starting
expansion factor).  You must either set this option or preset the
@command{a} code variable before calling @command{grafic}.
@c For large grids (> 10^7) one may wish to decrease
@c this; for small refinement grids (< 10^6) and multiple levels one may
@c wish to increase it. 

If the power spectrum is sufficiently steep so that k^2 P(k) converges
slowly at low wavenumbers (consider for example the scale-free free
power spectrum with @var{ns}=-3) the internal romberg integrator
fails; use the @command{-trlowk} option to avoid this problem.

@item @command{-ns}=@var{float32}
Long-wave spectral index (scale-invariant is @var{ns} = 1.), a
floating point number; defines the shape of the primordial power
spectrum. This option is mandsatory when @command{-fnl} is used or
@command{-table} is used with @command{choice}:@command{transfer_function}.

@item @command{-seed}=@var{uns32}
Mandatory option. The random number @command{seed}, - a positive integer smaller than
2^32-5=4294967291.

@item @command{-power-plot}=@var{kmin}:@var{kmax}
Use this option to make a plot of the power spectrum used to generate
the initial conditions. Floating point numbers @var{kmin} and
@var{kmax} define the range of the wave vectors (in the units of
inverse comoving Megaparsecs) covered by the plot. The text file
@command{power.dat} is produced in the working directory as the initial
conditions are generated. The first row in this file shows @var{ns}
and @var{pnorm}; the remaining rows are

@smallexample
k/h,    P(k)*h^3, @r{and} 4*pi*P(k)*k^3
@end smallexample
where h equals @command{h0}/100; k is the wave vector in the units of
inverse comoving Megaparsecs; and P(k) is the power spectrum in
comoving Megaparsecs cubed. 

In addition to the data we described each rows may contain additional
data appended at the end of each line. If this is the case, the
specification for the additional data is outside the scope of this
reference.

@item @command{-unscalable}

Using this option gives the advantage that the particle realization
does not depend on the number of processes used while running
@emph{gracos}.  Thus one is able to completely replicate the initial
particle data realization of the original @emph{grafic-2_101} code,
@xref{Serial Fortran N-body Codes}, @xref{sec_cgf}. The expense is somewhat slower performance
because the last process has to replay the complete sequence of random
numbers almost for the entire box in order to arrive to its starting
position in the sequence of random numbers.

Rundom number generators are very fast, and we would generally
recommend using this option for any runs smaller than 800^3 particles.

@item @command{-ic-vel-files} 

Generate files @command{ic_deltab}, @command{ic_velbx}, @command{ic_velby},
@command{ic_velby}, @command{ic_deltac}, @command{ic_velcx}, @command{ic_velcy},
@command{ic_velcy}; replicating the output of the original
@emph{grafic1} code.  If @emph{gracos} is ran in parallel, then for
each of the @var{file} of the above list, a set of files
@var{file-rank} are generated instead, where @var{rank} goes from 0 to
@var{Nproc-1}. This option may be useful if you are familiar with the
traditional @emph{grafic1} output and, perhaps, have some data
analysis routines that take these file as the input. It may as well be
useful as a way to extract the generated baryonic initial conditions
in case you have chose option @command{-enable-baryons}. Note if you are
using the @command{-fnl} or @command{-glass} extensions (See the following
table): their effect does not extend to the generated @command{ic_*}
files.

@item @command{-disable-domain-decomposition}
After having completely computed the overdensities, and velocities
while calling the @command{grafic} command, @command{gracos} by default uses
this information to completely initialize the particle data array and
perform the parallel volume domain decomposition, sending particles
between the processes; the latter procedure results in the complete
setup, such that the commands @command{pmpower} or @command{integ} may be
used, for example without much additional work. The domain
decomposition may take a considerable amount of time for very large
problem sizes, but is necessary for the followup with most of the other
@command{gracos} commands. Using the @command{-disable-domain-decomposition}
option will disable particle domain decomposition.

This option combined with the @command{-ic-vel-files} and
@command{-unscalable} for @command{icase}>0 will just result in the very close
replication of the result as computed by the serial @emph{grafic1}
code, @xref{Serial Fortran N-body Codes}. 

@end table

@vskip 20pt

@noindent
@emph{If only the above options are used (excluding the case
@command{icase} = 0) the produced result will be identical to the
roundoff error to the equivalent serial run with the
@command{grafic-2_101} code, @xref{Serial Fortran N-body Codes}.  The
following list adds important optimization options and a few new
options compared to those provided by the original
@command{grafic-2_101} code.}

@vskip 10pt

@table @code

@item @command{-table}=@command{path}:@var{string},@command{choice}:@var{string}
This option is applicable for the case @command{icase}=0 only.  Both
@command{path} and @command{choice} suboptions are mandatory.  @code{path}
gives the name of the two column ASCII file, whose first column is the
wave vector expressed in inverse comoving Megaparsecs. The
interpretation of the second column depends on the other suboption. If
@code{choice} equals ``@command{power_spectrum}'' then the
second column gives the primary power spectrum to be used to setup the
initial conditions; if @code{choice} instead equals
``@command{transfer_function}'' then the second column yields the
transfer function. The normalization of the data in the second column
is irrelevant as the power spectrum is normalized intrinsically.

The rows in file @code{path} can be given in an arbitrary order.
However, the wavevector values given in the table should cover the
complete wavevector range used in the code, since no automatic
extrapolation in the wavevector space is performed. A table with a
limited wavevector range should be manually appended with one or two
entries to extend the wavevector range as needed at both high and low
wavevector values.

@item @command{-disable-hanning}

Disable the use of Hanning filter, particular to the @command{grafic1} code.

@item @command{-fnl}=@var{float32}

Apply non-Gaussianity correction to the generated initial conditions,
@command{fnl} is the second order correction to the primordial proper
potential @mathenvd{\phi,phi}.  The motivation for the @emph{fnl}
model of initial conditions, pioneered, among others, by
@emph{E. Komatsu} is to explore the possibility that the initial
conditions in our universe are slightly non-gaussian, in which the
true initial conditions potential @mathenvd{\phi,phi} can be expanded
in terms of the gaussian potential @mathenvd{\phi_0,phi0} as
@mathenvd{\phi = \phi_0+f_{nl}(\phi^2-\langle \phi_0^2 \rangle ),phi =
phi0 + fnl ( phi^2 - < phi0^2 > )} when expressed in the units of the
speed of light squared. The @command{-ns} option to @command{grafic} must
also be set.

@item @command{-rsmoo}=@var{float32}
Apply the gaussian filter; the power spectrum used for the generation
of the initial conditions is multiplied by @mathenvd{\exp(-k^2\; {\tt
rsmoo}^2),exp(-k^2 rsmoo^2)}; @mathenvd{rsmoo,rsmoo} is expressed in
comoving Mpc.

@item @command{-trlowk}
Limit the integration to the frequencies above
pi/(@command{dx}*@command{nx}), while computing power spectrum normalization
to find starting expansion factor using @command{-sigstart} option; see
the description above for the @command{-sigstart} option.

@item @command{-glass}=@command{seed}:@var{uns32},@command{nexp}:@var{float}
Generate glass initial conditions, instead of those on the regular
grid. The @command{seed} suboption is mandatory and is used to initialize
the initial poisson distribution, necessary for making the glass
particle distribution. The @command{nexp} suboption is the number of
expansion factors to evolve the poisson distribution with negative
gravity requested to arrive to the target glass distribution.  The
quelity of the latter improves with the increase @command{nexp}; if no
option is given, this number is automatically initialized to the
theoretical estimate @mathenvd{ nmax^12, nmax^12 }, where
@mathenvd{nmax,nmax} equals the maximum of @command{nx}, @command{ny} and
@command{nz}; even though the theoretically estimated numerical value of
@command{nexp} is rather large, the nbody simulation will proceed rapidly
under the repulsive gravity. One may run the power spectrum tests like
those shown in @ref{Example: generating the glass Distribution} in order to show that
the generated glass is reliable.

@end table

@node Example: Tabulated Transfer Function
@subsubsection Example: Tabulated Transfer Function

In this section, we show how to start an arbitrarily large simulation
starting from the values of cosmological parameters and tabulation for
the power spectrum P(k) of matter density perturbations; we present
the readily executable @command{gracos} script below.

Let us assume that we are having a table of the power spectrum P(k):
the values of @mathenv{P_i}, tabulated at the wave numbers @mathenv{k_i};
where @mathenv{P} is expressed in arbitrary units (consistent for
different @mathenv{i}) and @mathenv{k_i} is expressed in the units of h/(Mpc
comov), where h is the present time Hubble constant, in the units of
100*km/s/Mpc. The power spectrum table should cover the entire range
of the wave vectors used during the @command{gracos} run because
@command{gracos} does not do any extrapolation of power spectrum in this
case. The @emph{GRACOS} installation provides a sample of such power
spectrum table for use in this example of the @command{gracos-examples}, but you should
use your own power spectrum table.

At the end of this subsection we provide the complete details on how
we generated this power spectrum table using @emph{CMBFAST}).

First, we define the @emph{WMAP3} cosmological parameters (we remind
that the ``.'' at the beginning of the line in @emph{bash} means
sourcing the path as if it were included into the script). The
@command{gracos-config} executable is a part of @emph{GRACOS}
installation and is used to locate files provided with @emph{GRACOS}
installaion; type

@smallexample
cat `gracos-config info pkgdatadir`/CosmoParams/WMAP3
@end smallexample
@noindent
in your shell prompt to see the @emph{WMAP3} parameters used.

The following line in the script defines the path to the power
spectrum file that we generated using @emph{CMBFAST}, also the part
of the installation for the purpose of this reference; again, the user
should generally use their own power spectrum table as the power
spectrum used in our example is in not suitable for a general
situation (as we see at the end of this section). 

Next, we specify the random number seed @command{seed} used to
generate the sequence of random numbers used by the initial conditions
generator and the size of the particle mesh @command{Ngrid} the user
is free to change this number. The @command{Ngrid} parameter is low
for the purpose of this example; the user is free to change this
number as needed to change the resolution of the N-body simulation (if
you significantly increase this number you should also increase the
number of processes for the @command{mpirun} command line
prefix).  Next, we arbitrarily set the simulation boxsize @command{Lbox},
expresses in comoving Megaparsec (not the Megaparsec/h!) and compute
the cell spacing @command{dx} using @command{bc} (the @command{bc} is a
simple calculator, usually provided with any standard Unix
distribution).

Next, we set the Plummer force siftening length @command{epsilon}
(expressed in the units of @command{dx}) and the integration timestepping
parameter @command{etat}; the @command{0.05} is a good value; using lower
value will result in more timesteps needed to complete the
simulation. Finally, we provide the comma separated list of of output
specifiers with @command{aouts} (only one in our example); each
formatted as @var{aout[:modifier]}, where @var{aouts} is the output
expansion factor and the modifier is a character specifying what to do
at the given expansion factor, See @ref{integ: Running an N-body simulation} for more explanations.

Now that we have defined all the simulation parameters, we begin
writing the body of the @command{gracos} script into a temporary file
whose name is generated with a standard @command{mktemp} Unix command.

First we set the necessary simulation parameters for @command{gracos}
with a set of @command{varset} commands, See @ref{Variables},
@ref{Variables} and @ref{gracos Variables}. The list of all the necessary
parameters for this run can easily be found by the trial and error
method (this is indeed how we knew which parameters are needed to be
set): @command{gracos} terminates with a useful error message any time a
required variable is uninitialized.

Next, we generate the particle initial conditions by calling the
@command{grafic} command using the @command{icase}=0 selection,
@xref{grafic: the Initial Conditions Generator}. The arguments of this command completely specify al
the additional information needed to generate the particle realization
of the given power spectrum. The @command{-power-plot} flag provedes the
option to output the power spectrum plot in the given range of wave
vectors.

Next, we call @command{image} to generate and image of the simulation
box, @xref{image: Particle Data Imaging}. The @command{-set-defaults} flag automatically
sets up a default selection for all the image parameters. The result is a
new @emph{GIF} file @file{i.gif} in the current working directory.

Next, using @command{dataman} (See @ref{File Input and Output with dataman}) we save
the initial conditions data into file using the distributed file
input-output, @xref{udf: Unified Datafile Format}; call @command{integ}
to start an N-body simulation, make an image using @command{image}, and
call @command{dataman} again to save the data in the evolved state.

Finally, we make another image of the simulation box, producing file
@file{f.gif}.

The entire @command{gracos} session is written into a temporary file
@command{$TmpFile}, whose name is passed as the argument to the
@command{-i} option to @command{gracos}; @command{gracos} is ran in parallel
with @command{mpirun} command on four processes. 

The user is free to increase the problem size by increasing
@command{Ngrid} and the number of processes, up to the system
resource limitation.

@example
@include gracos/gracos-examples/problems/pktable.sh.c
@end example

@heading Imaging

Shown below are the images in files @command{ini.gif} and @command{fin.gif}
(the latter is at redshift zero) produced by the @command{image} command
just before and after the particles are evolved with the @command{integ}
@emph{GRACOS} command as shown in the script above.  If we makes a
sequence of these images as we evolve the box we can actually observe
that the large scale structure is smoothly transformed from that of
shown on the left panel to that on the right (please See
@ref{Runtime Administration} for the available options on executing arbitrary
commands, such as @command{image} for making images, in the runtime).

@ifhtml 
@sp 2
@image{gracos/gracos-examples/problems/pktable/ini,,,,gif}
@image{gracos/gracos-examples/problems/pktable/fin,,,,gif}
@sp 2
@end ifhtml
@iftex @c .eps file
@image{gracos/gracos-examples/problems/pktable/inifin}
@end iftex

@heading An Example of Run Time Administration

It is generlly possible to make the images on the fly, using the
technique described in @xref{Runtime Administration}. Here we
demonstrate an example of this. By copying the example below into file
@command{inc.tcl} within the working directory @emph{during the run}, we
execute the imaging command (see the example) as soon as the runtime
control reaches the @command{integ} checkpoint; the current images are
then created and this file is automatically erased.  We do not show
the image generated; because they apparently depend on the exact time
at which the @command{inc.tcl} file is created. You should try this copy
command while running the above script.

@example
@include gracos/gracos-examples/problems/admin-image.tcl.c
@end example

@heading The Power Spectrum

Now, typing 
@smallexample
pmpower('pmpower-fin.dat')
pmpower('pmpower-ini.dat')
icpower
@end smallexample

@noindent
in @emph{MATLAB} in the same directory yields the plot below of the
power spectrum of density perturbations as a function of the wave
number, spanning the range of @mathenvd{[2 \pi /(@command{nx} \cdot
@command{dx}), [2 @command{pi} /(@command{nx} * @command{dx}) },@mathenvd{
\pi/@command{dx}), @command{pi}/@command{dx}) }
@c
as shown by the two vertical dash-dotted green lines.
@c

The red line at the bottom is the power spectrum used for the
generation of particle initial conditions. The oscillations of this
curve above the Nyquist frequency are caused by the use of the Hanning
filter and should not worry the reader, as they are outside the
wavevector range used. Passing the @command{-disable-hanning} option to
@command{grafic} disables the Hanning filter; the black dashed line
shows the unfiltered power spectrum.

The lower broken blue line is the measurement of the power spectrum
based on the unevolved particle positions (See the first
@command{pmpower} call in the @emph{GRACOS} script above).  The
deviations from the red curve are due to the following effects: 1)
Cosmic variance at low wavenumbers; 2) the effect of the randomness of
the current particle realization given the power spectrum; 3) coarse
graining near the Nyquist frequency.

The upper blue broken line is the power spectrum of matter
perturbations in the evolved state (the second @command{pmpower} call)
and the black dash-dotted line shows the initial power spectrum given
by the red line corrected multiplied by the linear growth rate
characteristic for the expansion from the starting expansion factor to
redshift zero, the final point of our simulation. The systematic
deviation of the black dash dotted line with the upper blue broken
line is due to the nonlinear evolution.

@float
@ifhtml @c .jpeg file
@sp 2
@image{gracos/gracos-examples/problems/pktable/power,,,,jpeg}
@sp 2
@end ifhtml
@iftex @c .eps file
@image{gracos/gracos-examples/problems/pktable/power}
@end iftex
@end float

@node Example: lingers Initial conditions
@subsubsection Example: @command{lingers} Initial conditions

In this section we show how to start an arbitrarily large simulation
of CDM with @command{gracos} directly, starting from the values of the
cosmological parameters using the included @command{lingers} package, a
subpackage by @emph{C. Ma, E. Bertschinger, 1995} included into
@emph{GRACOS}, @xref{Serial Fortran N-body Codes}.  @command{lingers} is automatically
installed during the @ref{GRACOS Package Installation}. The transfer function tabulated
in both redshift and k-space is included into the @command{linger.dat}
file automatically generated using @emph{lingers}.

As we know, the baryonic, CDM and matter transfer functions agree to
within the percent level close to @math{z=0}. For the power spectrum
normalization we use sigma8 of the power spectrum of all matter
linearly evolved to @math{z=0}. @mathenvd{\sigma_8, sigma8} is
measured as an integral with respect to @math{k} over the power
spectrum at a given @math{a}. Usually it is assumed that the power
spectrum grows as @mathenvd{D_+, D+}. With lingers however this
dependence is intrinsic and is not factored out.

If we plot the power spectrum at a=0.02 (normalized at z=0), then we go
back by the intrinsic growth factor and by the D+ in the two cases.  In
the result we get a good agreement between the curves. The ehu curve
goes right through the middle of the matter power spectrum
oscillations. But the CDM particle displacements are generated with the
CDM power spectrum, not matter. The CDM and MATTER power spectra, while
almost matching exactly each other at z=0, are very far apart at a=0.02
(about 12

Now, if we assume that Om=Oc+Ob, we imply that the power spectra of Oc
and Ob exactly match each other at a=0.02. The correct thing to do
would be to generate the displacements using the matter power spectrum
at astart.

The evolution factor of CDM is different than that for MATTER.

@example
@include gracos/gracos-examples/problems/lingers.sh.c
@end example

Typing @command{icpower} and @command{pmpower} now, in @emph{MATLAB},
within the working directory yields the power spectrum plot. The
resulting plot is distinguishable from the plot in the previous
section, by the presence of the baryonic oscillations.

@ifhtml @c .jpeg file
@sp 2
@image{gracos/gracos-examples/problems/lingers/power,,,,jpeg}
@sp 2
@end ifhtml
@iftex @c .eps file
@image{gracos/gracos-examples/problems/lingers/power}
@end iftex

@node Example: BBKS Initial Conditions
@subsubsection Example: BBKS Initial Conditions

An analytical approximation to the initial power spectrum by Bardeen,
Bond, Kaiser and Szalay named thus the @emph{BBKS approximation} is 
directly applicable in @command{gracos}.

File @command{gracos-examples/problems/bbks.sh} 
In this section we show how to start an arbitrarily large cosmological
N-body simulation starting with values for a few cosmological and
simulation parameters, and assuming that the initial power spectrum is
described by the BBKS approximation.

The user should be familiar with the content of the script shown below
from the discussion in the two previous sections. The script below
does not make calls to executables rather than @command{gracos}: using
the BBKS approximation is reserved for @command{icase}=2:
@c @example
@c @include gracos/gracos-examples/problems/bbks.sh.c
@c @end example

Typing @command{icpower} in @emph{MATLAB} within the working
directory yields the power spectrum plot below. As we see by
comparison with the power spectrum plot generated with @emph{CMBFAST}
and @emph{lingers}, the @emph{BBKS} approximation is slightly
underestimates power for lower wave vectors, while slightly
exaggerating for larger @math{k}'s.

@node Example: Scale-Free Initial Conditions
@subsubsection Example: Scale-Free Initial Conditions

Sometimes it is useful to start a simulation using scale-free initial
conditions with P(k) proportional to k^ns.

This section shows how, given a few required parameters, to start an
arbitrarily large simulation with scale-free power spectrum of matter
perturbations.  The @command{icase}=3 option to @command{grafic} is
dedicated to scale free initial conditions. The script below is very
similar to the one in the previous section, which, similarly to the
present case, uses an analytical expression for the initial power
spectrum of CDM density perturbations.

@example
@include gracos/gracos-examples/problems/scalefree.sh.c
@end example

@node poisson: Poisson Distribution
@subsection @command{poisson}: Poisson Distribution

This routine may be a good starting point if you want to write your own
initial conditions generator that can not be expressed as an
extension of @command{grafic}.

This command generates @var{N} Poisson-distributed particles across
the box.  The syntax is
@smallexample
poisson -seed=@var{seed}
@end smallexample
@noindent
where @var{seed} is a random number seed.

Variables @command{nx}, @command{ny} and @command{nz} must be set.  If
@command{npartall} code variable is unset, then the total number of
generated particles along with the @command{npartall} code variable are
set to @command{nx}*@command{ny}*@command{nz}; otherwise @var{N} equals
@command{npartall}.

The internal procedure is as follows.  The volume domain decomposition
is performed so that approximately equal volumes are assigned to each
process. The Poisson random number generator is called on each process
with exactly the same random number seed so that each process scans
the identical sequence of randomly distributed particles.  If a
particle belongs to a given process, that process assigns that
particle to itself; the entire seqence of particles is thus assigned
to all @var{Nproc} processes.

The Poisson random number generation is relatively fast. However, the
procedure described above, is not scalable.  In principle, by
generating particles with varying random number seed on each process,
one can make this procedure scalable: each process would try to assign
each particle in the sequence by a certain mapping from the uniform
distribution within simulation volume into the irregular volume
particular to each process.  The Jacobian of the mapping must equal to
one; we are not certain at this point however on how to setup such a
mapping, suggestions are welcome however. Hence, we are currently
employing the above non-scalable procedure which is safe against any
tiling effects.

@node Example: Poisson Distribution
@subsubsection Example: Poisson Distribution
In the example below we generate the power spectrum of the Poisson
particle distribution (See @ref{poisson: Poisson Distribution}) and
compare it with the analytic slope (zero).
@example
@include gracos/gracos-examples/tests/powersp/poisson.sh.c
@end example

Typing 
@smallexample
pmpower
yline( log10(1/(2*pi)^3) )
@end smallexample 
in @emph{MATLAB} yields the image below.  The theoretical power
spectrum of the Poisson distribution is the constant 1/(2*pi)^3, shown
with the horizontal slash-dotted line; this prediction is mostly
observed in the plot generated. The large error bars on large scales
(low wavevectors) are due to cosmic variance or low statistical sample
of these modes; while at the very high wavenumbers there is a
characteristic upturn, whose nature is not entirely obvious.
The plot shows the bounds of the validity of the power spectrum
generated with @command{pmpower}.
@ifhtml @c .jpeg file
@sp 2
@image{gracos/gracos-examples/tests/powersp/poiss,,,,jpeg}
@sp 2
@end ifhtml
@iftex @c .eps file
@image{gracos/gracos-examples/tests/powersp/poiss}
@end iftex

@node Example: generating the glass Distribution
@subsubsection Example: generating the glass Distribution

This command generates @var{N} Glass-distributed particles across the
box.  The syntax is
@smallexample
glass -seed=@var{seed} -nexp=@var{nexp}
@end smallexample
@noindent
where @var{seed} is a random number seed and @var{nexp} is the number
of expansion factors requested to evolve glass. Further in this
Section @ref{Example: generating the glass Distribution} we provide an example of using this command.

In @emph{GRACOS}, the procedure for making glass distribution is defined
as follows: first we distribute particles uniformly across the box,
using the Poisson distribution; then we evolve the particles with
negative gravity. By measuring the power spectrum, we find that the
power spectrum, initially dominated by the shot noise, settles down to
form the single power law power spectrum characteristic for glass
@mathenv{P(k)}.

@dfn{Glass distribution} for particles, introduced by
@emph{S. White}, is similar to the distribution of moleculae in a
liquid, (we know from the statistical mechanics background that the
glass material is liquid). @emph{GRACOS} provides the utility for
generating glass distribution and glass initial conditions, See
@ref{Example: generating the glass Distribution} for more details.

As a good test of @emph{GRACOS} one can run a simulation to produce
the glass structure and measure the power spectrum during the inversed
gravity run. Shown below, as usual is the @emph{GRACOS} script that
does the job. This is also, the first script in this reference that
makes use of the runtime administration technique to measure the power
spectrum on the fly (See further below for more details).
@example
@include gracos/gracos-examples/tests/powersp/glass/generate.sh.c
@end example

The above script as is will just generate the assumed glass particle
distribution. But in order to test wether a reliable glass
distribution is indeed generated we must measure the power spectrum
and compare it with the power spectrum of glass distribution which is
shown theoretically to be a fourth power low of the wave number
@math{k}. We can just place the @command{pmpower} command at the end of
the above script. But it would be useful to see how the power spectrum
really evolves as we evolve particles with negative gravity for this
purpose we make use of the repeated runtime administration method
described in Section @ref{Repeatable Administration}. Shown below is
the @emph{Python} runtime administration file that should be placed
into the same directory as the script above. The runtime
administration script conditionally invokes a set of @emph{GRACOS}
commands to execute in a @emph{GRACOS} subsession.  The script mostly
consists of the useful comments but is essentially just a few lines
long.

@example
@include gracos/gracos-examples/tests/powersp/glass/generate-admin.py.c
@end example
Using the Pythons flexibility it is quite easy to make any other
selection in the above script, say, for example, to make power
spectrum measurements each specified number of timesteps, make images
of the particle distribution, save the data into files, or any other
operation allowed by the set of the @emph{GRACOS} commands. One can
modify the content of this file @emph{during the run} to change the
administration mode as needed. Using runtime administration thus
yields a very powerful method of control.

Now, if you run the @command{bash} shell script at the beginning of
this section, and issue the following commands in @emph{MATLAB} (in
the same directory)

@smallexample
>> pmpower('glass-0.dat')
>> pmpower('glass-22.dat')
>> pmpower('glass-62.dat')
>> pmpower('glass-116.dat')
>> pmpower('glass-149.dat')
>> pmpower('glass-194.dat')
>> pmpower('glass-fin.dat')
@end smallexample
@noindent 
you will see the following plot of the power spectra at the timesteps
specified in the runtime administration script.

@ifhtml @c .jpeg file
@sp 2
@image{gracos/gracos-examples/tests/powersp/glass/glass,,,,jpeg}
@sp 2
@end ifhtml
@iftex @c .eps file
@image{gracos/gracos-examples/tests/powersp/glass/glass}
@end iftex

The upper curve on the plot just shows the white noise of the initial
Poisson distribution (glass generation starts with uniformly
distributed random poisson particle distribution, See @ref{Example: generating the glass Distribution}
for more details). The white noise curve is in agreement with the
theoretically predicted zero white noise power slope. As we evolve
matter with the negative gravity to form glass distribution the matter
becomes less and less chaotic as particles move under the repulsive
force from each other; the power spectrum thus settles down on all
scales. Since the theoretical slope of the glass distribution is
approximately the fourth power of the wave vector, the modes with the
highest wave number hit their theoretical glass power spoectrum curve,
in agreement with the behavior shoun, where for the timestep 62, for
example, the small scale perturbations are smoothly distributed along
the theoretical curve and the largest scale perturbations are still
settling down. Finally, as we evolve matter for 15 powers of the
expansion factor, the matter settles down on all scales sampled by the
simulation desolution. We have thus reached the accurate distribution
on all scales.

@node Glass Extension
@subsection Glass Extension

This option is under development; please, email @emph{GRACOS} authors
if you have any specific suggestions.

@node halo_finder: Halo Finder
@section @command{halo_finder}: Halo Finder

This command is currently under development; email @emph{GRACOS}
authors to indicate your interest in this command or any suggestions
you may have.

@node Tables of Reference
@section  Tables of Reference

This section provides the tables of all the available @command{gracos}
commands, variables and modes. A short description (a @dfn{docstring})
is supplied for each entry; the detailed descriptions may be found in
more specialized sections of this reference.

@node gracos Commands
@subsection @command{gracos} Commands

The following table lists all the available @command{gracos} commands and
options. The internal @emph{Tcl} procedures and the procedures
sourced in the @command{gracos} startup file @command{init.tcl} (See
@ref{Commands}) are not listed here.

@include commands.t

@node gracos Variables
@subsection @command{gracos} Variables

The following table lists of all the @command{gracos} variables for
version @value{VERSION}. The description of bitwise variables also
lists the available modes.

@include variables.t

@node Additional Programs and Libraries
@chapter Additional Programs and Libraries

@node gracos-pkgs: Required Package Installer
@section @command{gracos-pkgs}: Required Package Installer
The fastest way to assure the proper installation of the required
packages listed in @ref{Required and Recommended Packages} is probably
using the @command{gracos-pkgs} script-executable (only 70+ lines)
included with @emph{GRACOS} source code distribution, located under
the @command{bin} subdirectory of the @emph{GRACOS} top source code
directory tree. This script-executable is just a sequence of simple
shell commands sufficient for completing all the necessary
downloads and installations for the required packages,
@c
and is entirely self-sufficient, - there are no prerequisites except
for the bare Unix-flavored computer system, an internet connection and
the reliability of our server storing the packages.

@c
The @command{gracos-pkgs} executable can be copied to any machine and
should simply be executed (no root password is needed). The automatic
procedure that follows, involving all the downloads and installations
takes about ten minutes. In the result, all the packages required for
@emph{GRACOS} installation are automatically downloaded under
@command{$HOME/downloads.tmp} and installed under @command{$HOME/local} with
no other system paths involved. After the installation procedure has
completed for all the packages, the @command{$HOME/downloads.tmp}
directory is automatically removed.
@c
There are a few important points to be aware of before you actually
run the executable:
@itemize @bullet
@item
Your home directory should have at least 24Mb permanently available
space for installations and about 200Mb temporarily available space
for package compilation products
@item
Beware if you are already using either the @command{$HOME/local} or
@command{$HOME/downloads.tmp} directory paths for any other purpose; -
please modify the script if necessary.
@item
The @command{configure} commands used to build the makefiles for the
compiled packages assume your current environment settings.  Beware if
you are using your special user specific settings for such
environment variables as @command{LDFLAGS} or @command{CPPFLAGS}. The
user environment is imported by the package specific configure scripts
invoked within @command{gracos-pkgs}. It is safest to run
@command{gracos-pkgs} on the default unmodified user system environment.
@item
Even if some or all of them are already installed on your system, it
is all right to install them again. Not every systemwide @var{FFTW}
installation, follows our selection for the @var{FFTW}
@command{./configure} flags, for example.
@item
If you invoke the executable on the same system (including the CPU
type) more than once, all the previous effects of this executable are
automatically erased.  If you invoke the executable on the same file
on the machines of different CPU types (for example, when your home
directry is NFS mounted across many machines) the effects of running
this executable will add up since the produced installation path is
specific to the CPU type (@command{i686}, @command{x86_64}, etc) of
the target machine.
@end itemize
@c
The installation locations of the packages are
@command{$HOME/local/arch/`uname -m`/@var{PACKAGE_NAME}}. Architecture
specific directory paths are absolutely required when installation
paths are accessible via NFS (Network File System) by the machines of
different CPU type. Generally, the installation procedure should be
repeated for each architecture type.

@node gracos-clean: Directory Cleanup Utility
@section @command{gracos-clean}: Directory Cleanup Utility

As with any other package, the users who look into the source code of
will have to deal with many files automatically generated as various
utilities or applications (for example @command{configure} or
@command{make}) are ran.  When the number of these files is large it
generally becomes difficult to distinguish the automatically generated
files from those manually updated. A user who modifies a source code
will generally want to package the directory into another
@command{.tar.gz} file and give it to a collaborator; in order to do that
he or she will have to first clean the directory of all the
unnecessary files, such as the executable, the object files or even
the @command{Makefile}s, which are automatically generated by the
@command{configure}.

The @emph{Python} @command{gracos-clean} utility we are introducing does
not have any prerequisites, except for the @command{.rules} files (see
below) properly maintained by the user, which is usually not hard to
do.

@example
@include gracos-clean-help.atexi
@end example

In the list below we classify all the possible procedures that locally
automatically generate new files; the characteristic patterns of the
these files can be grouped by the types of the generating
procedure. The @command{.rules} text files present in most @emph{GRACOS}
source code directories are used to specify the group selection.
Because the first three of the procedures in the list are not as
portable as the the more standard @command{configure} and
@command{make} procedures that follow, the portable @emph{GRACOS}
package distribution includes the files generated by the former.

@itemize @bullet
@item 
The @command{bootstrap} procedure (label '@var{b}'). 

In @emph{GRACOS}, we run @command{./bootstrap} in the top
@emph{GRACOS} source code directory; the package distribution
includes all the files generated by this procedure.

@item 
The generated source code files (label '@var{g}'). 

In @emph{GRACOS}, we run @command{./srcgen.sh} in the top source code
directory; the package distribution includes all the files generated
by this procedure.

@item 
Generated document files (label '@var{d}').

In @emph{GRACOS}, we use the directions in @command{doc/src/README}
under the top source code directory; the package distribution includes
all the files generated by this procedure.

@item 
The @command{configure} procedure (label '@var{c}').

The procedure is described for @emph{GRACOS} in @ref{Configuration}.

@item 
The @command{make} procudere (label '@var{m}').

The @command{make} procedure is described for @emph{GRACOS} in
@ref{Compilation and Installation}.

@item 
Running @command{gracos} or other executables, label '@var{r}'.

In @emph{GRACOS}, the procedure is to run any of the examples in
@ref{Reference to gracos Executable}.

@item 
Any other procedure introduced by a user (label '@var{u}').

@end itemize

@node gracos-config: Utility for Linking Programs with GRACOS
@section  @command{gracos-config}: Utility for Linking Programs with @emph{GRACOS}

@command{gracos-config} is a utility for linking programs with
@emph{GRACOS}.  By installation procedure (See @ref{Installation Procedure}), the
@command{PATH} shell environment variable directly points to the
location of the @command{gracos-config} executable. Now, using this
executable allows one to immediately retrieve the locations for any
other components of @emph{GRACOS} installation (the location of
include files and libraries, package data, etc). @command{gracos-config}
is used in most of the cases presented in @ref{Reference to gracos Executable}. Typing
@command{gracos-config --help} yields the following message:

@example
@include gracos-config.atexi
@end example

The @command{info} argument to @command{gracos-config} is used to retrieve
the standard locations; the values of the following
@command{VAR}iables are currently displayed by typing
@command{gracos-config info}:

@smallexample
@include  gracos-config-vars.atexi
@end smallexample

@noindent
See Section @emph{Variables for Installation Directories} in 
@uref{http://www.gnu.org/prep/standards/standards.html, GNU coding standards}
for explanations.

@node GRACOS Library
@section @command{GRACOS} Library
A number of special purpose auxiliary sub packages were written to for
applications with @command{gracos} and packaged as the so called
@dfn{GRACOS library} 
@itemize @bullet
@item
@command{libgracos.a}
@item
@command{libgracos_mpi.a}
@end itemize
(the paths are lower-cased for simplicity).  The libraries and the
corresponding header files get installed (unless the
@command{--disable-libs} option is used for @command{configure}) during
@command{make install} into one of the installation directories.

The following table lists the C-headers for the sub packages whose
compiled object files are merged to create the @command{libgracos.a} and
@command{libgracos_mpi.a} libraries. 

@include subpackages.t

The second column shows the use of MPI in the sub packages.  The
sub packages marked 'Serial' are compiled with a serial @emph{C}
compiler and are included into both @command{libgracos.a} and
@command{libgracos_mpi.a}. 

The sub packages marked 'Joint MPI' are compiled twice: with serial and
MPI compiler, using different @emph{C} preprocessing directive to
separate group in and out the calls to @emph{MPI} functions in the
source code; these packages are also included into both libraries
providing both serial and parallel implementations of their
declarations.

The sub packages marked 'MPI' contain
unavoidable @emph{MPI} library function calls. They are compiled with
@command{mpirun} and are included only into the @command{libgracos_mpi.a}
library.

@node gracos-rw: Cosmographic Quantities for FLRW Cosmology
@section @command{gracos-rw}: Cosmographic Quantities for FLRW Cosmology

@example
@include gracos-rw-help.atexi
@end example

@node gracos-ehu: Eisenstein-Hu Transfer Function
@section @command{gracos-ehu}: Eisenstein-Hu Transfer Function

A well-established fit for the matter transfer function (TF) by Daniel
J. Eisenstein & Wayne Hu yields the precision of matter transfer
function within a few percent level across wide range of the
cosmological parameters. For convenience, @emph{GRACOS} installation
includes the executable, - the generator of the transfer function.

@example
@include gracos-ehu-help.atexi
@end example

@node gracos-hist: Utility for Plotting Histograms
@section @command{gracos-hist}: Utility for Plotting Histograms

@example
@include gracos-hist-help.atexi
@end example

@node Serial Fortran N-body Codes
@section Serial Fortran N-body Codes
The @command{src}/@command{f77} subdirectory in @emph{GRACOS} contains the
serial @emph{Fortran 77} N-body codes written in 1993-1994 that have
been well known to the cosmological community (especially the
@command{grafic-2_101} and @command{p3m2-93} packages) and have served as
the development base for @emph{GRACOS}. The complete list is shown
below, these codes are installed as a part of @emph{GRACOS} package;
they are used for testing and verification purpose in @ref{sec_cgf}
and for generating the matter transfer function, @xref{grafic: the Initial Conditions Generator}.

The following is the complete list of the Fortran packages

@table @code{}

@item @command{grafic-2_101}
Installed executables: @command{grafic1} and @command{lingers}. 

@command{lingers} generates the transfer function and/or the initial
particle velocities from a given power spectrum of matter density
perturbations. Only minor differences from the original version,
available at @url{http://web.mit.edu/edbert/grafic-2.101.tar.gz};
except for the new @command{random9} random number generator that
fixes the portability bug found in the original version.

@item @command{p3m2-93}

Installed executables: @command{iniftab} and @command{p3m2-93}.

@command{iniftab} is used to generate force tables, @command{p3m2-93} is
used to run an N-body simulation using the generated force tables.
The @command{p3m2-93} code accepts files written in the @ref{p3mdat: Fortran p3m Datafile Format}
as the initial conditions and outputs the particle data in the same
format at the universe expansion factors passed to the standard input
by the user.

@item @command{ic2dat}

A C-code converting the velocities of the particles stored in the
@command{ic_vel} files produced by the @command{grafic-2_101} code into the
particle data file suitable for input for @command{p3m2-93} executable,
@xref{p3mdat: Fortran p3m Datafile Format}.

@item @command{swapb}
Convert the @command{float32} and @command{int32} data from big
endian to little and vice versa.

@end table

The @command{grafic-2_101} and @command{p3m2-93} packages distributed as the
part of @emph{GRACOS} slightly differ from their original versions. A
very limited number of changes were introduced, See
the @file{ChangeLog} files.  These changes do not introduce any
errors: the output of a run with the original version of the codes is
identical to the equivalent run accomplished with the modified
versions; the latter however introduce some additional options and
minor bug fixes.

@node bits
@section   @command{bits}

@include bits.t

@node System Tests
@section System Tests

This chapter describes a few non-standard system tests that have been
successfully used for identification of some particular system-related
problems running @command{gracos}. The executables are installed during a
regular @emph{gracos} installation, @xref{Compilation and Installation}. If your
system fails on any of these tests, then you should either reconfigure
@emph{gracos} with some specific @command{./configure} options to work
around limitations, or replace the faulty hardware. We hope that the
need for this chapter will diminish as the Linux systems improve over
time.

@node ibtest: System Call Portability Test
@subsection @command{ibtest}: System Call Portability Test

Some networking systems are known to have
@command{fork}/@command{system} call portability issues for parallel
applications.  @emph{Infiniband}, is the example: a parallel
application using system calls terminates randomly without an error
message, leaving the impression that the errors are due to erroneous
handling of dynamic memory allocation. This problem has initially been
submitted as a bug report to the TACC computer center and according to
@emph{Karl W. Schulz} of the support staff, as of January 1, 2007
there is no available fix.

@command{ibtest} is the utility for testing your system for this issue.
The following usage applies
@example
@include ibtest.atexi
@end example
@noindent
To test your networking system, run @command{ibtest} on
your system in parallel on four processes. The job should complete
within a few minutes. The standard output of process zero indicates
the test result. If it is positive you should not use
@command{--disable-system-calls} for @emph{GRACOS} configuration.

@node memtest: Hardware Memory Test
@subsection @command{memtest}: Hardware Memory Test

@command{memtest} is a testing utility for memory hardware on your
cluster and can be used to test memory on all the nodes of your
cluster simultaneously. The following usage applies

@example
@include memtest.atexi
@end example

Run @command{memtest} in parallel on all the nodes available in your
cluster.  The message shown in the standard output of process zero
indicates the hostnames of the nodes (if any) that have faulty memory
chips. 

The memory is tested by allocation of the total memory fraction given
by the @command{-f} option on all nodes, filling it with a prescribed
sequence of data and later verifying that the sequence of data as
written in the machine is authentic. 

The default value used for @command{--mem-frac} option is sufficiently
low to avoid swapping, increase this number cautiously to improve the
quality of the test result. Also, use larger value for
@command{--number-iterations} than the default at your discretion using
larger numbers increases the intrinsic number of iterations and
therefore also increases the quality of the diagnostics.

Generally, only the manufacturer supplied diagnostic tools can
reliably test the RAM chips on your system. Running manufacturers
diagnostics tools is often hardware specific and running it in
parallel may be tricky. @command{memtest} on the other hand is portable
and will diagnose only serious problems if they exist.
@c
The probability of a failure of any one memory RAM is quite low but is
considerable for large clusters of computers.  We suggest running
memtest from time to time to ensure no serious hardware problems.

The need to write this testing utility originated when one of our
early @command{gracos} runs crashed repeatedly with an error message
indicating a segmentation fault. The error message always appeared on
one of the nodes, which lead us to believe that the problem was caused
by a faulty memory chip as it was wearing down. @command{memtest} has
been used to prove this and was used to identify many more faulty
memory chips on our cluster.

@node Installation Procedure
@chapter Installation Procedure

@node System Requirements
@section System Requirements

@emph{GRACOS} is ported to most @emph{UNIX} flavored systems, except
for the @emph{Mac OS}, whose native C-compiler does not support
@command{ntptimeval}. It may be possible to compile @emph{GRACOS} on
this system using a different C-compiler (either @emph{gcc} or
@emph{icc}), however this has not been tested.

@node Required and Recommended Packages
@section Required and Recommended Packages
@emph{GRACOS} installation @emph{requires} the following freely
downloadable open source packages to be installed on your system

@itemize @bullet
@item
A Message Passing Interface (MPI) implementation must be
installed; the list of freely available implementations
includes @uref{http://www.lam-mpi.org,LAM}, @uref{http://www-unix.mcs.anl.gov/mpi/mpich/,MPICH},
 and @uref{http://www.open-mpi.org,Open MPI}.

@item
The @uref{http://www.fftw.org,FFTW} package @uref{http://www.fftw.org/fftw-2.1.5.tar.gz,fftw-2.1.5}
must be installed, configured with an @code{--enable-mpi}
@code{--enable-float --enable-type-prefix} options to enable support
for MPI, single precision mode and type prefix specification for the
@abbr{FFTW} header files. 

@item
@emph{Tcl} library @uref{http://www.tcl.tk/man/tcl8.4,version 8.4}
must be installed.

@end itemize

The following packages are @emph{recommended}
@itemize @bullet
@item
@emph{MATLAB}; @emph{GRACOS} provides a few @emph{MATLAB} scripts
to analyze various data from the datafiles produced in the
output.  Those scripts, located upon finishing the installation
procedure at @command{@var{$PREFIX}/matlab} (See @ref{Environment Setup}
for the definition of @var{$PREFIX}), can easily be rewritten for
other plotting software.
@end itemize

@noindent

@node GRACOS Package Installation
@section @emph{GRACOS} Package Installation

@emph{GRACOS} installation procedure below follows the standard path
widely adopted in the open source software world.  Once the package is
installed it is not necessary to keep the source code directory;
keeping it may be useful however for reference or revision purposes.

If you have any problems with the GRACOS installation procedure below please
consult the @uref{http://www.gracos.org/faq/index.html,GRACOS FAQ} webpage.
Then, if that does not help please let us know.

@node Configuration
@subsection Configuration

The @emph{GRACOS} installation procedure starts with the conventional

@smallexample
./configure [@var{OPTION}]... [@var{VAR=VALUE}]...
@end smallexample
@noindent
assuming that the required packages are already installed.  The
setting of @command{configure} @var{VAR}iables is probably simpler via
the use of shell environment variables rather than the
@command{configure} command line (type @command{./configure -h} for more
information).  One should make careful considerations for the settings
for @var{OPTION}s and @var{VAR}iables before proceeding with the
@command{configure} command. We provide useful guidelines below.

First, one has to make a choice of the installation path for
@emph{GRACOS} and specify it with the @command{--prefix} option to
@command{configure}.  It is recommended to install @emph{GRACOS} into
a unique directory by using
@command{--prefix}@command{=}@var{@dots{}SPECIFY}@dots{}@command{/gracos-@value{VERSION}}.
If you skip the @command{--prefix} @command{configure} option the default
systemwide setting (@command{/usr/local}) will be automatically used
and you will be required to have a root password for install.

Second, one has to make sure that the required packages are found by
@command{configure}.  The @command{PATH} shell environment variable must
be set appropriately so that the location of the compiled MPI
executables (usually @command{mpicc} and @command{mpirun}) is automatically
detected. If the location of any of the required @emph{Tcl} and
@emph{FFTW} packages is not standard, the @command{LDFLAGS} and
@command{CPPFLAGS} configure variables must be set accordingly.

Third, one has to make a wise decision for the choice of the
@emph{C}-compiler flags, particularly the optimization options. 
@c
Setting @command{CFLAGS="-Wall -O3 -fno-strict-aliasing"} is a good
initial guess. If you are using the @command{icc} compiler, you
might be able to use @command{'-fast'} istead of @command{'-O3'}.
@c
If you plan running a massive N-body simulation, it may very well be
worth researching the compiler documentation in order to choose the
most efficient optimization options.
@c
For example, if you are using the @command{icc} compiler, read the
``Optimization Levels'' and ``Automatic Processor-specific
Optimization'' sections in the manual page and find options for
optimizing by @emph{vectorization}. Vectorization yields the runtime
speedup by a significant fraction (roughly 20-30
of slightly increased compilation and installation time.

You may switch between different C-compilers by using the @command{CC}
@command{configure} variable.

Finally, some important non-standard @command{configure} options are
listed below (type @command{./configure -h} for the complete list of
@command{configure} options).

@c
Using @command{--enable-tests} option results in production of a number
of light weight executables used for testing in the subpackages at the
cost of slightly increased installation time. Those executables may
serve as the hello world examples for the users who link their codes
against @emph{GRACOS} libraries.

@c
The @command{--disable-slibs} option disables compilation of the serial
libraries. The @command{--disable-libs} options disables the installation of
all libraries and header files at @command{make install}.

@c
The @command{--disable-system-calls} @command{configure} option must be
set on the systems that have @command{fork/system} support issues.  If
you do not use @emph{Infiniband} networking system in your cluster
(ask your system administrator) you are not likely to have this
problem as this is the only system of the currently known to us having
this problem. Use @command{ibtest} utility at any time to test your
system for this issue, @xref{ibtest: System Call Portability Test}.

@node Compilation and Installation
@subsection  Compilation and Installation

After the @command{configure} command has finished
successfully, type

@smallexample
make
@end smallexample
@noindent
to compile and, if succeeded,

@smallexample
make install
@end smallexample
@noindent
to install all the @emph{GRACOS} components on your system. 

@node Environment Setup
@subsection  Environment Setup
Once @command{make install} is finished, the user should append the @command{PATH}
and @command{MATLABPATH} environment variable in their shell profile
appropriately. For example, the @command{sh} or @command{bash} shell
users should put the following into their @command{~/.bashrc} file

@smallexample
export PATH=@var{$PREFIX}/bin:$PATH 
export MATLABPATH=@var{$PREFIX}/matlab:$MATLABPATH
@end smallexample
@noindent
where @command{@var{$PREFIX}} is normally the argument of the @command{--prefix}
option used for the @command{./configure}; the @command{tcsh} or
@command{csh} users should instead put

@smallexample
setenv PATH @var{$PREFIX}/bin:$PATH 
setenv MATLABPATH @var{$PREFIX}/matlab:$MATLABPATH
@end smallexample
@noindent
into their @command{~/.tcshrc} or @command{~/.cshrc} files (whichever
is applicable, or ask your system administrator).

@node Uninstallation
@subsection  Uninstallation

@emph{GRACOS} can be uninstalled either by typing @command{make
uninstall} within the source code directory or, if you did specify a
unique installation path for @command{@var{$PREFIX}}, by invoking @command{rm -rf}
for the installation path.

@node Source code Directory Cleanup
@section  Source code Directory Cleanup

As a side effect of the package installation procedure, many files are
automatically generated throughout the source code directory tree,
making it heavy sized and sometimes confusing for the users who may wish to
work with the source code. Type @command{gracos-clean --initial} anywhere
within the source code directory to bring all the subdirectories to
the initial pristine state of the package; See @ref{gracos-clean: Directory Cleanup Utility}
for more options and details.

@chapter Authors and Copyrights

The @command{gracos} package is written by
@itemize
@sp 1
@item
Alexander Shirokov, (CITA/MIT)
@item
Edmund Bertschinger (MIT)
@sp 1
@end itemize

@verbatim
Copyright (C) 2002-2007 Alexander Shirokov and Edmund Bertschinger

This program is free software: you can redistribute it and/or modify
it under the terms of the GNU General Public License as published by
the Free Software Foundation, either version 3 of the License, or
(at your option) any later version.

This program is distributed in the hope that it will be useful,
but WITHOUT ANY WARRANTY; without even the implied warranty of
MERCHANTABILITY or FITNESS FOR A PARTICULAR PURPOSE.  See the
GNU General Public License for more details.

You should have received a copy of the GNU General Public License
along with this program.  If not, see <http://www.gnu.org/licenses/>.

Software Included in GRACOS:

Hilbert Curve implementation: files hilbert.h and hilbert.c
	    Copyright 1998, Rice University

GIF maker:  based on the original files wgif.c, compress.c and cmap.h
	    Copyright 1989, 1990, 1991, 1992, 1993 by John Bradley

M4 Macros:  files gse_get_version.m4 and get_gse_version
	    based on the original files lam_get_version.m4 and get_lam_version
            Copyright (c) 2001-2006 The Trustees of Indiana University.  
            Copyright (c) 1998-2001 University of Notre Dame. All rights reserved.
	    Copyright (c) 1994-1998 The Ohio State University.  All rights reserved.

	    file acx_mpi.m4 by Steven G. Johnson
	    Copyright 1997-1999, 2003 Massachusetts Institute of Technology

Transfer Function Fitting Formula: file ehu-power.c
	    based on the original files power.c by
	    Daniel J. Eisenstein & Wayne Hu, Institute for Advanced Study

@end verbatim

@chapter Acknowledgements

Alexander Shirokov would like to thank Neal Dalal for suggestions that
helped to develop this work.

This material is based upon work supported by the National Science
Foundation under Grant No. 0407050.  Any opinions, findings, and
conclusions or recommendations expressed in this material are those of the
author(s) and do not necessarily reflect the views of the National Science
Foundation.

This work was financially supported by the Canadian Institute for
Theoretical Astrophysics (CITA) and the Natural Sciences and
Engineering Research Council of Canada (NSERC).

@c @printindex
@bye